# Software Design Document, Testing, and Deployment and Configuration Management

## Unified University Inventory System (UUIS)

Prepared for : Imaginary University of Arctica

Version 1.0

Submitted to: Mr. Serguei Mokhov

By: Team1, COMP 5541, Winter 2010

# Members of Team1:

| Name | Email |
|---|---|
| Abirami Sankaran | a_sankar@encs.concordia.ca |
| Andriy Samsonyuk | a_samso@encs.concordia.ca |
| Maab Attar | m_att@encs.concordia.ca |
| Mohammad Parham | m_parham@encs.concordia.ca |
| Olena Zayikina | o_zayik@encs.concordia.ca |
| Omar Jandali Rifai | o_jandal@encs.concordia.ca |
| Pavel Lepin | p_lepin@encs.concordia.ca |
| Rana Hassan | ra_hass@encs.concordia.ca |

# TABLE OF CONTENTS

## LIST OF TABLES

## LIST OF FIGURES

# 1. Introduction

The project is to create a unified inventory system for all the faculties of the Imaginary University of Arctica. The system should be able to allocate inventory to the authenticated users, keep track of changes, and be accessed only by the authorized personnel.

This design document presents the designs used or intended to be used in implementing the project. The designs described, follow the requirements specified in the Software Requirements Specifications document prepared for the project.

## 1.1. Purpose

The purpose of this document is to present a detailed description of the designs of the Unified University Inventory System, created for the Imaginary University of Arctica. Firstly, this document is intended for the programming group in Team 1, to use the designs as guidelines to implement the project. Equally, this document is also for the team's instructor, Mr. Serguei Mokhov, as it fulfils one of the requirements of the project. Lastly, this document could be used for designers who try to upgrade or modify the present design of the inventory system.

## 1.2. Scope

This document gives a detailed description of the software architecture of the inventory system. It specifies the structure and design of some of the modules discussed in the SRS. It also displays some of the use cases that had transformed into sequential and activity diagrams. The class diagrams show how the programming team would implement the specific module.

## 1.3. References

The user of this SDD may need the following documents for reference:

IEEE Standard 1016-1998, IEEE Recommended Practice for Software Requirements Specifications, IEEE Computer Society, 1998.

Mokhov, S. (2010). Selected Project Requirements. In *Concordia*. Retrieved from http://users.encs.concordia.ca/~c55414/selected-project-requirements.txt

Team 1, 2010. Software Requirements Specification, UUIS. Last modified: Apr. 29, 2010.

Team 1, 2010. Software Test Documentation, UUIS. Last modified: Apr. 29, 2010.

### 1.4. Overview

This document is written according to the standards for Software Design Documentation explained in "IEEE Recommended Practice for Software Design Documentation".
Sections 3 – 5 contain discussions of the designs for the project with diagrams, section 6 shows samples of UI from the system, and section 7 contains the class diagrams. The appendices contain the setup and configuration needed for the UUIS, a list of functions that are implemented in this version, and that are to be implemented in future version, and a list of tools and environment used in the entire project, along with the time contribution of team members. The appendices also include the test report and test cases.

## 2. Design considerations

### 2.1. Assumptions

The user of the inventory system is aware of basic operations of a computer and web pages. The user also understands the standard terms used for operation.

### 2.2. Constraints

The system is built accessible only through university's website. The system is implemented using Java and JSP technologies.

### 2.3. System environment

The web based unified inventory system is designed to work on all operating systems. The system is accessible through any laptop and desktop, that is connected to the IUofA server. It is accessible at all times.

### 2.4. Design methodology

The system is designed with flexibility for further development and/or modification. The system is divided into manageable processes that are grouped to sub-modules and modules that are built with abstraction.

## 3. Architecture

### 3.1. System design

The block diagram below shows the principal parts of the system and their interactions.

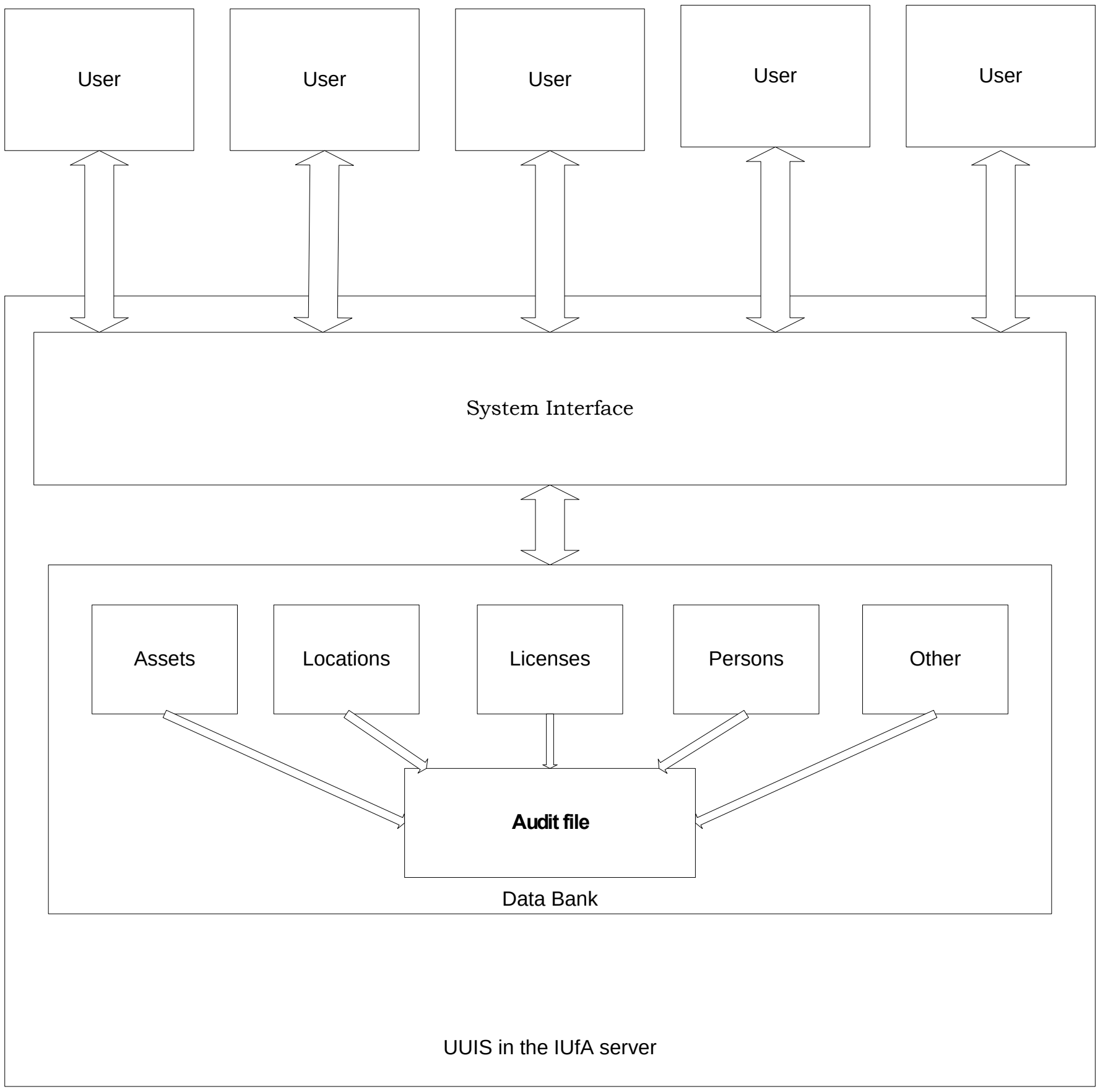

**Figure 1.** Block diagram for UUIS

The context diagram shows the main actors interacting with the system.

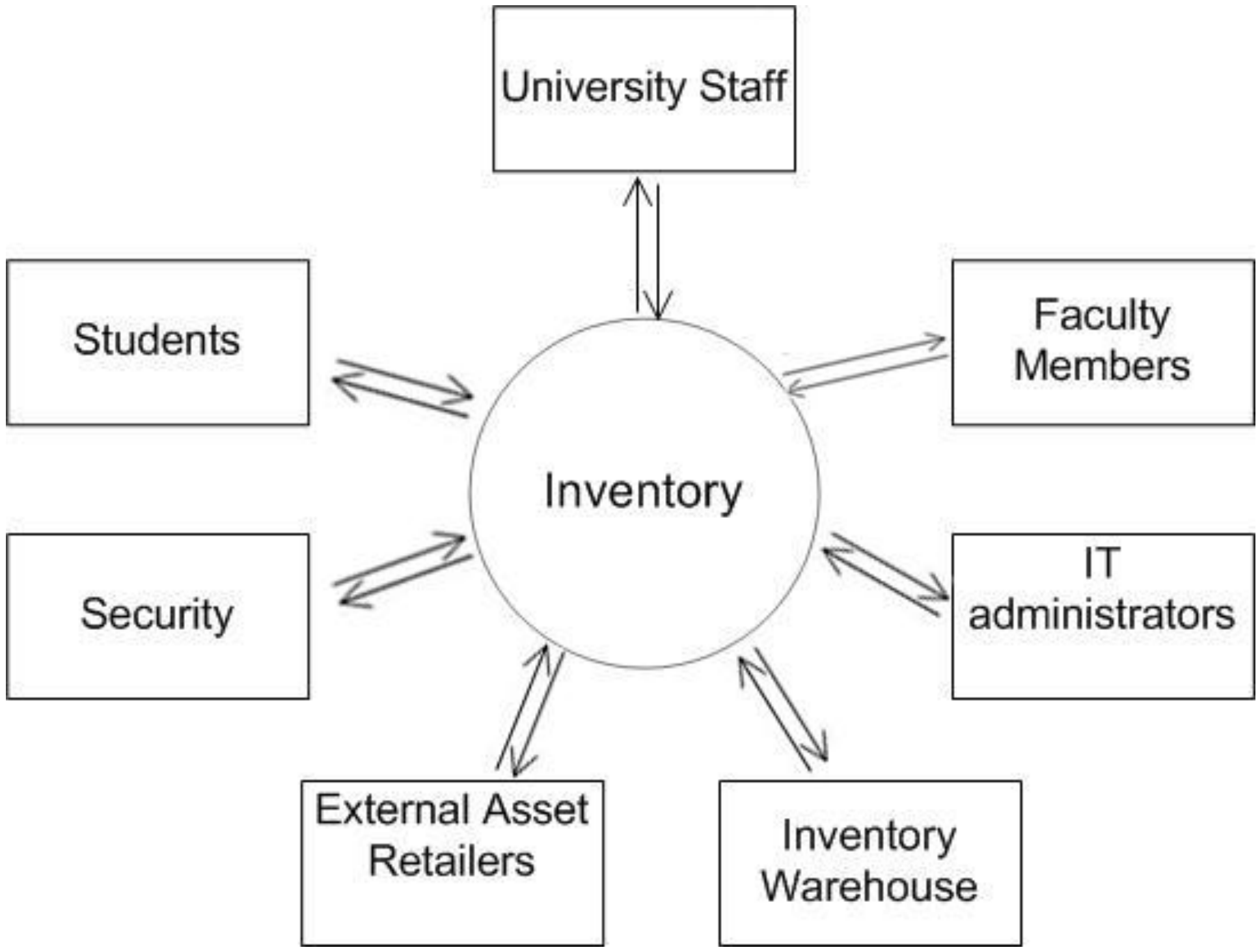

**Figure 2.** Context diagram for UUIS

## 3.2. System decomposition

### 3.2.1. Functional decomposition tree

Please note:
'All categories' refers to Assets, Locations, Licenses, Persons
Module 'Persons' doesn't have the process assign/reject
Module 'Borrow' belongs only to Assets and Licenses

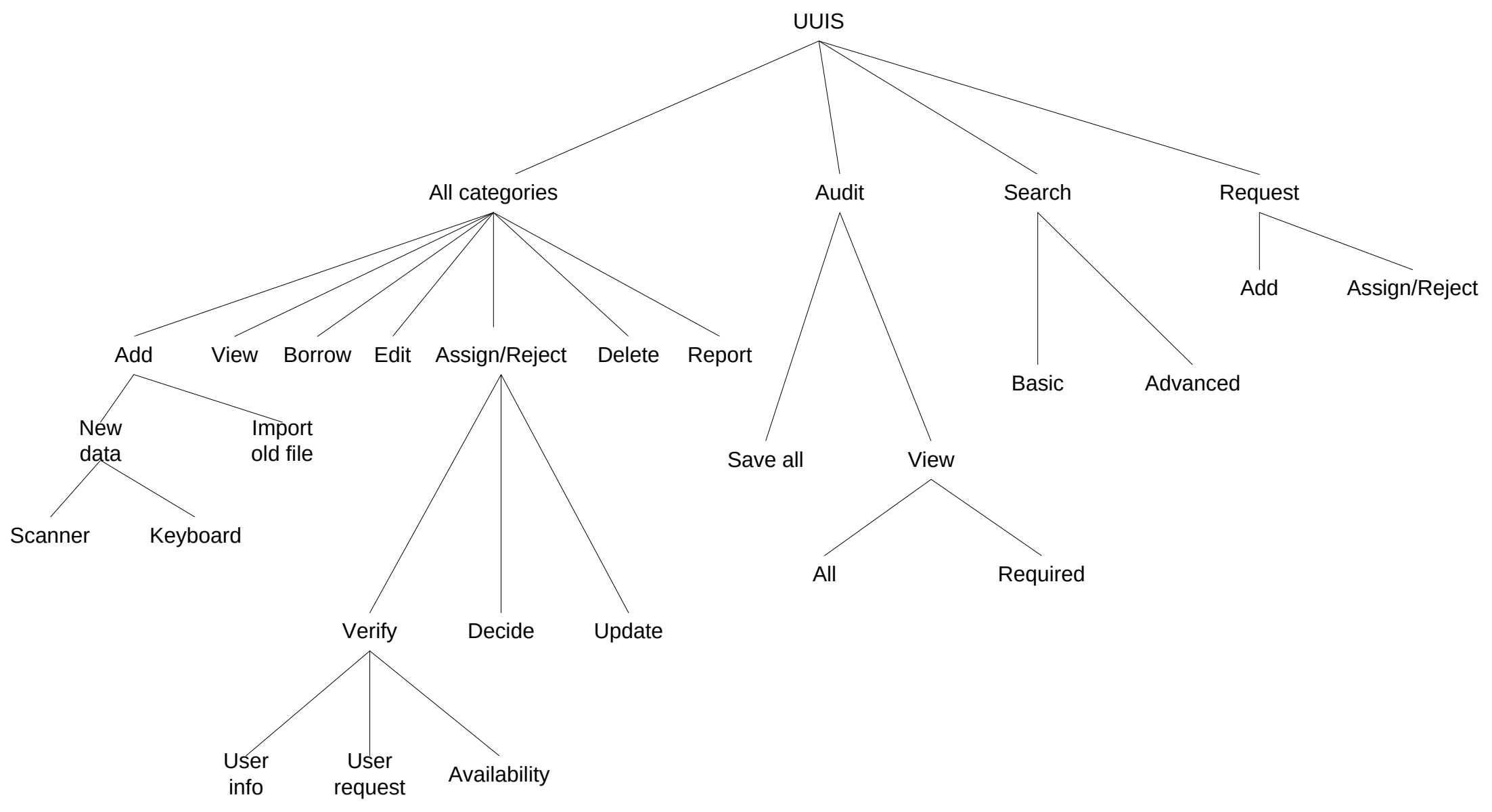

**Figure 3.** Decompositional Tree for UUIS

# 4. Data design

## 4.1. Data description

MySQL database and JDBC to communicate with the database that is installed locally on the server.

## 4.2. Data dictionary

**Table 1.** Data Dictionary

| | **Field** | **Type** | **Null** | **Default** |
|---|---|---|---|---|
| **approved_by** | Request_IDi | int(11) | Yes | *NULL* |
| | Person_ID | int(11) | Yes | *NULL* |
| | Approval_level | enum('Level1','Level2', 'Level3') | No | Level1 |
| | Approved_Date | timestamp | Yes | CURRENT_TIMESTAMP |
| | | | | |
| **assets** | Asset_ID | int(11) | No | |
| | BarcodeNum | varchar(50) | Yes | *NULL* |
| | SerialNum | varchar(50) | Yes | *NULL* |
| | Location_ID | int(11) | No | |
| | Type | enum('Desk','Computer','Academic_stuff','Other','Mouse','Keyboard','Printer','Monitor', 'Table','Chair','Projector','Software') | No | |
| | Description | | Yes s | *NULL* |
| | Status | enum('available','broken','not_available') | No | available |
| | Color | varchar(250) | Yes | *NULL* |
| | Material | varchar(250) | Yes | *NULL* |
| | Brand | varchar(250) | Yes | *NULL* |
| | Host | varchar(250) | Yes | *NULL* |
| | Created_date | timestamp | No | CURRENT_TIMESTAMP |
| | PurchaseNo | varchar(250) | Yes | not set |
| | RequestNo | varchar(250) | Yes | not set |
| | | | | |
| **assets_group** | Asset_master_ID | int(11) | Yes | *NULL* |
| | Asset_child_ID | int(11) | Yes | *NULL* |
| | Type | enum('Group','Work_station','Office_equipment') | No | group |
| | | | | |
| **assets_history** | Asset_ID | int(11) | No | *NULL* |
| | BarcodeNum | varchar(50) | Yes | *NULL* |
| | SerialNum | varchar(50) | Yes | *NULL* |
| | Location_ID | int(11) | Yes | *NULL* |
| | Type | varchar(50) | Yes | *NULL* |
| | Description | varchar(2000) | Yes | *NULL* |
| | Status | varchar(50) | Yes | *NULL* |
| | Color | varchar(250) | Yes | *NULL* |
| | Material | varchar(250) | Yes | *NULL* |
| | Brand | varchar(250) | Yes | *NULL* |
| | Host | varchar(250) | Yes | *NULL* |
| | Modified_by | varchar(250) | Yes | *NULL* |
| | PurchaseNo | varchar(250) | Yes | Not set |
| | RequestNo | varchar(250) | Yes | Not set |
| | | | | |
| **batch_request** | Request_ID | int(11) | Yes | *NULL* |
| | Asset_ID | int(11) | Yes | *NULL* |
| | Type | enum('Move','Buy','Repair','Delete') | No | Repair |
| | | | | |
| **building** | Building_ID | int(11) | No | *NULL* |
| | Address | varchar(250) | Yes | *NULL* |
| | Name | varchar(250) | Yes | |
| | Type | enum('Big','Medium', 'Small') | No | Medium |
| | FloorNum | int(3) | Yes | 1 |
| **fac_dep** | Fac_dep_ID | int(11) | No | |
| | Building | int(11) | Yes | *NULL* |
| | Name | varchar(250) | No | |
| | Type | enum('Faculty','Department') | No | Department |
| | Belong_to | int(11) | Yes | *NULL* |
| | | | | |
| **full_person_info** | Person_ID | int(11) | Yes | 0 |
| | FirstName | varchar(250) | Yes | *NULL* |
| | LastName | varchar(250) | Yes | *NULL* |
| | UserName | varchar(50) | Yes | *NULL* |
| | Password | varchar(50) | Yes | *NULL* |
| | Address | varchar(250) | Yes | *NULL* |
| | EmailAddress | varchar(50) | Yes | *NULL* |
| | MobileNumber | varchar(50) | Yes | *NULL* |
| | PersonCode | varchar(50) | Yes | *NULL* |
| | Status | enum('available','blocked','temporary','deleted') | Yes | available |
| | Type | enum('full_worker','temp_worker','grad_student','undergrad_student','temporary', 'visitor','undefined') | Yes | undefined |
| | Check_Biometric | tinyint(4) | Yes | 0 |
| | Created_date | timestamp | Yes | 0000-00-00 00:00:00 |
| | Delete_date | date | Yes | *NULL* |
| | Name | varchar(250) | Yes | *NULL* |
| | LocationNum | varchar(10) | Yes | 0 |
| | | | | |
| **licenses** | License_ID | int(11) | No | |
| | Asset_ID | int(11) | Yes | *NULL* |
| | Name | varchar(250) | Yes | *NULL* |
| | Type | enum('BSD','Open Source','Dual', 'Quantity') | Yes | Quantity |
| | Licence_counter | int(11) | Yes | *NULL* |
| | Price | decimal(10,0) | Yes | *NULL* |
| | Term | varchar(250) | Yes | *NULL* |
| | Licence_company | varchar(250) | Yes | *NULL* |
| | Creation_date | timestamp | Yes | CURRENT_TIMESTAMP |
| | Deleted_date | date | Yes | *NULL* |
| | PurchaseNo | varchar(250) | Yes | not set |
| | RequestNo | varchar(250) | Yes | not set |
| | | | | |
| **location_location** | Location_master_ID | int(11) | No | |
| | Location_child_ID | int(11) | No | |
| | Relation_type | enum('contain','unit') | No | contain |
| | | | | |
| **location_plan** | Plan_ID | int(11) | No | |
| | Plan_of_Location_ID | int(11) | Yes | *NULL* |
| | Plan | varchar(550) | Yes | *NULL* |
| | | | | |
| **locations** | Location_ID | int(11) | No | |
| | Capacity | int(3) | Yes | 0 |
| | Type | enum('storage','home', 'floor', 'holl','admin_office','suite', 'cubicle','atrium','teaching_lab','research_lab','grad_seat','teacher_office','drawer','printer_room') | No | drawer |
| | Belong_to | int(11) | Yes | *NULL* |
| | Description | varchar(1000) | Yes | No Description |
| | LocationNum | varchar(10) | No | 0 |
| | KeyNum | int(11) | Yes | 0 |
| | CodeNum | int(11) | Yes | 0 |
| | Width | varchar(250) | Yes | *NULL* |
| | Length | varchar(250) | Yes | *NULL* |
| | | | | |
| **person** | Person_ID | int(11) | No | |
| | FirstName | varchar(250) | Yes | *NULL* |
| | LastName | varchar(250) | Yes | *NULL* |
| | UserName | varchar(50) | No | |
| | Password | varchar(50) | Yes | *NULL* |
| | Address | varchar(250) | Yes | *NULL* |
| | EmailAddress | varchar(50) | No | |
| | MobileNumber | varchar(50) | Yes | *NULL* |
| | PersonCode | varchar(50) | Yes | *NULL* |
| | Status | enum('available','blocked', 'temporary', 'deleted') | No | available |
| | Type | enum('full_worker','temp_worker','grad_student','undergrad_student','temporary', 'visitor','undefined') | No | undefined |
| | Check_Biometric | tinyint(4) | Yes | 0 |
| | Created_date | timestamp | No | CURRENT_TIMESTAMP |
| | Delete_date | date | Yes | *NULL* |
| | | | | |
| **person_assets** | Asset_ID | int(11) | Yes | *NULL* |
| | Person_ID | int(11) | Yes | *NULL* |
| | Type | varchar(250) | Yes | *NULL* |
| | Check_in_Date | timestamp | Yes | CURRENT_TIMESTAMP |
| | | | | |
| **person_depart** | Fac_Dep_ID | int(11) | No | |
| | Person_ID | int(11) | No | |
| | Type | enum('works_for','study_in') | No | study_in |
| | | | | |
| **person_location** | Location_ID | int(11) | No | |
| | Person_ID | int(11) | No | |
| | Type | enum('grad_seat','research_seat','responsible','works_place', 'belong') | No | belong |
| | | | | |
| **person_roles** | Role_ID | varchar(250) | No | |
| | Person_ID | int(11) | No | |
| | | | | |
| **request** | Request_ID | int(11) | No | |
| | Requested_by_ID | int(11) | Yes | *NULL* |
| | Location_to_ID | int(11) | Yes | *NULL* |
| | Type | enum('Movement','Acquisition', 'Reparation', | No | Movement |
| | Description | varchar(2000) | Yes | *NULL* |
| | Status | enum('Not_Approved','Approved', 'Rejected','Compleated') | No | Not_Approved |
| | Creation_Date | timestamp | No | CURRENT_TIMESTAMP |
| | Delete_date | date | Yes | *NULL* |
| | Period | int(11) | Yes | *NULL* |
| | | | | |
| **roles** | Role_ID | varchar(250) | No | |
| | Description | varchar(2000) | Yes | *NULL* |
| | | | | |
| **roles_set** | Role_ID | varchar(250) | No | |
| | Level_Name | varchar(250) | No | |
| | | | | |
| **sub_group** | Asset_ID | int(11) | No | |
| | sub_type | varchar(250 | Yes | *NULL* |
| | | | | |
| **voice** | Voice_ID | int(11) | No | |
| | Person_ID | int(11) | Yes | *NULL* |
| | Voice | varchar(250) | Yes | *NULL* |

# 5. Component design

“Users with appropriate permission” in the diagram refers to the users who are given exemptions or/and users of a particular level. Precise permissions are listed in Section 2.3 of SRS.

For diagrams with multiple functions, the design is the same for those functions, except for parameters/methods/classes.

## 5.1. Login

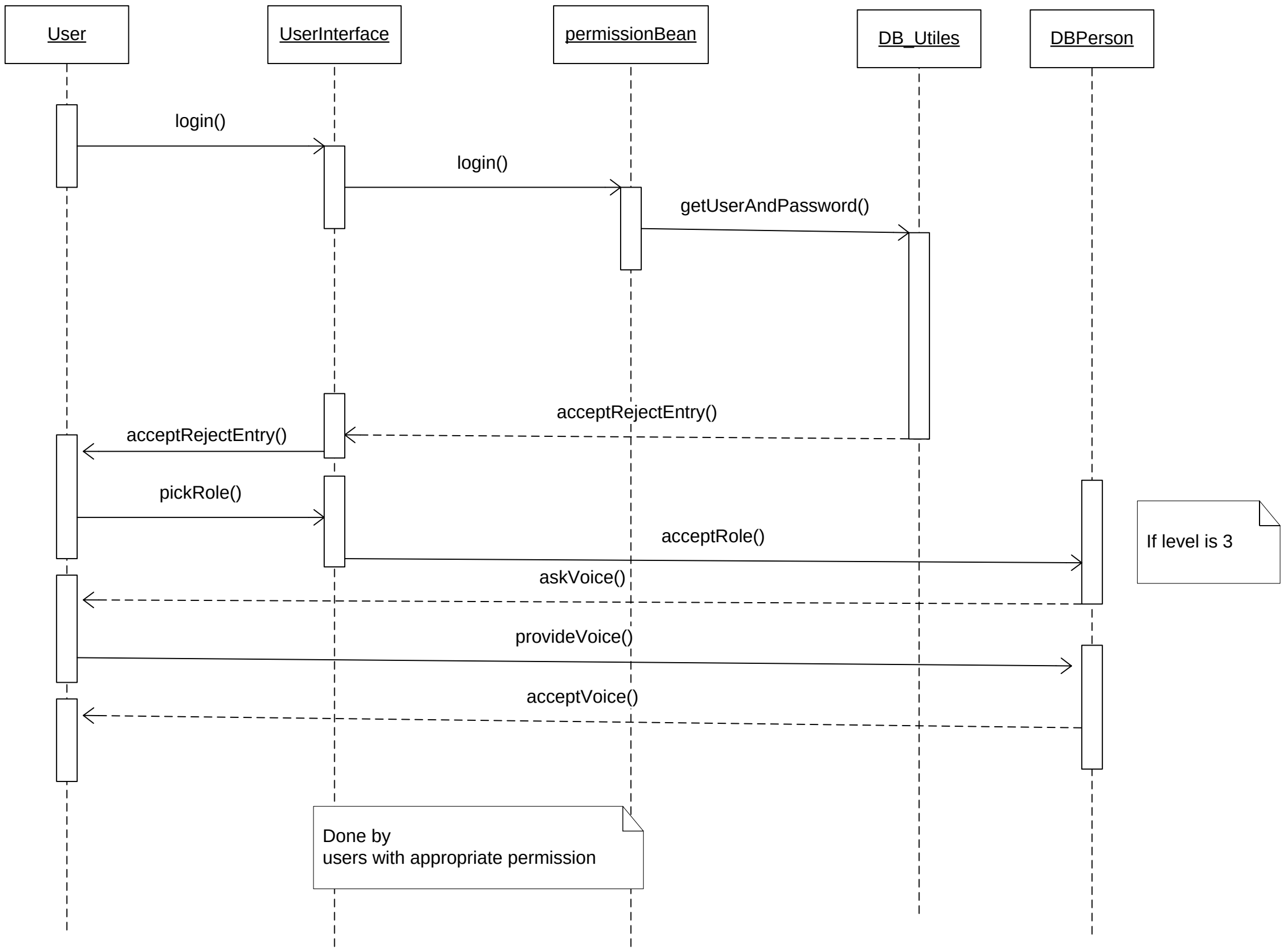

**Figure 4.** Sequence diagram for Login

## 5.2. Provide biometric characteristics

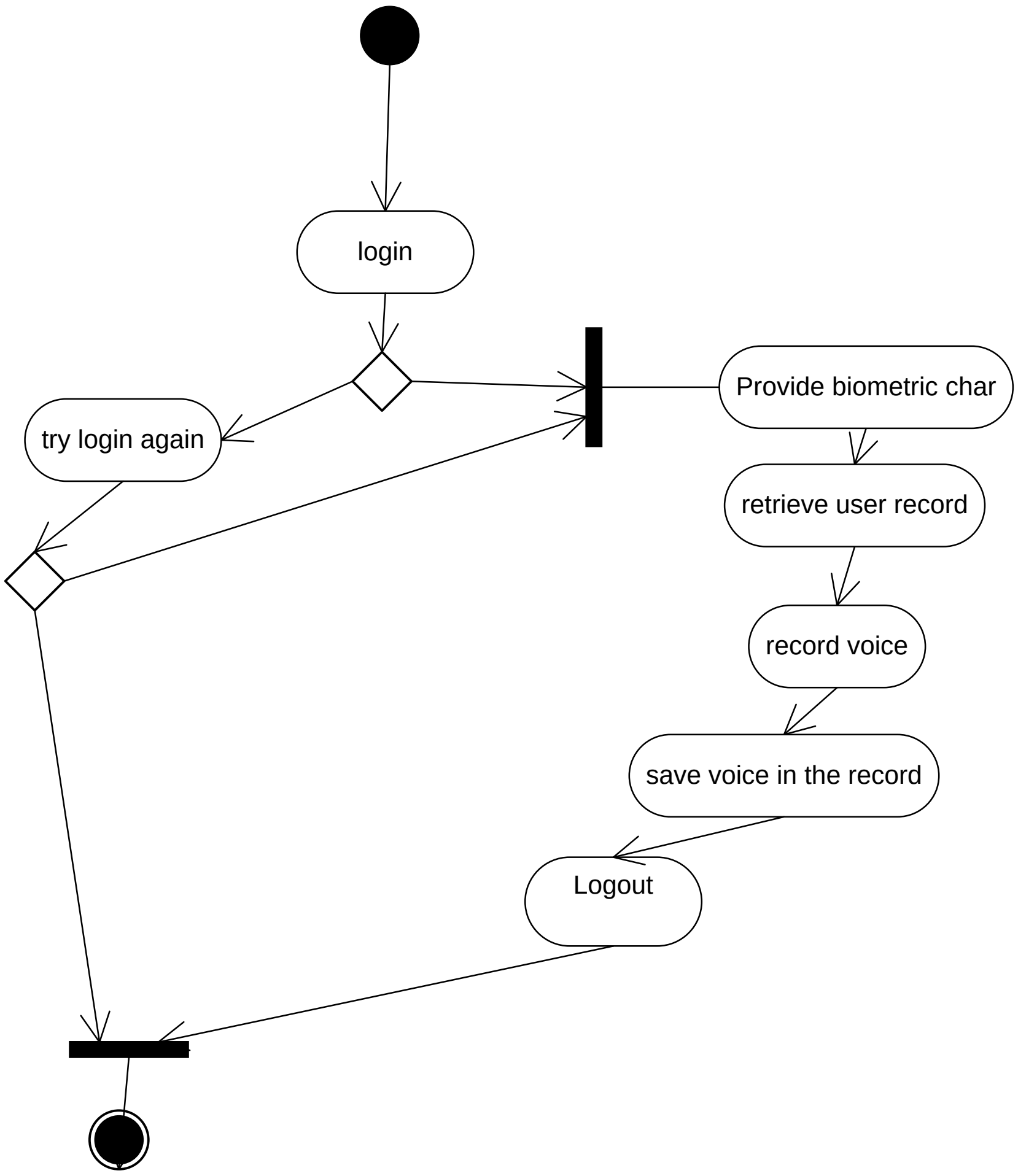

**Figure 5.** Activity diagram for Provide biometric characteristics

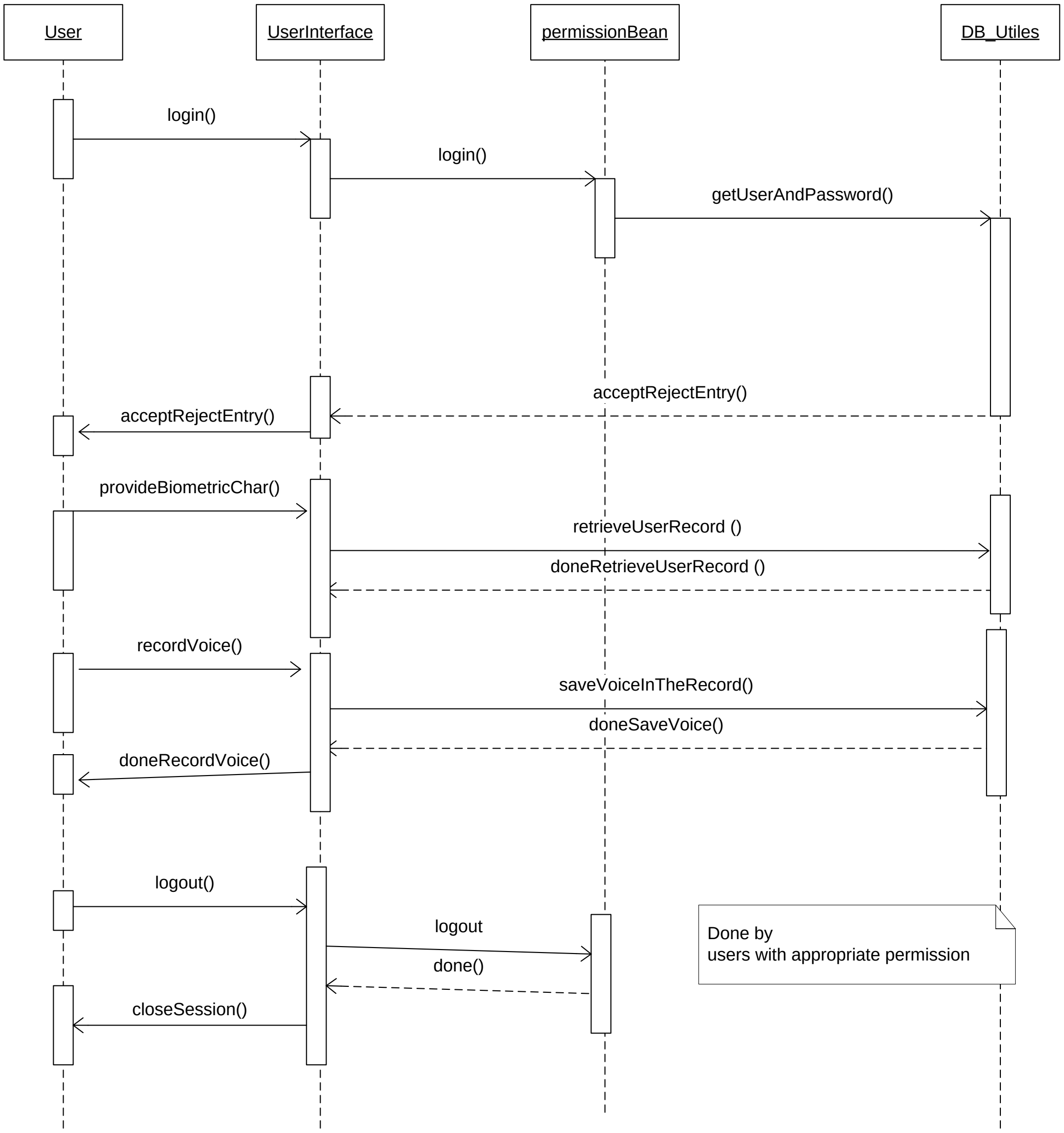

**Figure 6.** Sequence diagram for Provide biometric characteristics

## 5.3. Select language

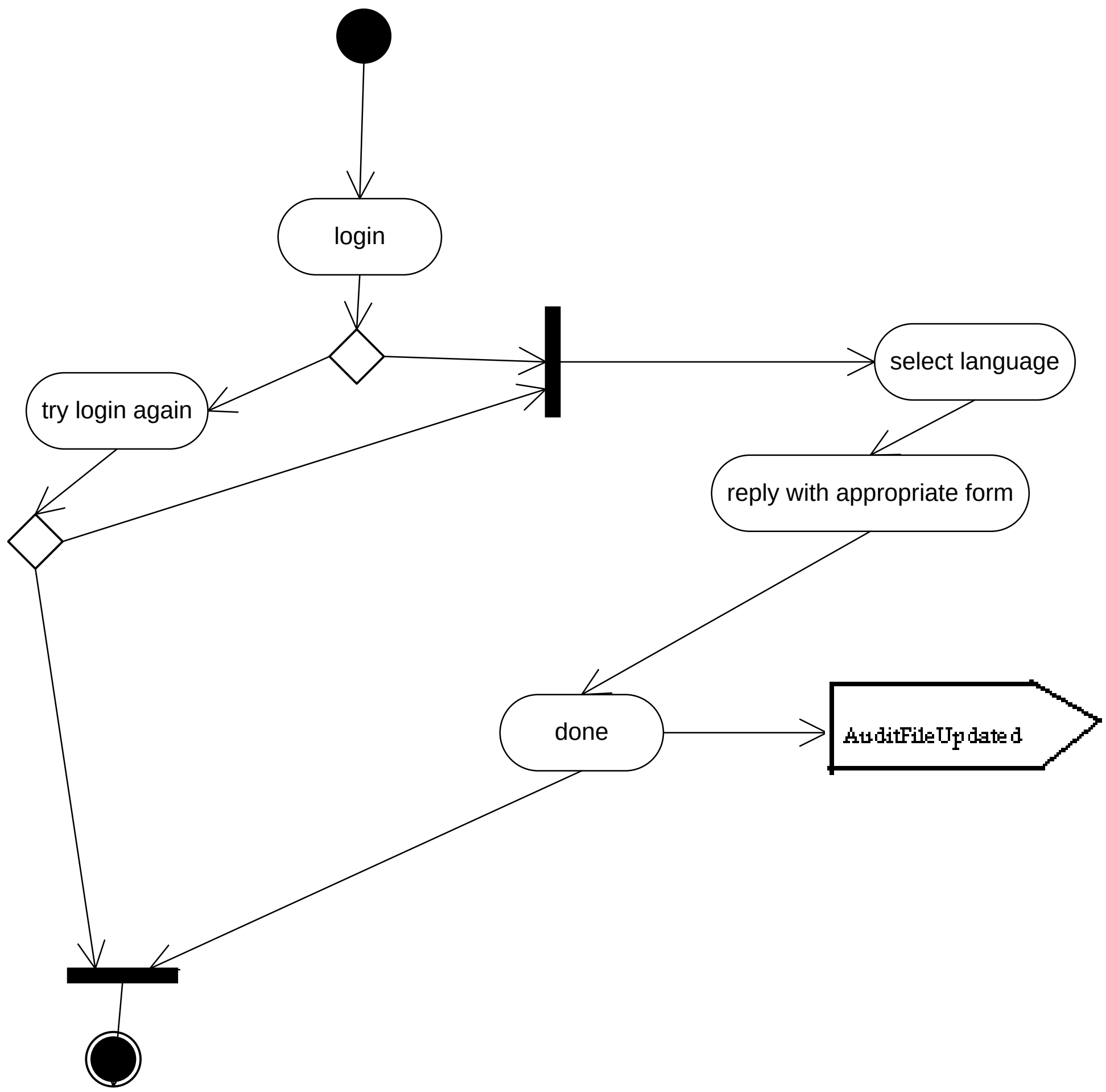

**Figure 7.** Activity diagram for Select language

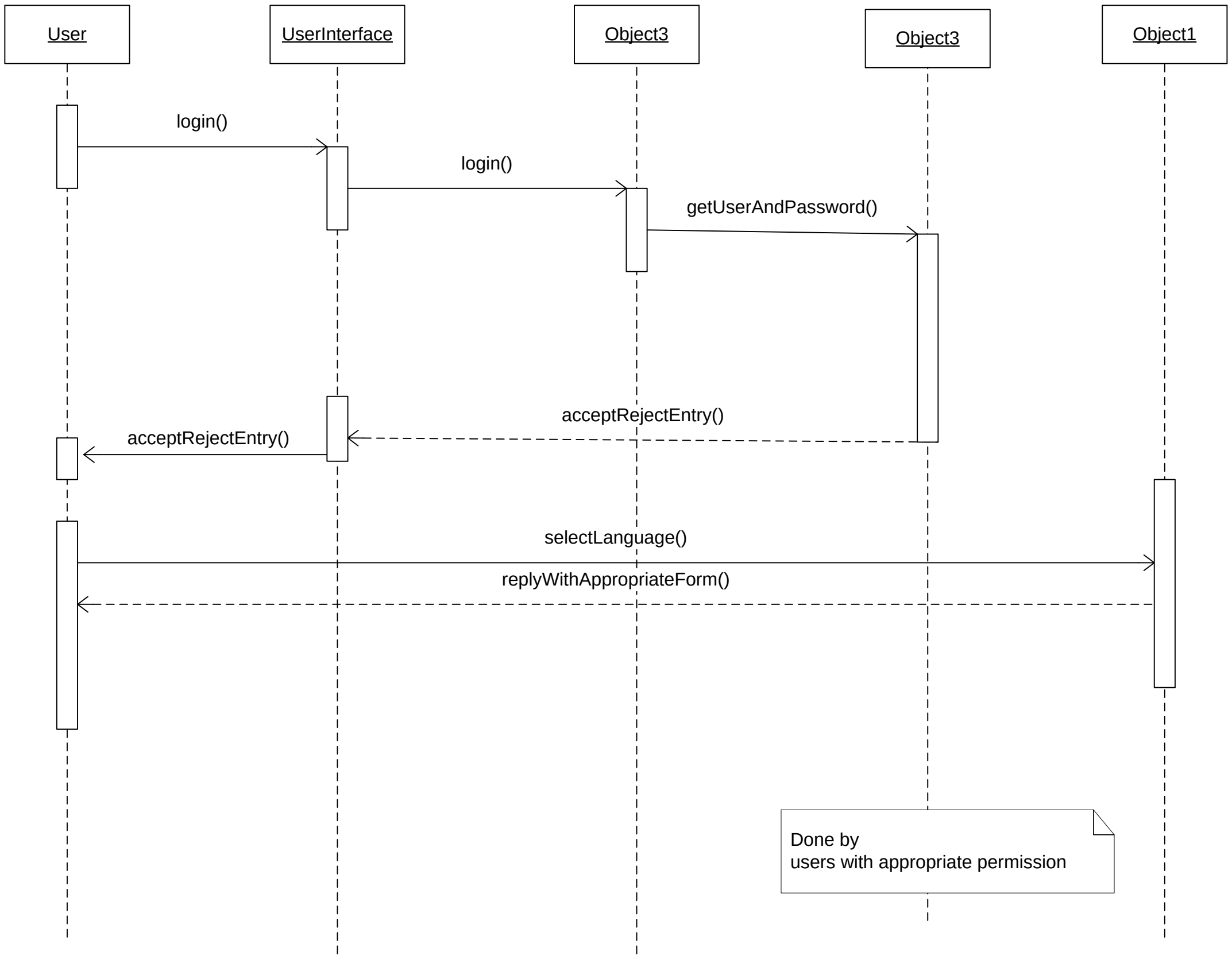

**Figure 8.** Sequence diagram for Select language

## 5.4. Create new group / subgroup / type

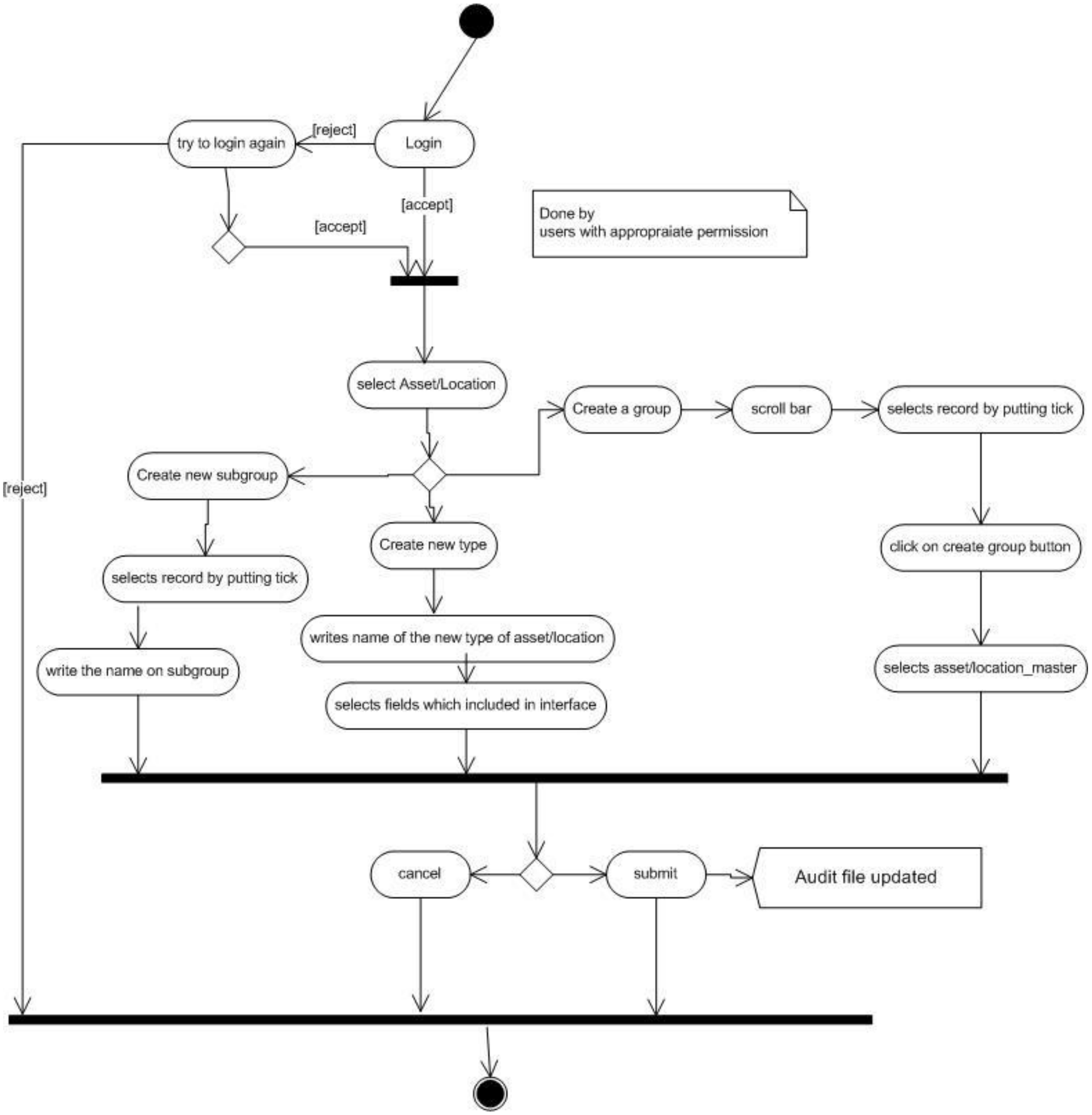

**Figure 9.** Activity diagram for Create new group / subgroup / type

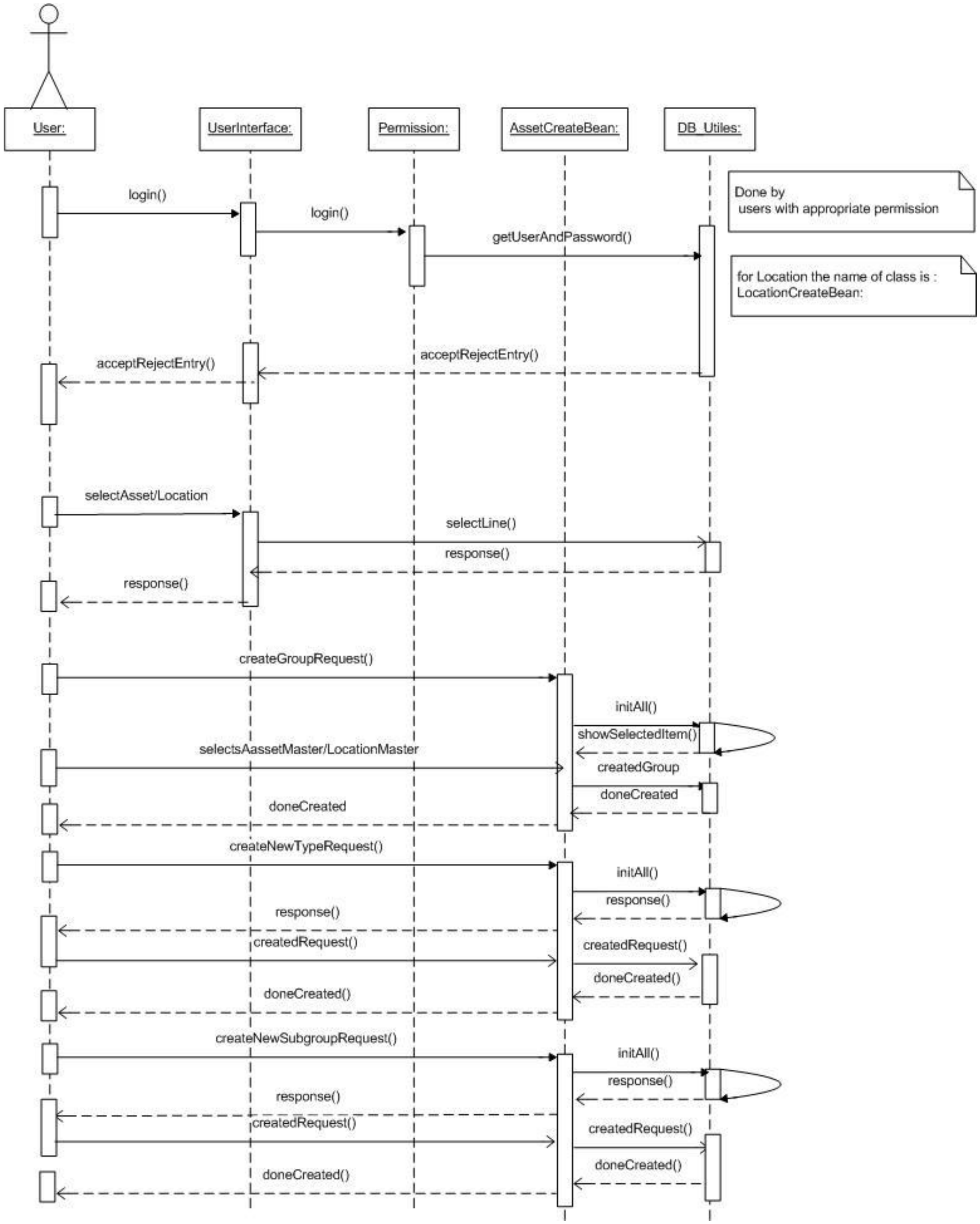

**Figure 10.** Sequence diagram for Create New group / subgroup / type

## 5.5. Import of assets / licenses / locations / persons

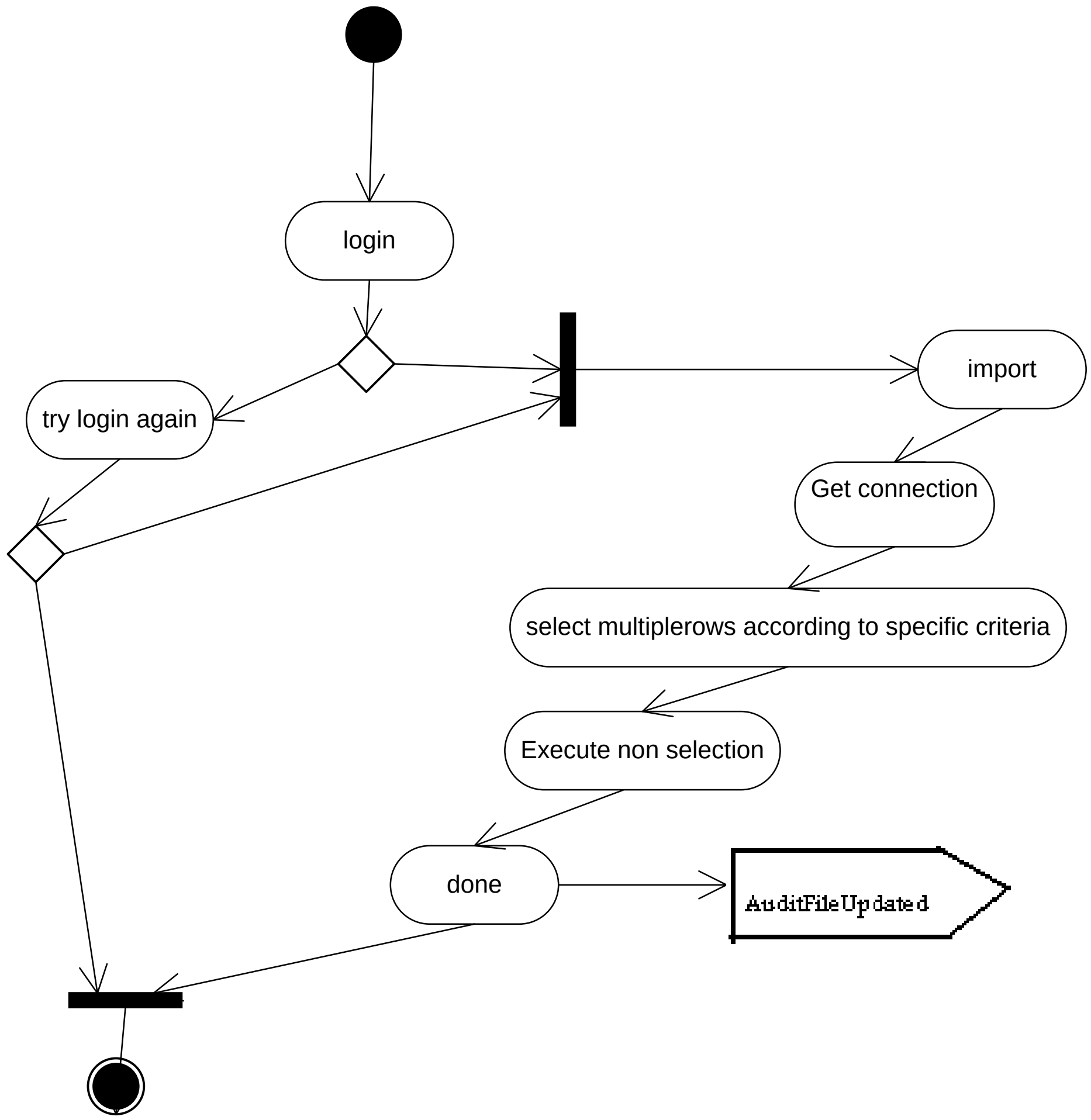

**Figure 11.** Activity diagram for Import of assets / licenses / locations / persons

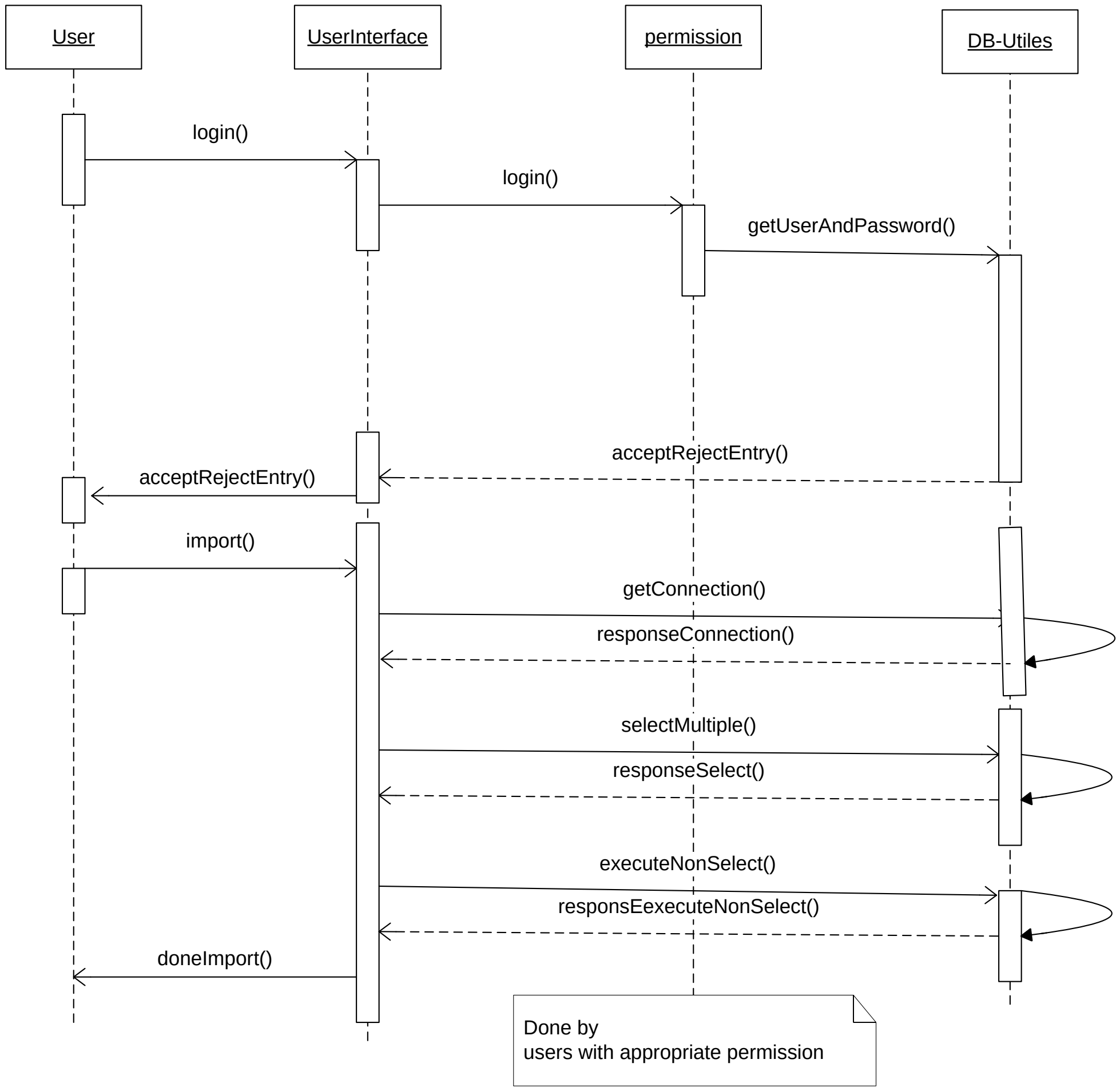

**Figure 12.** Sequence diagram for Import of assets / licences / locations

## 5.6. View/Add/Modify – assets / licenses / locations

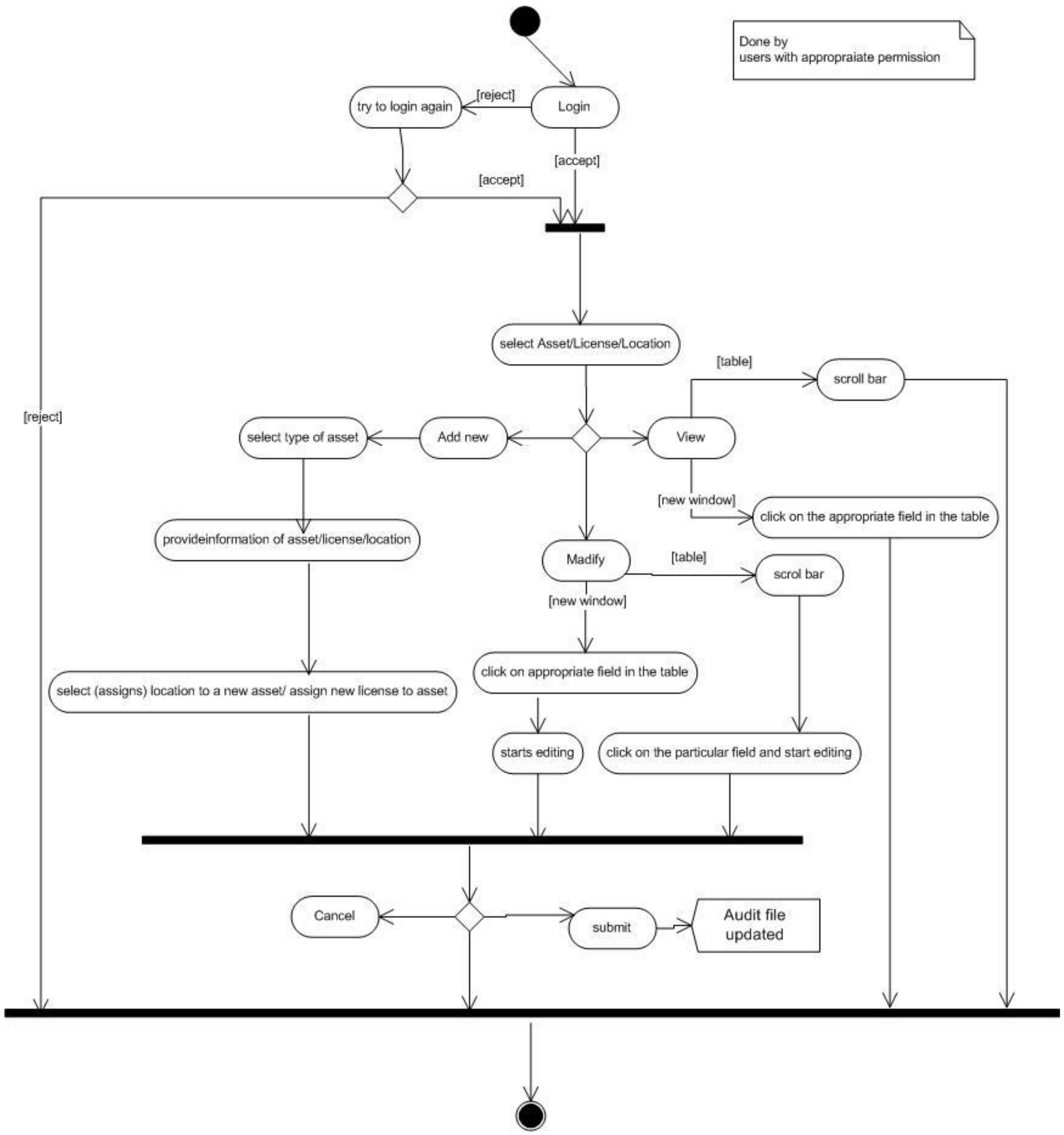

**Figure 13**. Activity diagram for View / modify – assets / licenses / locations

## 5.7. Assign/Modify - role

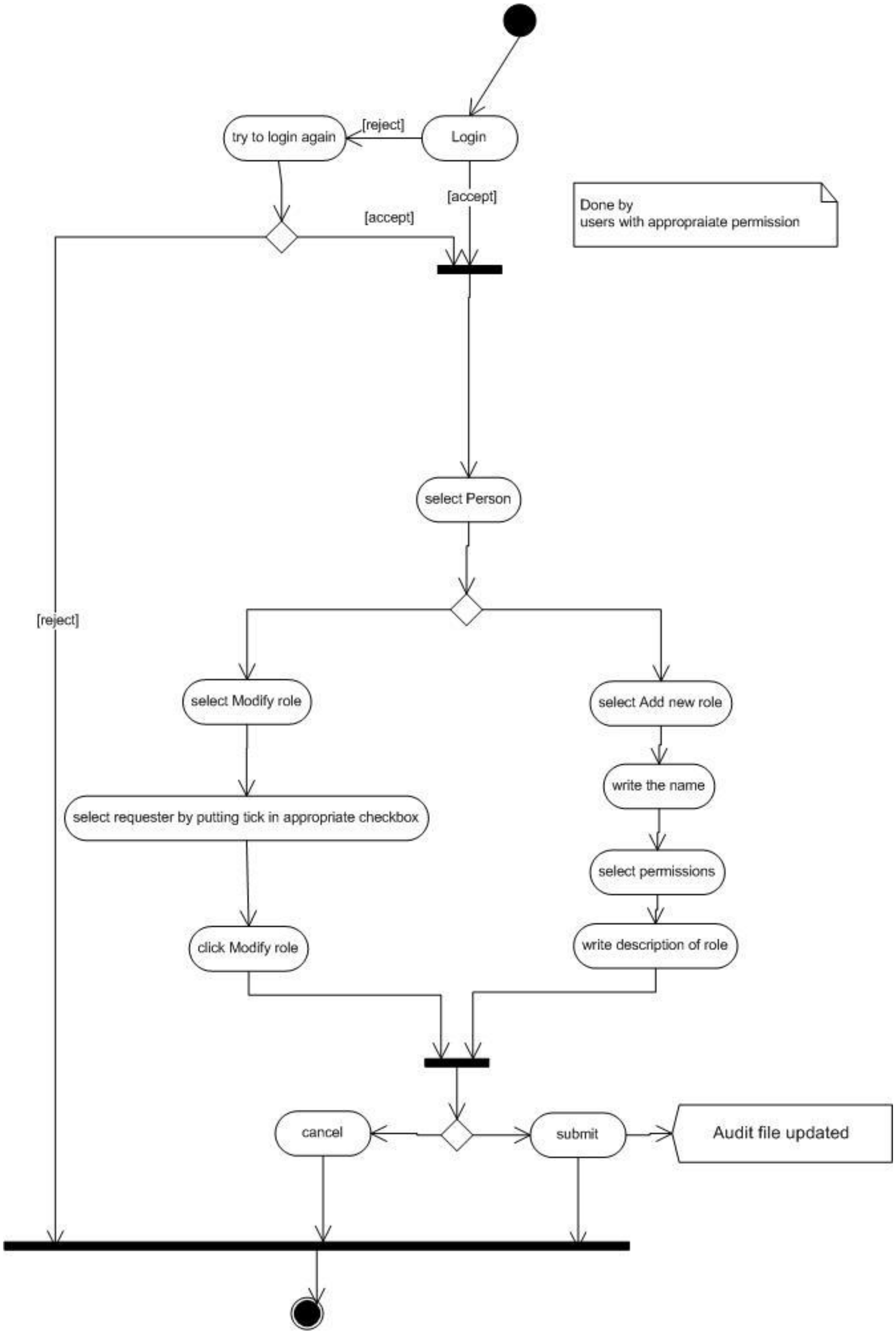

**Figure 14.** Activity diagram for assign / modify - role

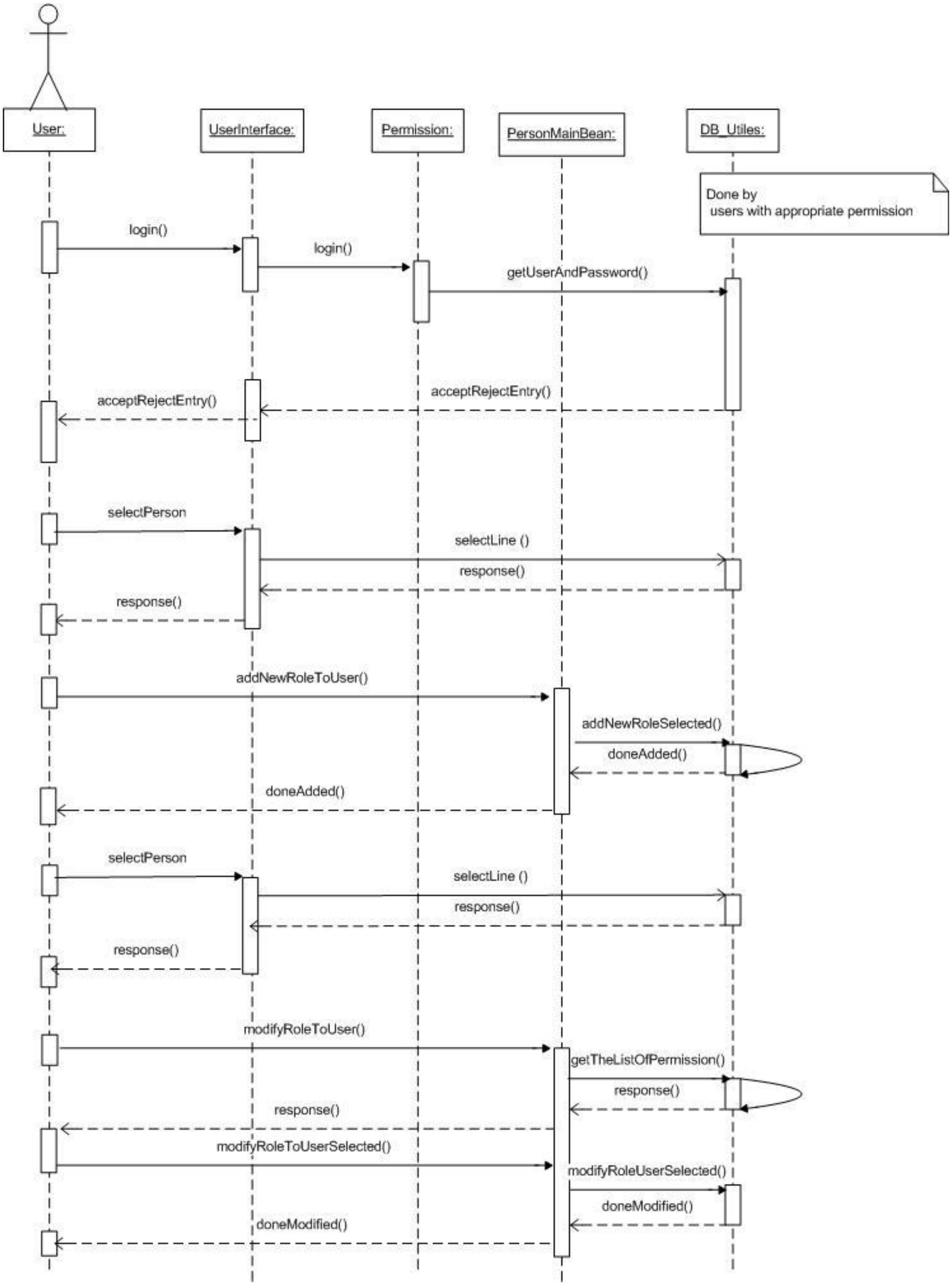

**Figure 15.** Sequence diagram for Assign / Modify - role

## 5.8. Add request

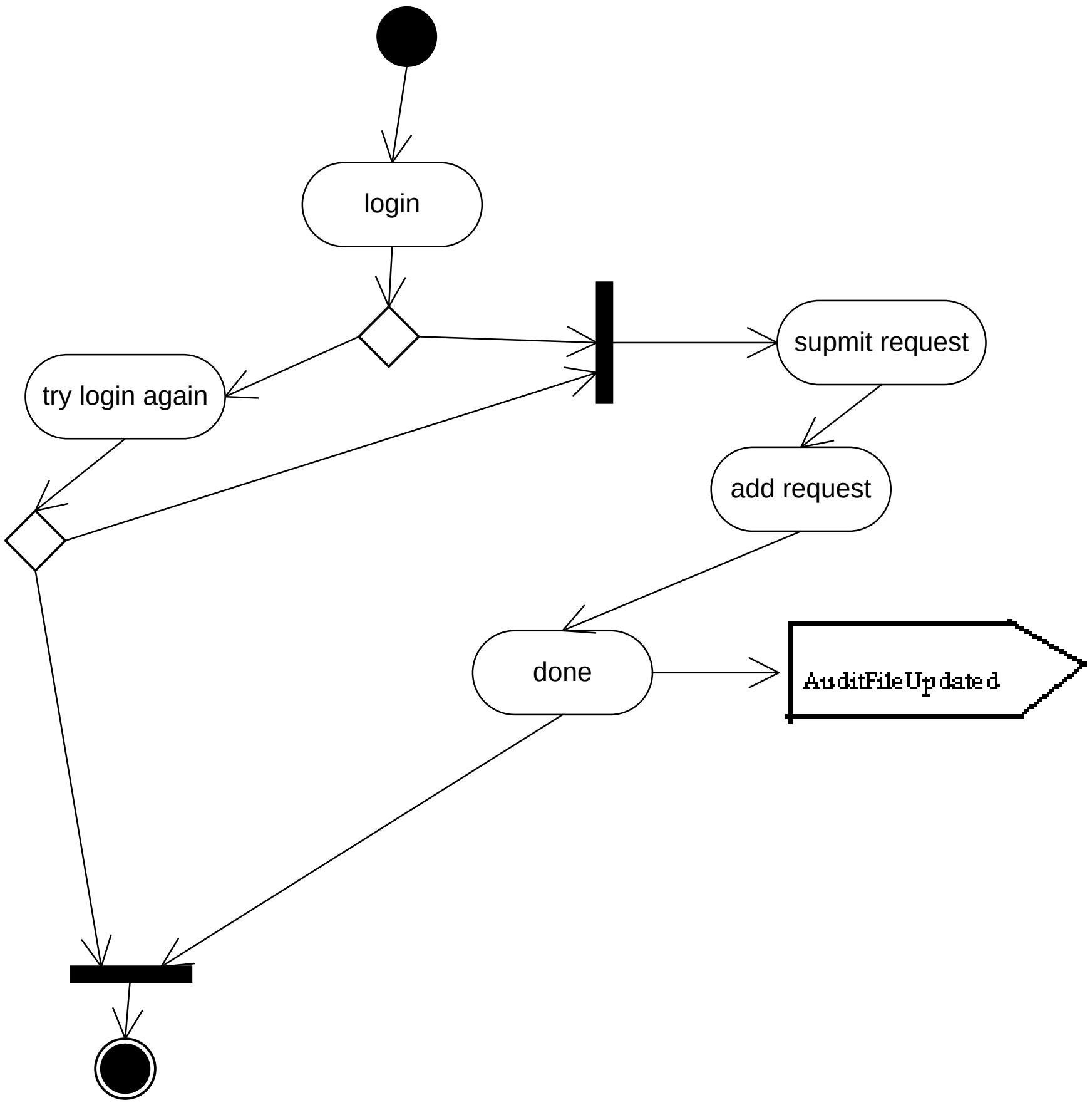

**Figure 16.** Activity diagram for Add request

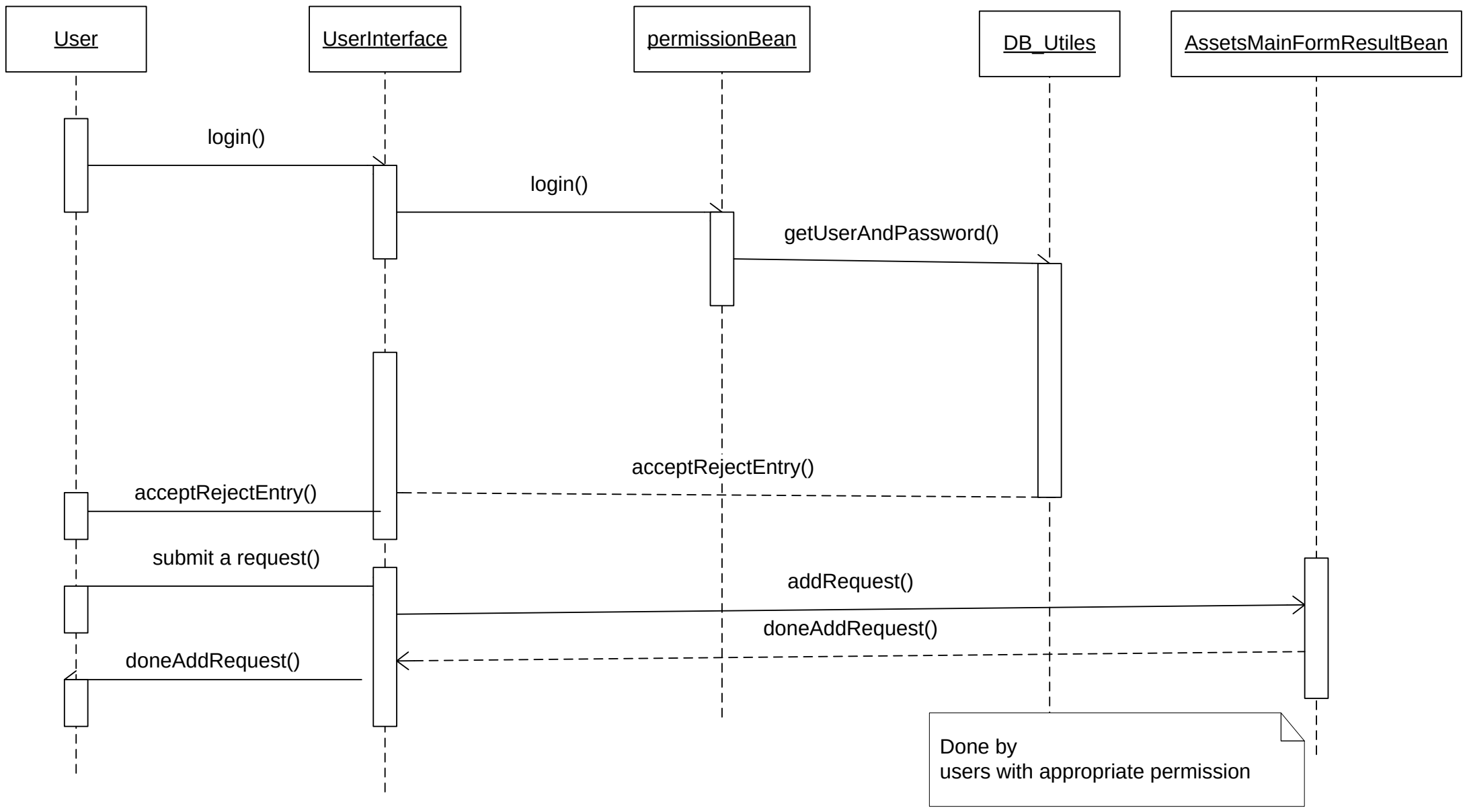

**Figure 17.** Sequence diagram for Add request

## 5.9. View/Approve/Reject request

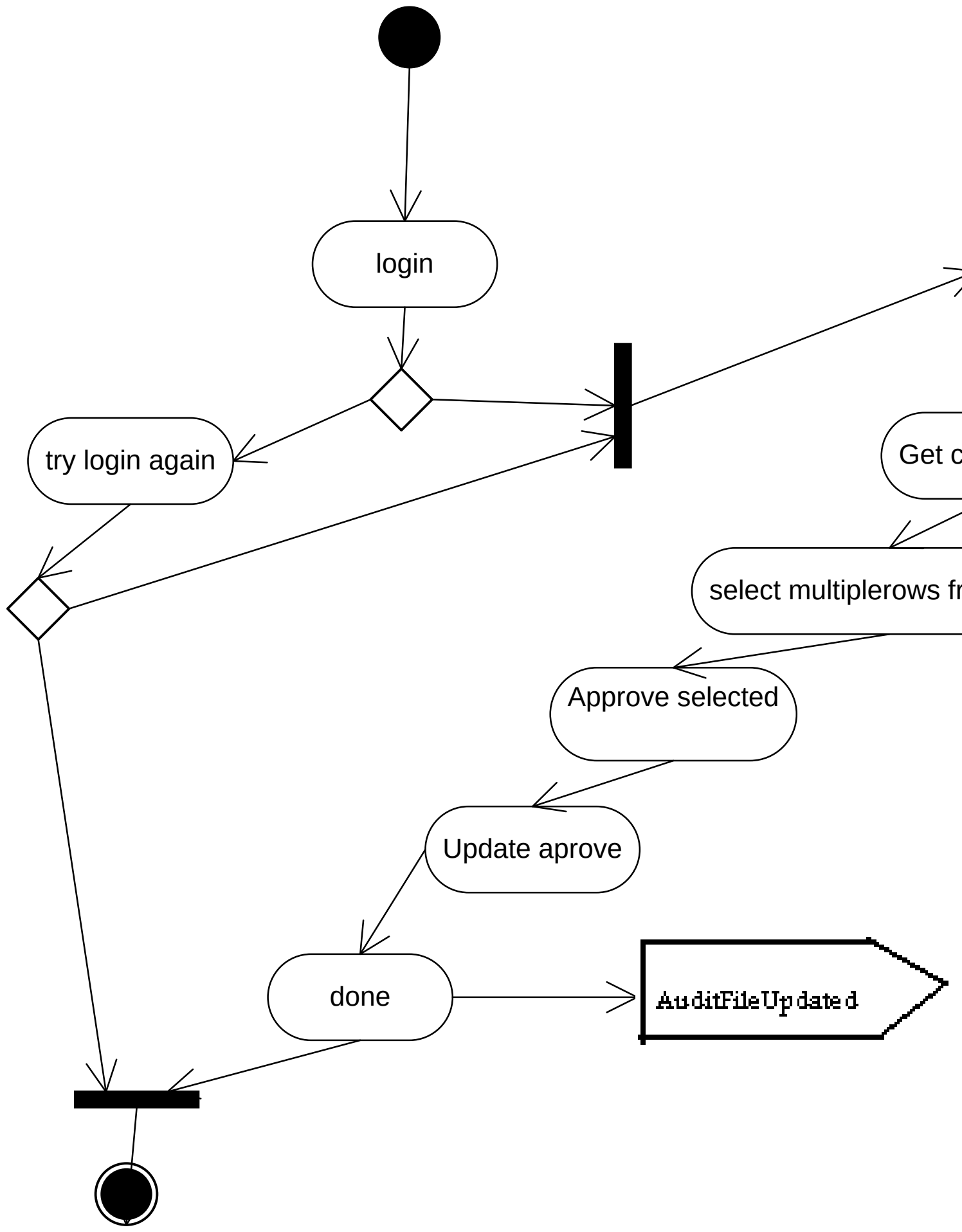

**Figure 18.** Activity diagram for View / Approve / Reject request

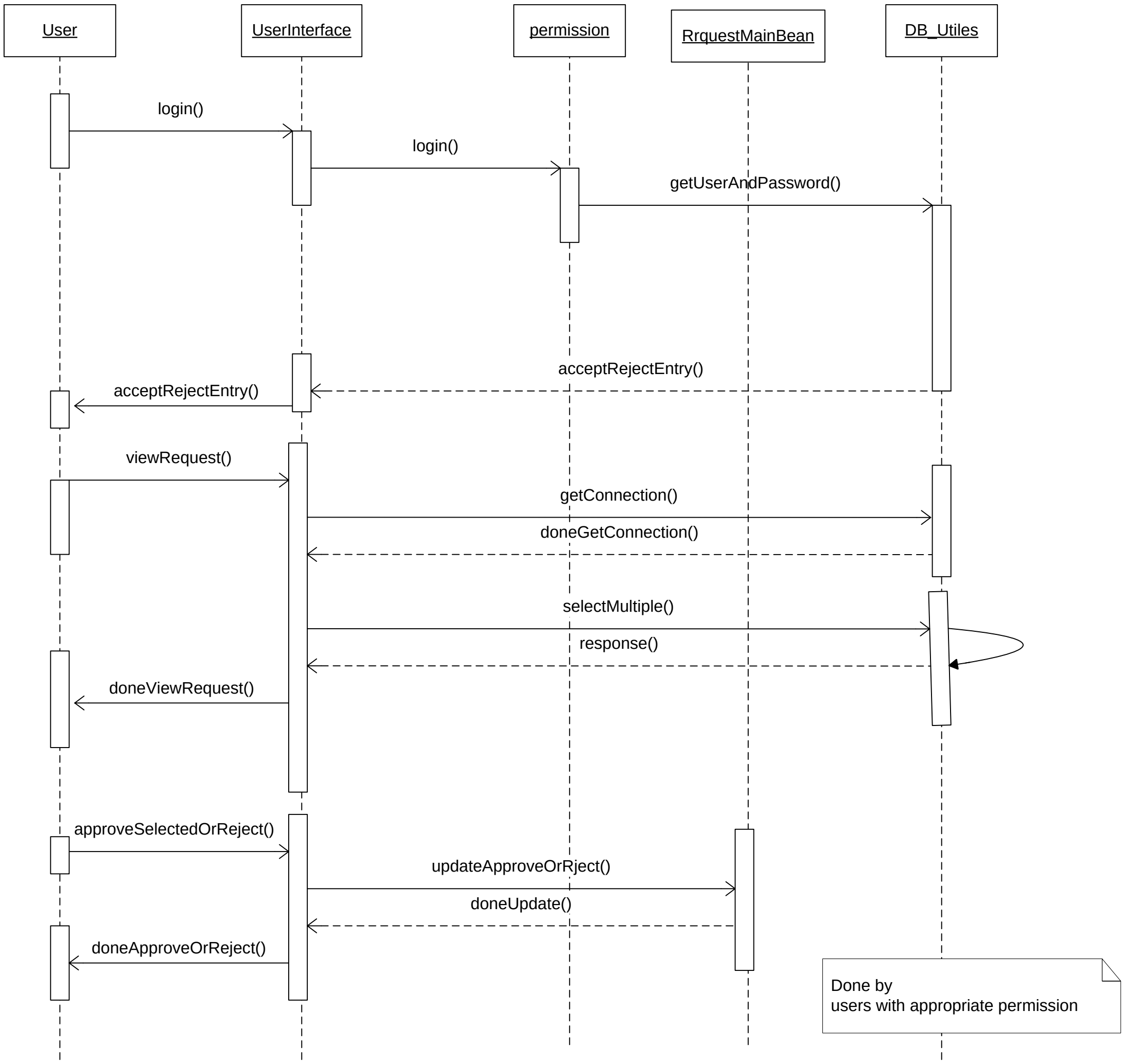

**Figure 19**. Sequence diagram for View / Approve / Reject request

## 5.10. Assign assets / licenses / locations

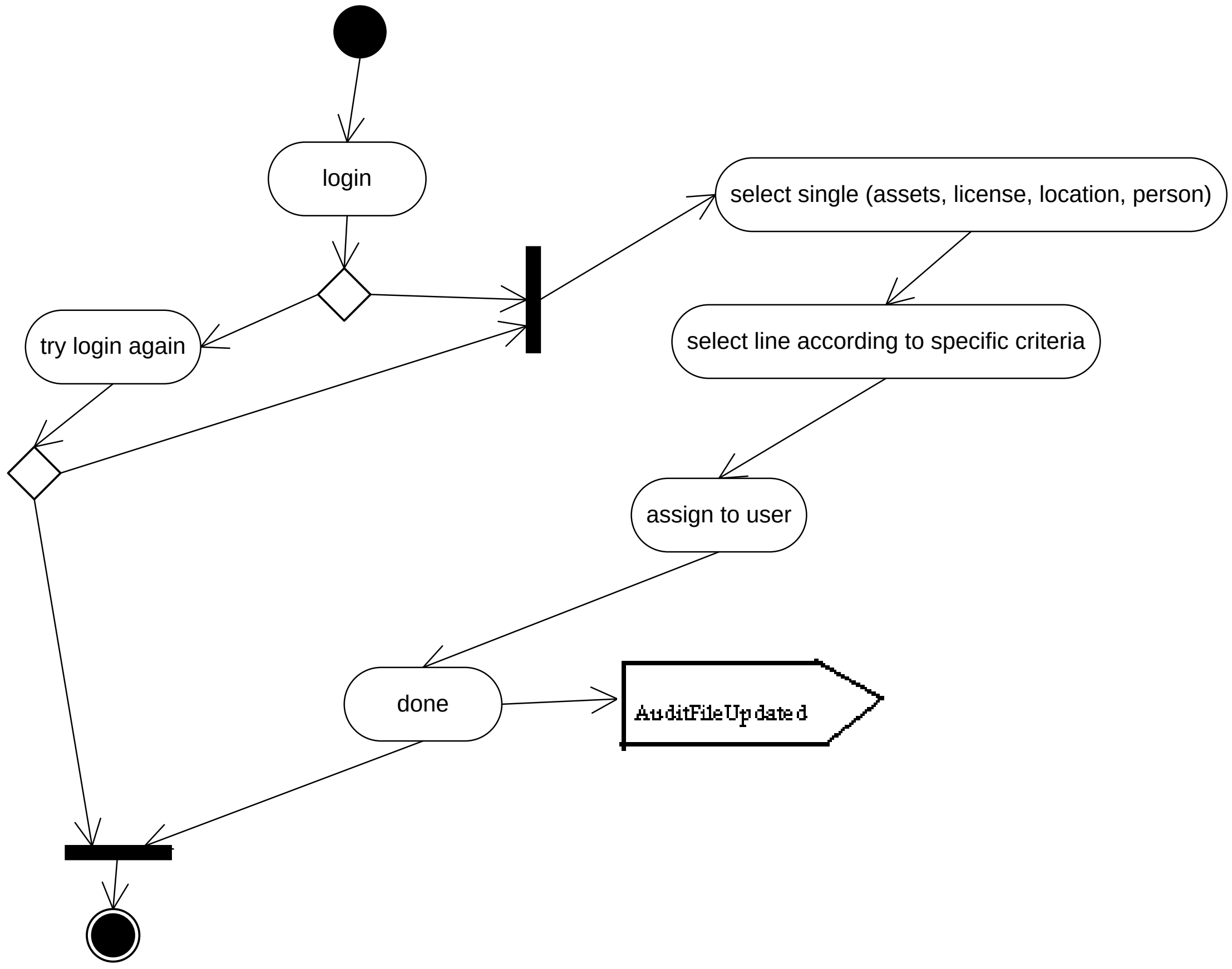

**Figure 20.** Activity diagram for Assign assets / licenses / locations

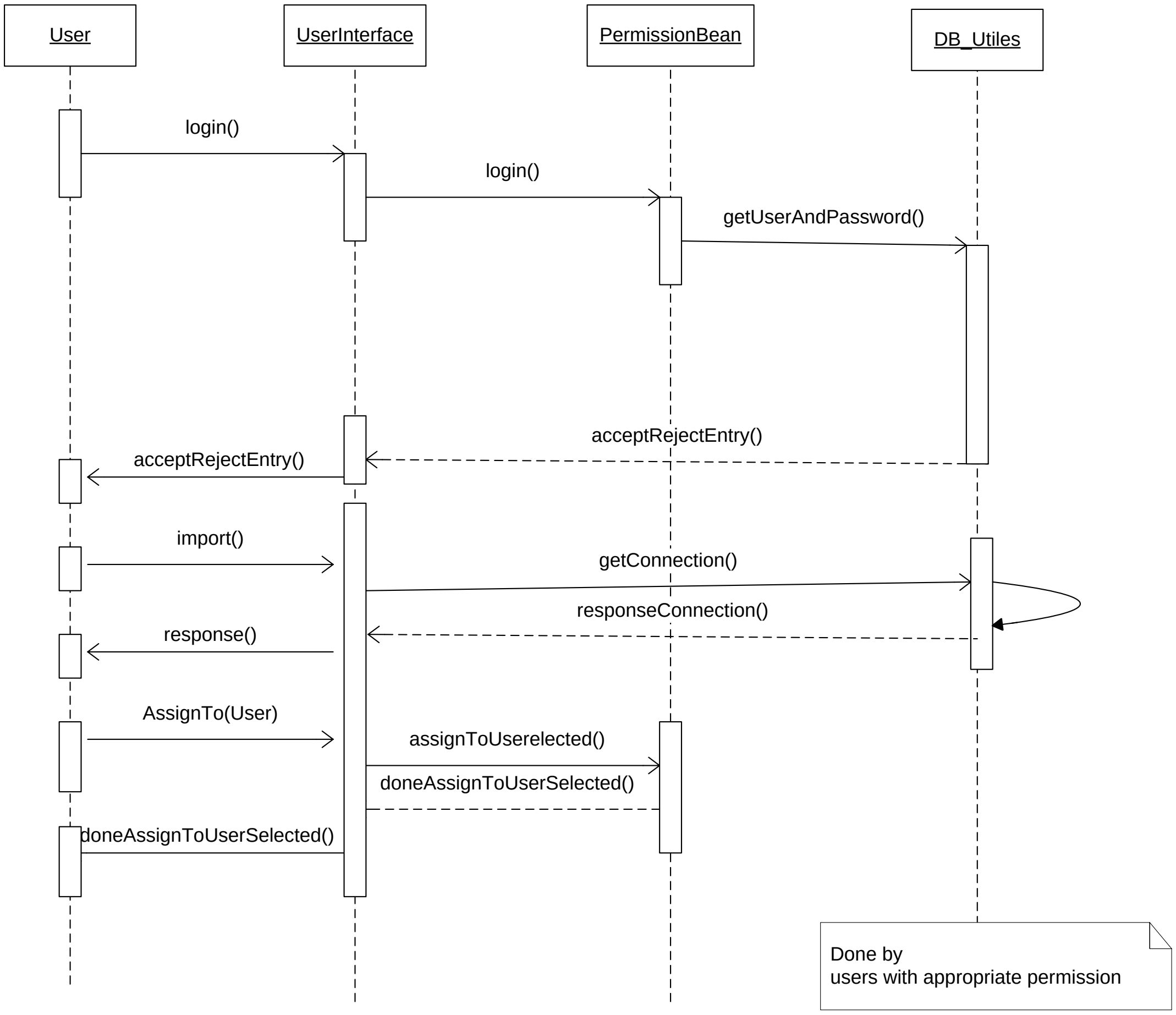

**Figure 21.** Sequence diagram for Assign assets / licenses / locations

## 5.11. Delete assets / licenses / locations / persons

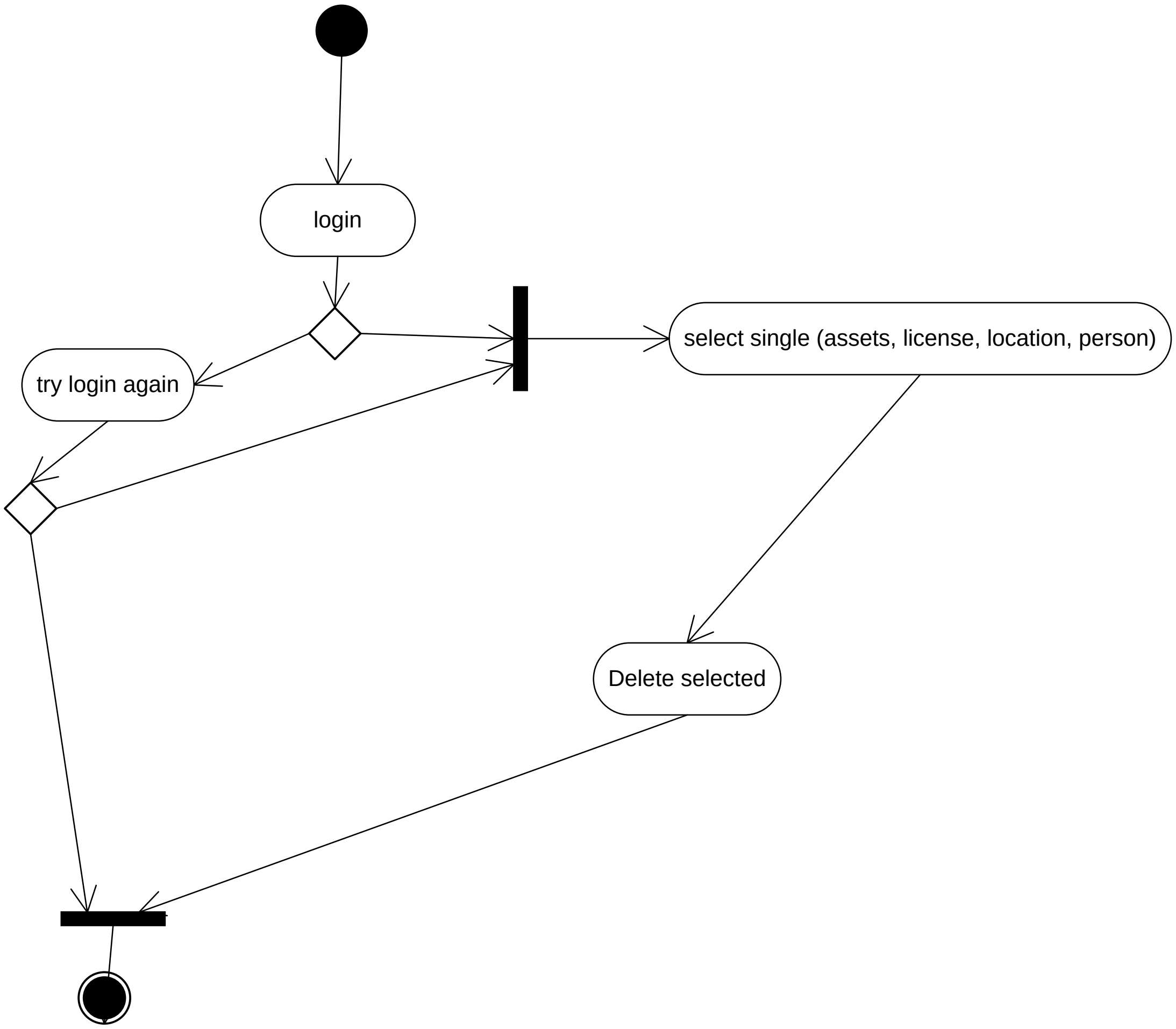

**Figure 22.** Activity diagram for Delete assets / licenses / locations / persons

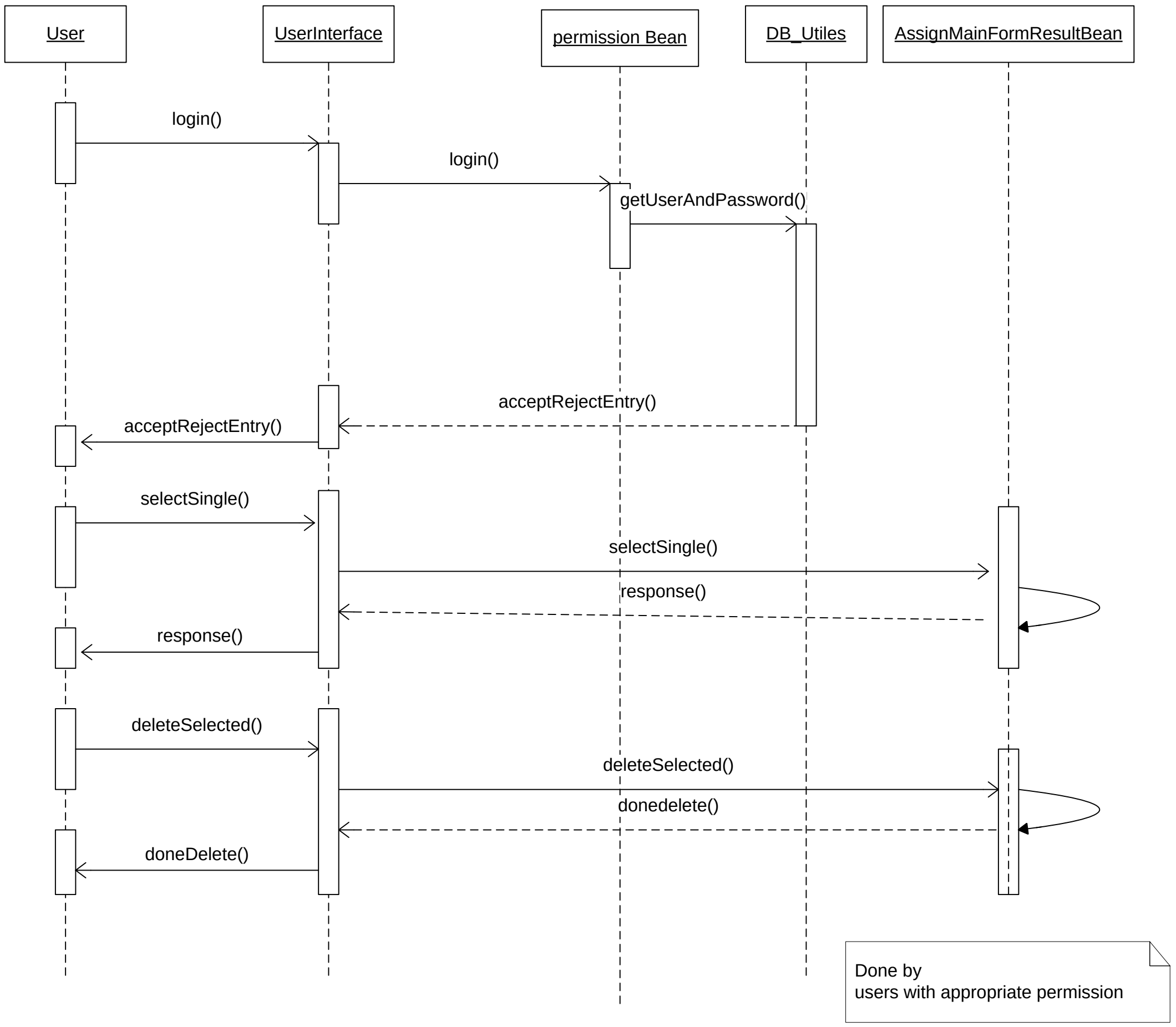

**Figure 23.** Sequence diagram for Delete assets / licenses / locations / persons

## 5.12. View/Modify - person

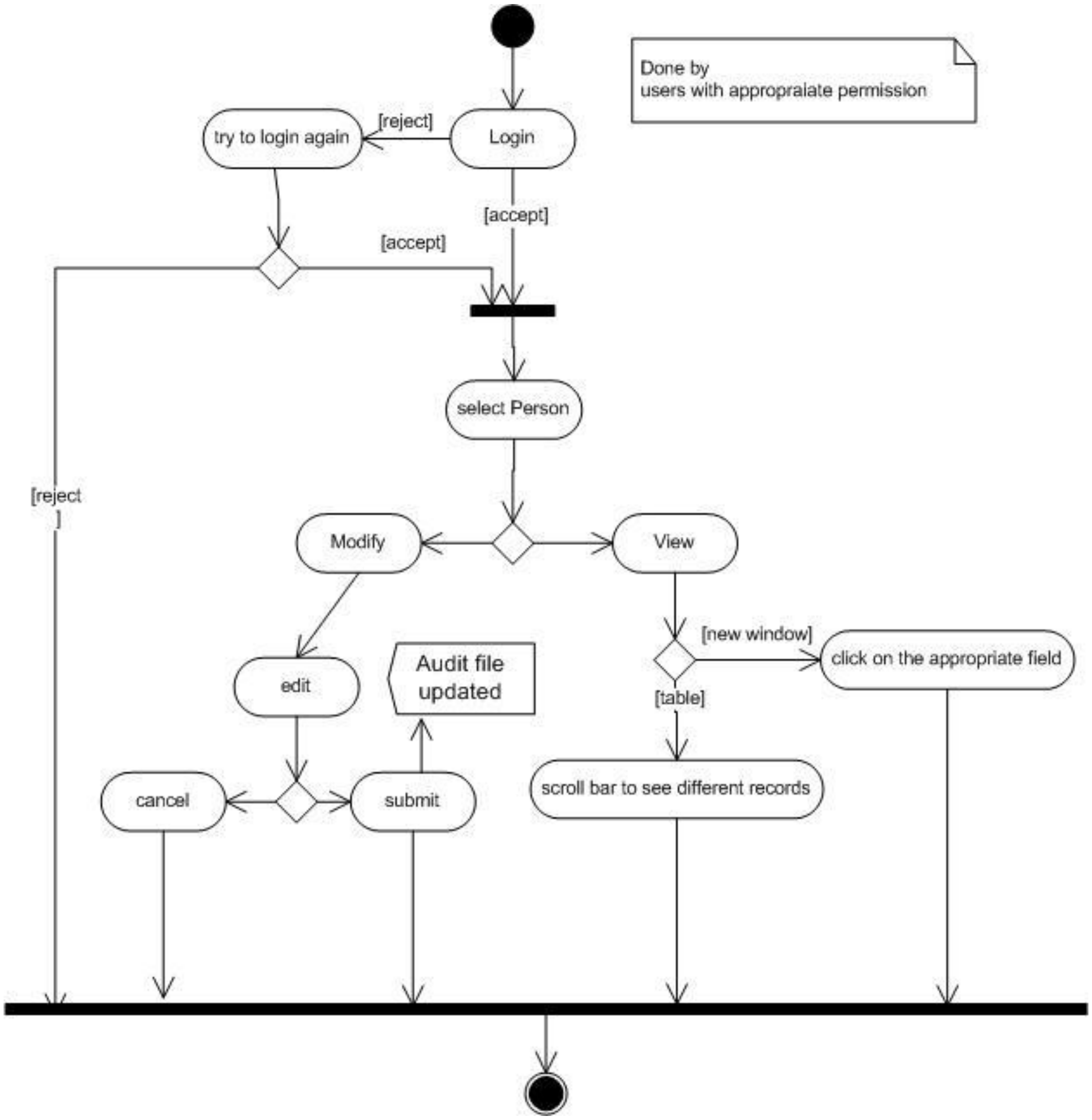

**Figure 24.** Activity diagram for View / Modify - person

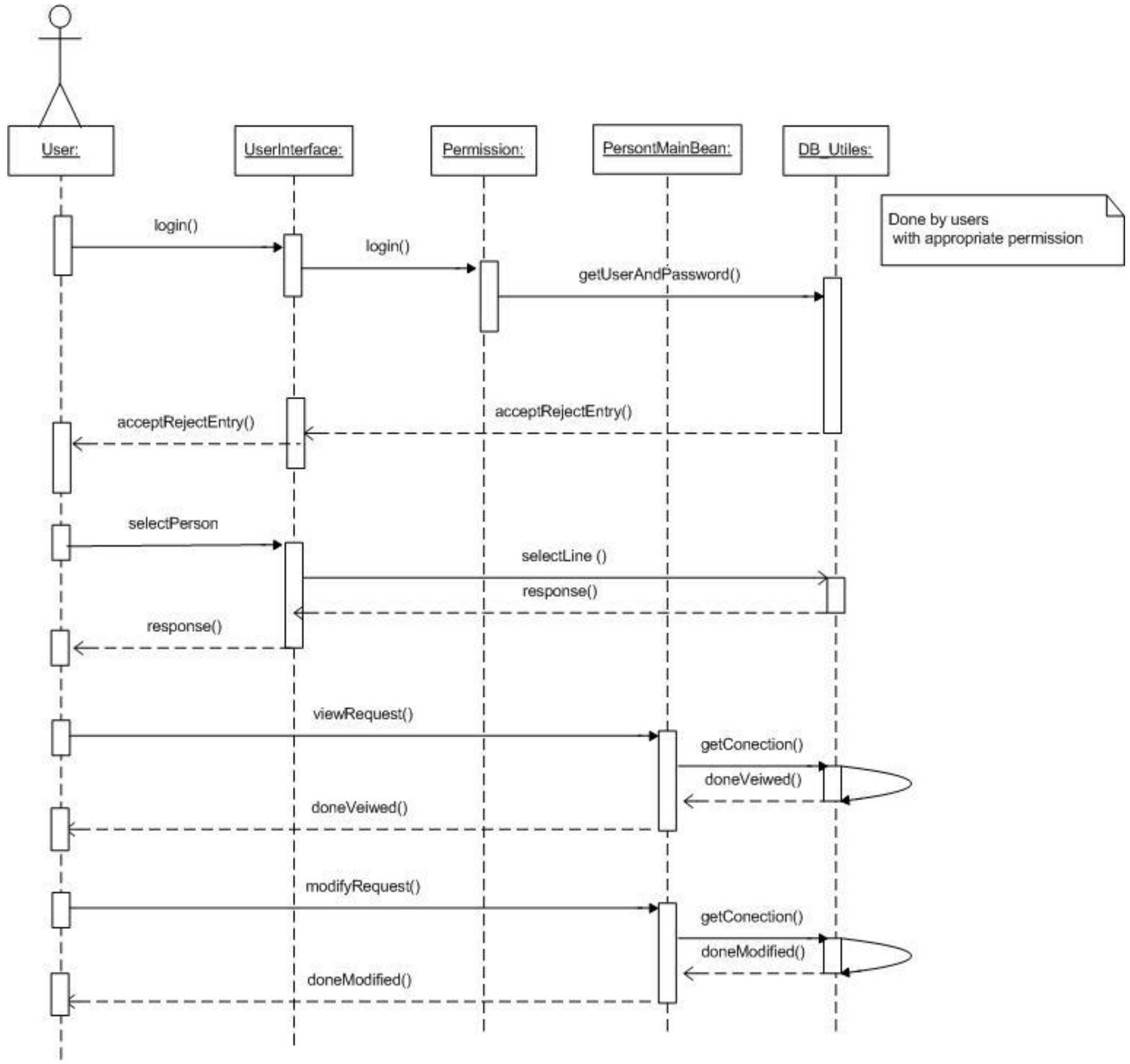

**Figure 25.** Sequence diagram for View / Modify - person

## 5.13. Add / View / Modify – faculty / department

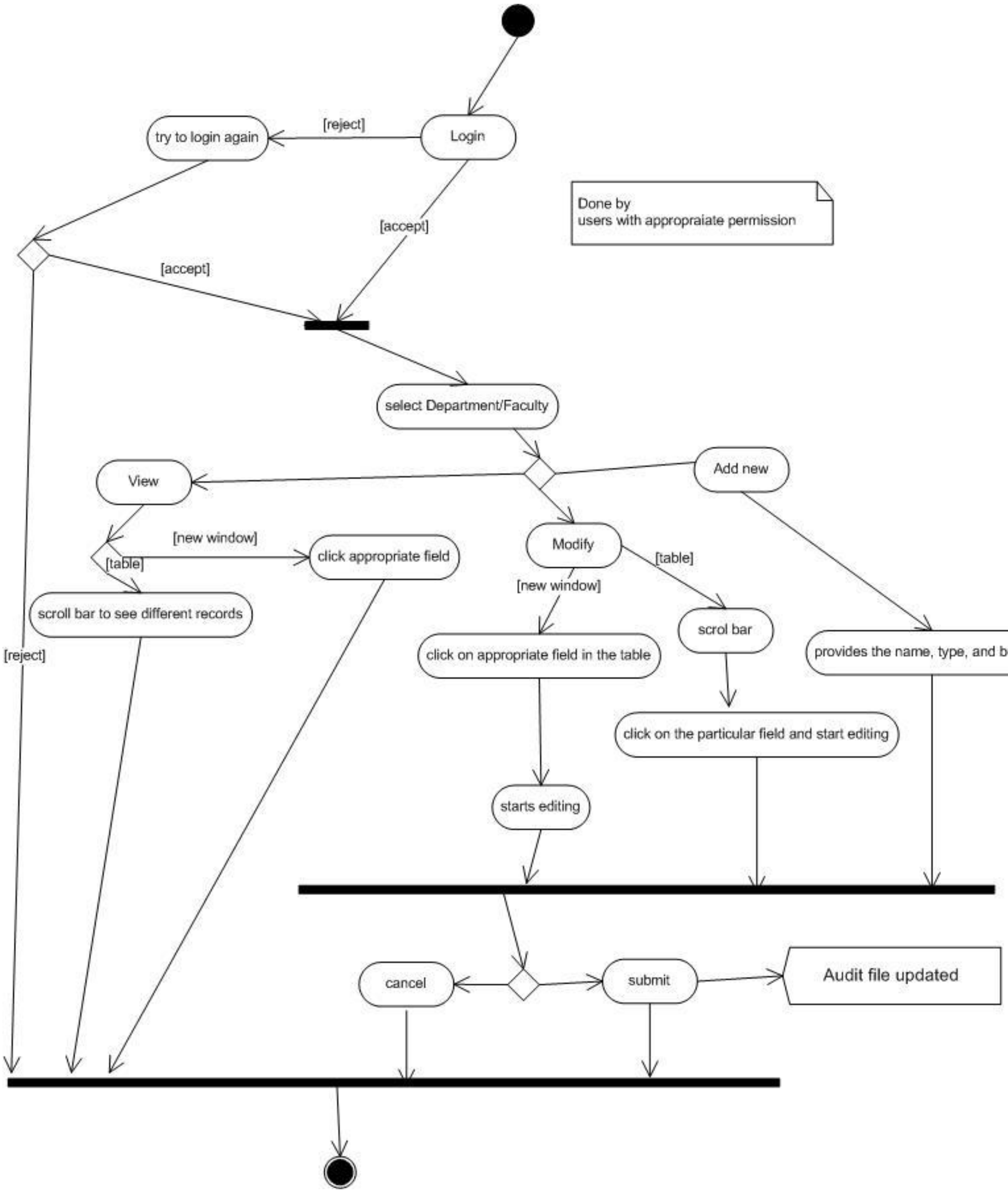

**Figure 26.** Activity diagram for Add / View / Modify – faculty / department

## 5.14. Report

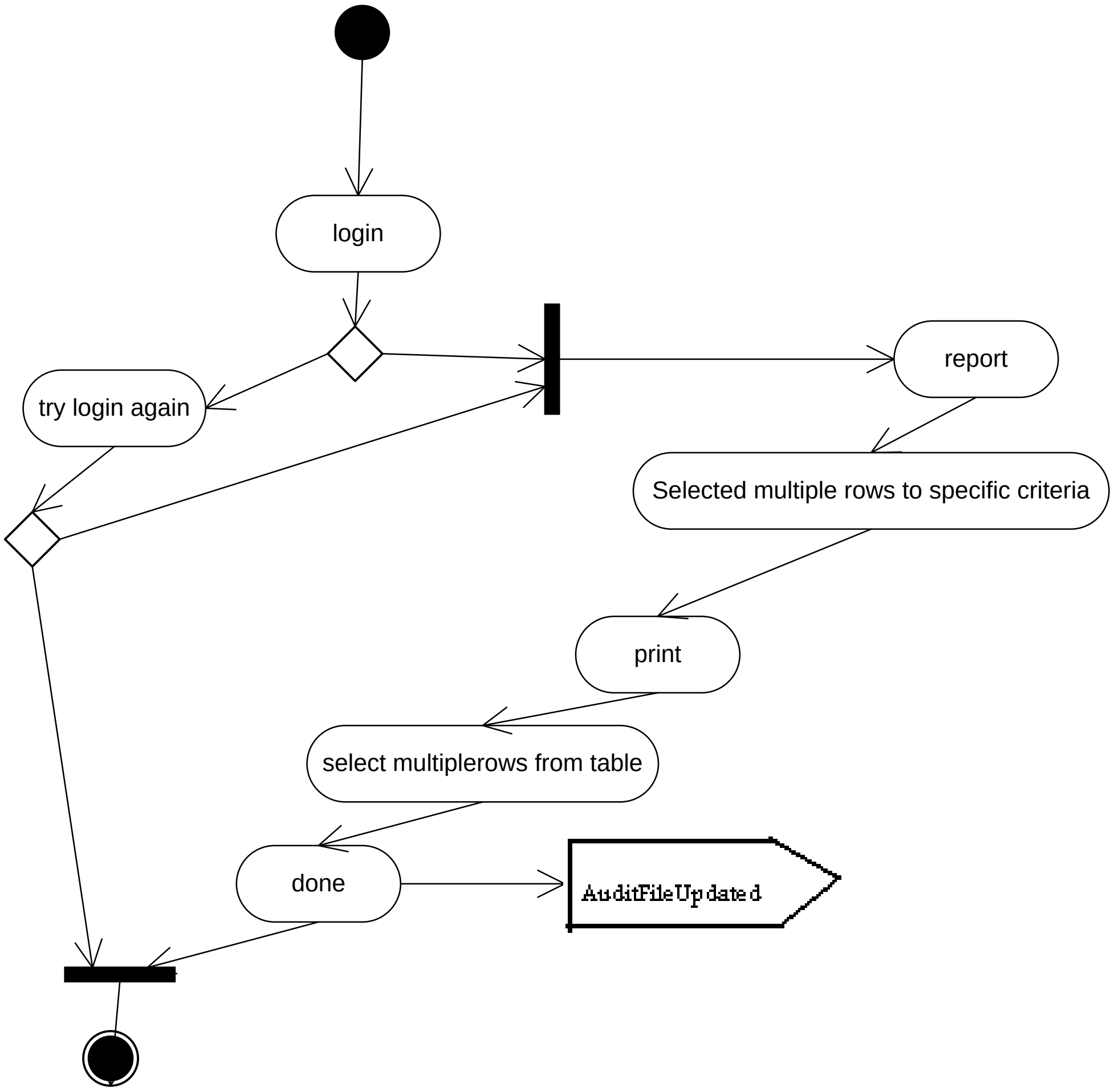

**Figure 27.** Activity diagram for Report

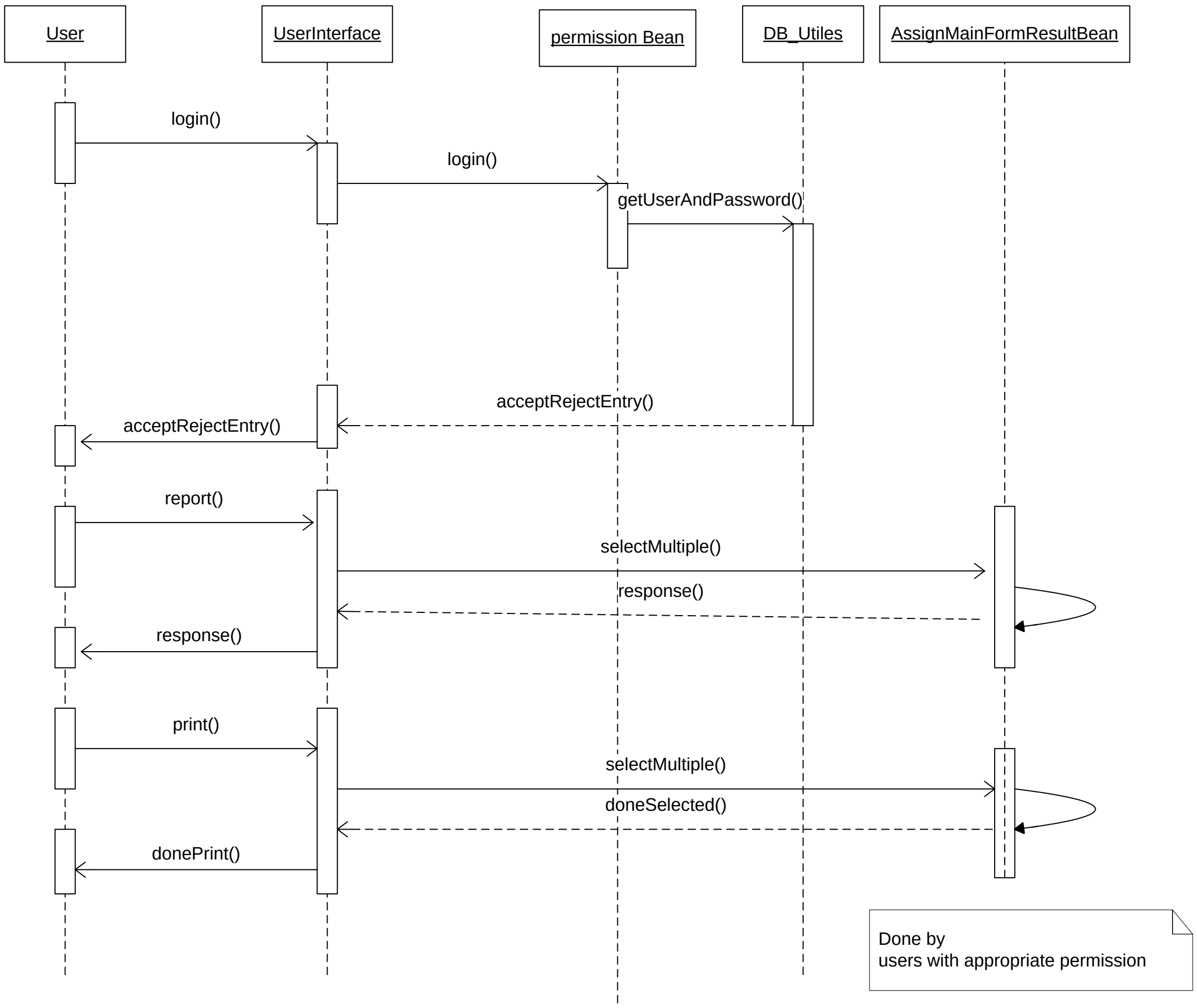

**Figure 28.** Sequence diagram for Report

## 5.15. Borrow – asset / location

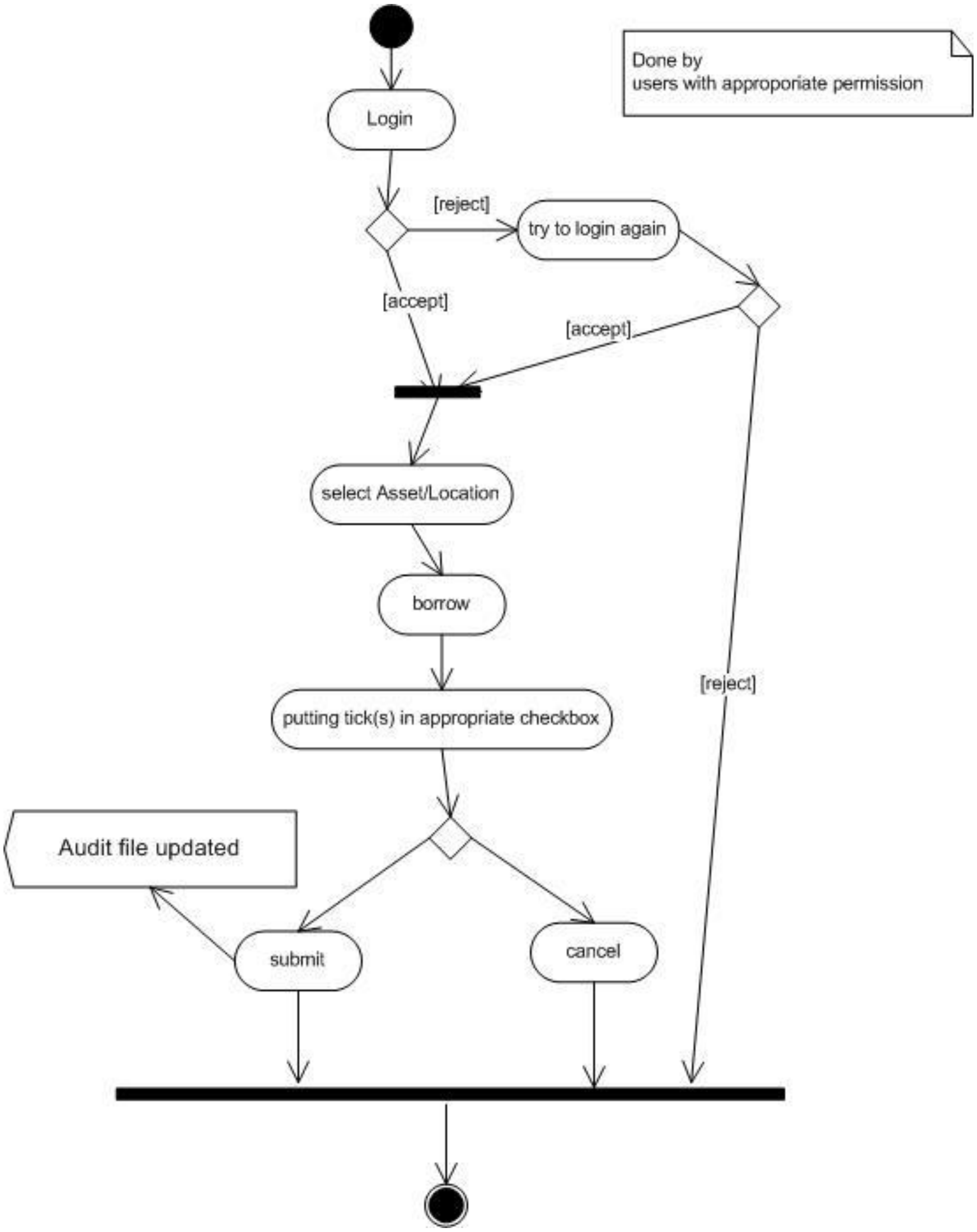

**Figure 29.** Activity diagram for Borrow – asset / location

## 5.16. View/Print – plan of big location

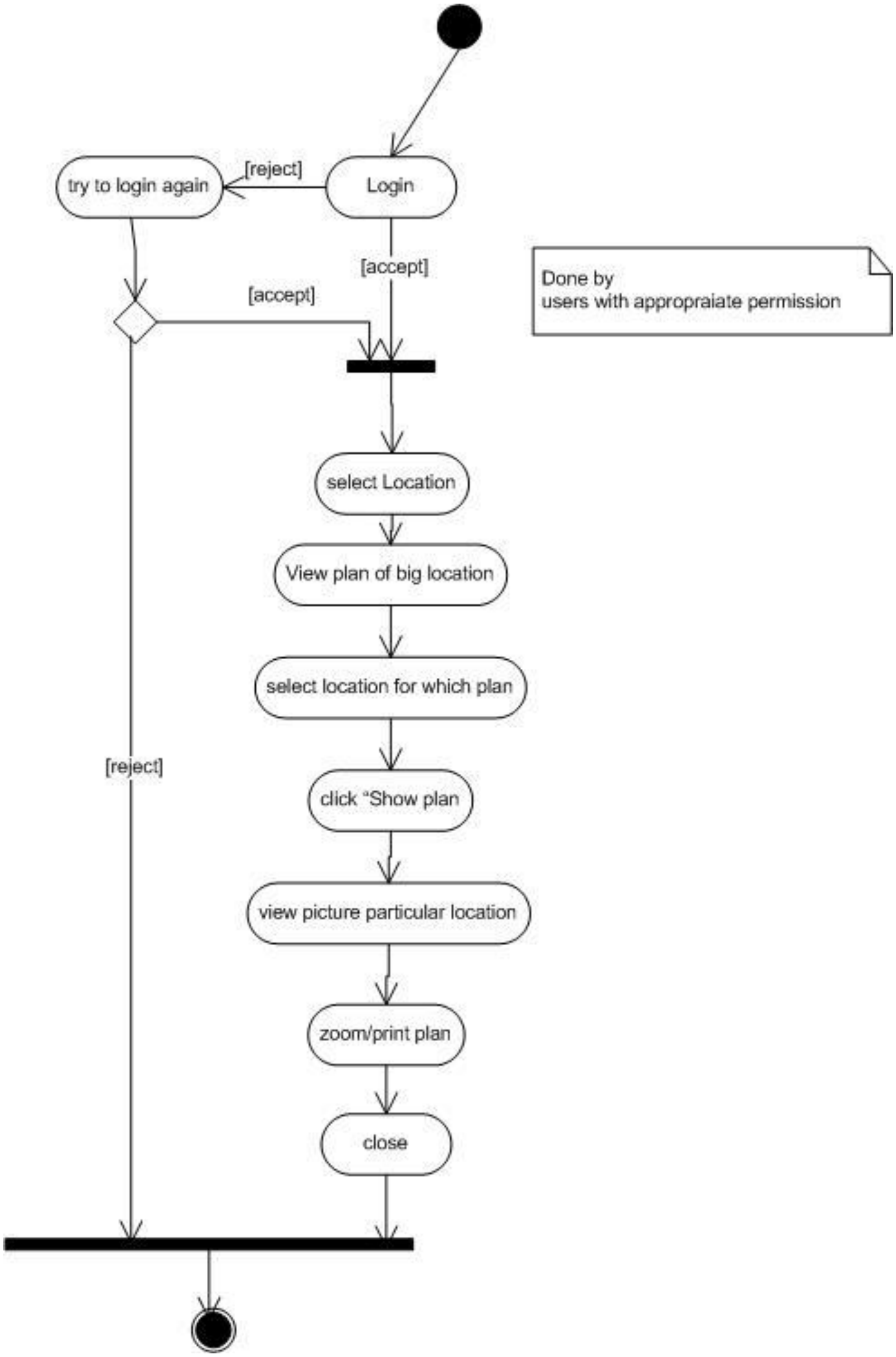

**Figure 30.** Activity diagram for View / Print – plan of big location

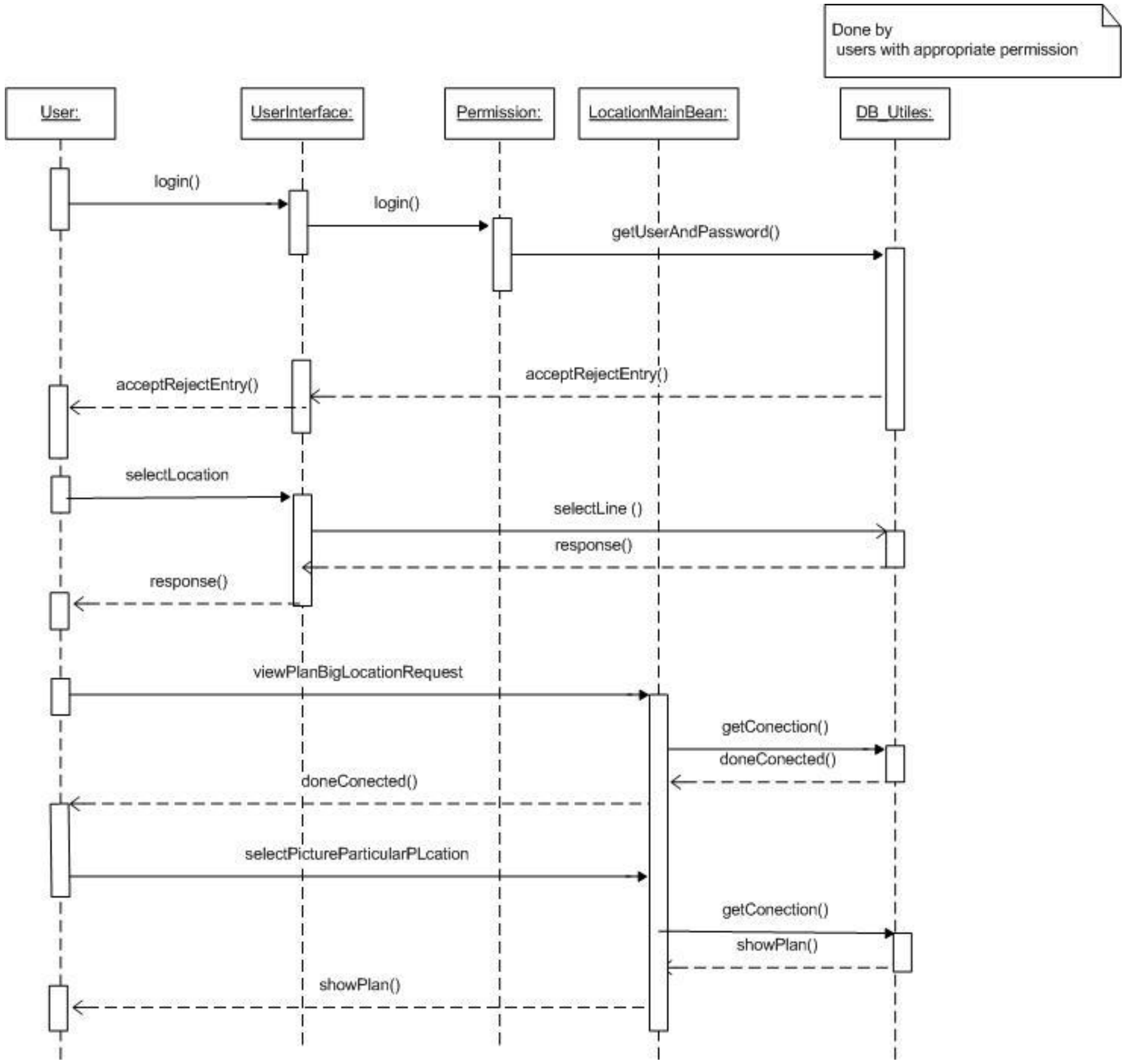

**Figure 31.** Sequence diagram for View / Print – plan of big location

# 6. Software interface design

## 6.1. User interface design

UI is designed according to UI design principles.

**The structure principle**: UI is organized in such a way that related things are combined together and unrelated things are separated.

**The simplicity principle:** It is easy to follow the provided interface. In the case of mistake, system displays error message.

**The visibility principle**: All system's functions are available through UI. It does not overwhelm users with too many alternatives.

**The feedback principle**: Through the system of messages, the design keeps users informed of actions, errors, or exceptions.

**The reuse principle**: In design, same names were used to perform the same operations with different objects in order to reduce ambiguity.

### 6.1.1. Web pages in a tree

The system's web pages are presented in a tree in **Error! Reference source not found.**31. From "Welcome" page user can reach "Main" page. From "Main" page user can reach following pages: "Asset", "License", "Location", "Person", "Administration", "Faculty and Department", "Requests", "Search", "Report". All these pages cover necessary functionality of system. It is easy to navigate between these pages. User constantly has access to it through the menu on the left side of page. Note: user has access to welcome page only after login, he can't come back to it, because this page has descriptive characters and doesn't have influence on functionalities of the system.

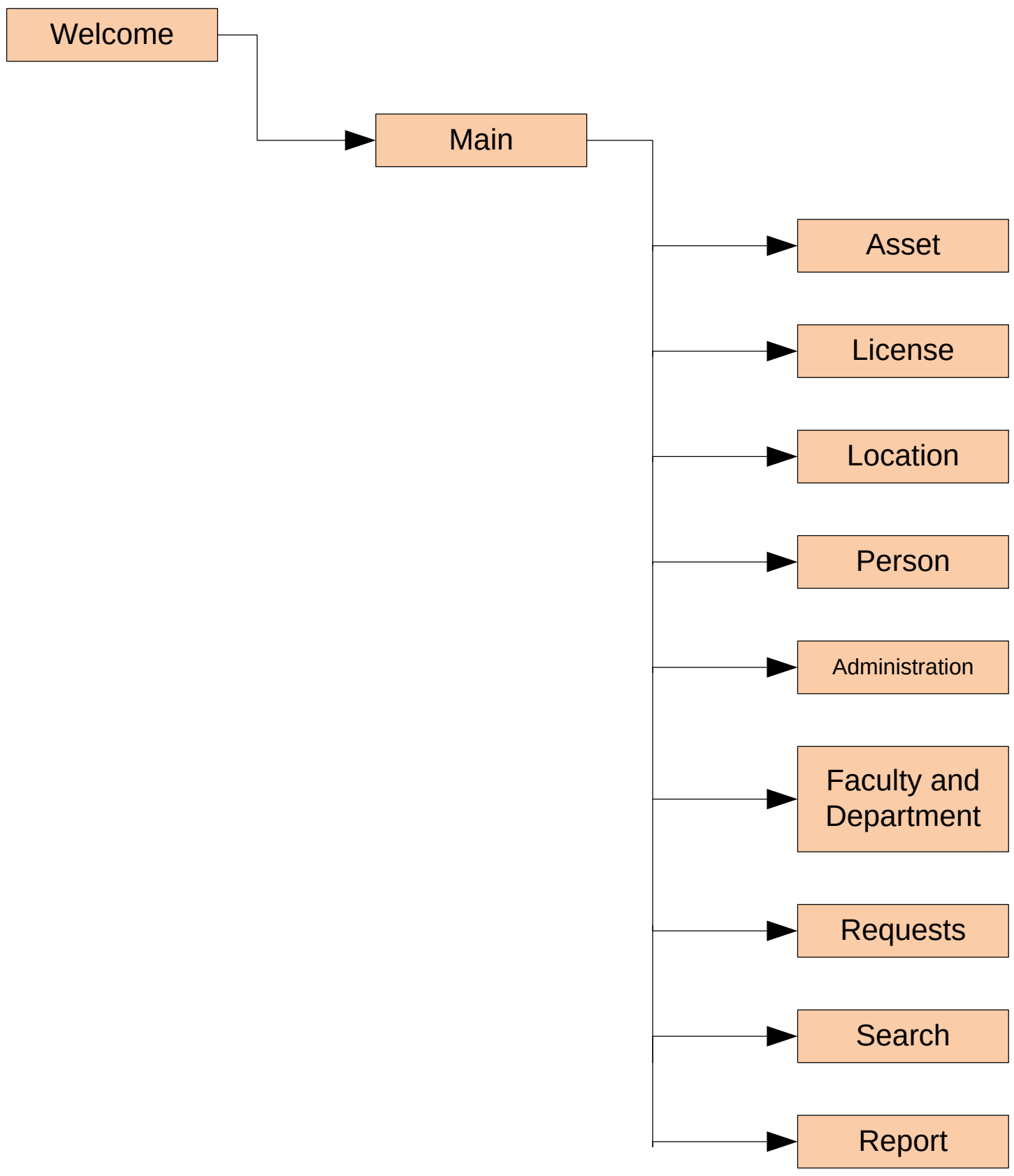

**Figure 32.** A tree of web pages

### 6.1.1.1. Description

**"Welcome"** page has descriptive characters; it contains a list of main system's functionality and contact information. After login "Welcome" page changes and guides user on how to work with system.

**"Main"** page is constantly present on the left side of the screen and contains menu, which covers main functionalities of the system.

Page **"Asset"** has its own menu on the top of the page, which contains all required operations that could be performed with assets.

Page **"License"** has its own menu on the top of the page, which contains all required operations that could be performed with licenses.

Page **"Location"** has its own menu on the top of the page, which contains all required operations that could be performed with locations.

Page **"Person"** has its own menu on the top of the page, which contains all required operations that could be performed with persons. Note: In inventory system, operations with DB "Person" are reduced according to requirements.

Page **"Administration"** has its own menu on the top of the page, which contains all required operations that could be performed with roles and permissions.

Page **"Faculty and Department"** has its own menu on the top of the page, which contains all required operations that could be performed with Faculties and Departments. Note: In inventory system operations with Faculties and Departments are reduced according to requirements.

Page **"Requests"** has its own menu on the top of the page, which contains all required operations that could be performed with requests.

Page **"Search"** has its own menu on the top of the page, which allows the user to perform basic and advanced search.

Page **"Report"** has its own menu on the top of the page, which contains all required operations that could be performed with reports.

### 6.1.1.2. Objects and Actions

In the **"Welcome"** page user provides login and password in appropriate text boxes and confirms this operation, clicking on button "Submit". For high privileged users system will ask to provide biometric characteristic (voice). Also user has opportunity to select language from down drop menu.

After logging **"Main"** page is available to user. It contains following menu: "Asset", "License", "Location", "Person", "Administration", "Faculty and Department", "Requests", "Search", "Report" and link "Logout". User clicks on necessary item in the menu in order to transfer in the next page.

Page **"Asset"** has menu with submenus on the top of the page:

1. Asset
    - Add new
    - View

- Delete
- Borrow
- Create group
- Create new type
- Create new subgroup

2. Import
    - Import from *.csv file/scanner
3. Assign to
    - Assign to person
    - Assign to location
4. My profile
    - View asset(s) assigned to me
    - View asset(s) borrowed by me

User clicks on necessary item in the menu in order to perform operation. Description of operation can be found in the Use Cases.

In addition user can perform next actions:

- see all available fields and records in the table, using scroll bar;
- sort records in the table by clicking on the button ↓ next to the name of field;
- set a filter selecting type of asset from drop down menu in the group “Filter” and then clicking button “Apply”;
- hide/show columns in the table by clicking link “Hide/Show” next to the name of field;
- select number of records per page by clicking available links “15”, “50”, “100”, “150”, “200”;
- navigate between pages with table DB “Asset” by clicking available links “1”, “2”, … ;
- perform basic search for current page putting string in appropriate text box and clicking button “Search”;
- edit record in the table by clicking link “Edit” next to the appropriate record;

Page **“License”** has menu with submenus on the top of the page:

1. License
    - Add new
    - View all licenses
    - View asset’s licenses
    - Delete
    - Borrow
    - Create new type
2. Import
    - Import from *.csv file/scanner
3. Assign to
    - Assign to asset

4. My profile
    - View license(s) assigned to me
    - View license(s) borrowed by me

User clicks on necessary item in the menu in order to perform operation. Description of operation can be found in the Use Cases.
The list of addition operations is the same as for “Asset” (see above).

Page **“Location”** has menu with submenus on the top of the page:

1. Location
    - Add new
    - View
    - Delete
    - Create group
    - Create new type
    - View plan of big location
    - Print plan of big location
2. Import
    - Import  from *.csv file
3. Assign to
    - Assign to person
    - Assign to another location
    - Assign to Department
4. My profile
    - View location(s) assigned to me

User clicks on necessary item in the menu in order to perform operation. Description of operation can be found in the Use Cases.
The list of addition operations is the same as for “Asset” (see above).

Page **“Person”** has menu with submenus on the top of the page:

1. Person
    - View
    - Delete
    - Create new type
    - Provide biometrical characteristic
2. Import
    - Import  from *.csv file

User clicks on necessary item in the menu in order to perform operation. Description of operation can be found in the Use Cases.
The list of addition operations is the same as for “Asset” (see above).

Page **"Administration"** has menu with submenus on the top of the page:

1. Role
   - Add new role (package of permissions)
   - Edit role
2. Permission
   - Add new permission
   - Edit user's permission
3. Assign to
   - Assign person(s) to role
   - Assign person to location(s)
   - Assign role to person(s)
   - Assign permission to person(s)
4. My profile
   - View my role
   - View my permission

User clicks on necessary item in the menu in order to perform operation. Description of operation can be found in the Use Cases.
The list of addition operations is the same as for "Asset" (see above).

Page **"Faculty and Department"** has menu with submenus on the top of the page:

1. Faculty
   - Add new
   - View
   - Edit
2. Department
   - Add new
   - View
   - Edit

User clicks on necessary item in the menu in order to perform operation. Description of operation can be found in the Use Cases.
The list of addition operations is the same as for "Asset" (see above).

Page **"Requests"** has menu with submenus on the top of the page:

1. Add new
2. Approve/Reject
3. View list of all requests in the system

User clicks on necessary item in the menu in order to perform operation. Description of operation can be found in the Use Cases.

Page **"Search"** allows to user perform Basic search and Advanced search. Page contains text box for searched string, button "Search", table with names of databases and fields where search can be performed. User should put tick(s) in appropriate checkbox(es) in order to select databases. User should press "Ctrl" and make a click by mouse in the names of fields in order to specify name of fields where search should be performed.

Page **"Report"** has menu with submenus on the top of the page:

1. Create report
2. Print report
3. Auditing
4. My profile

User clicks on necessary item in the menu in order to perform operation. Description of operation can be found in the Use Cases.
A scheme presented, in figure below, shows the main web pages (pink rectangles) and main actions (blue rectangles) that can be performed on each page.

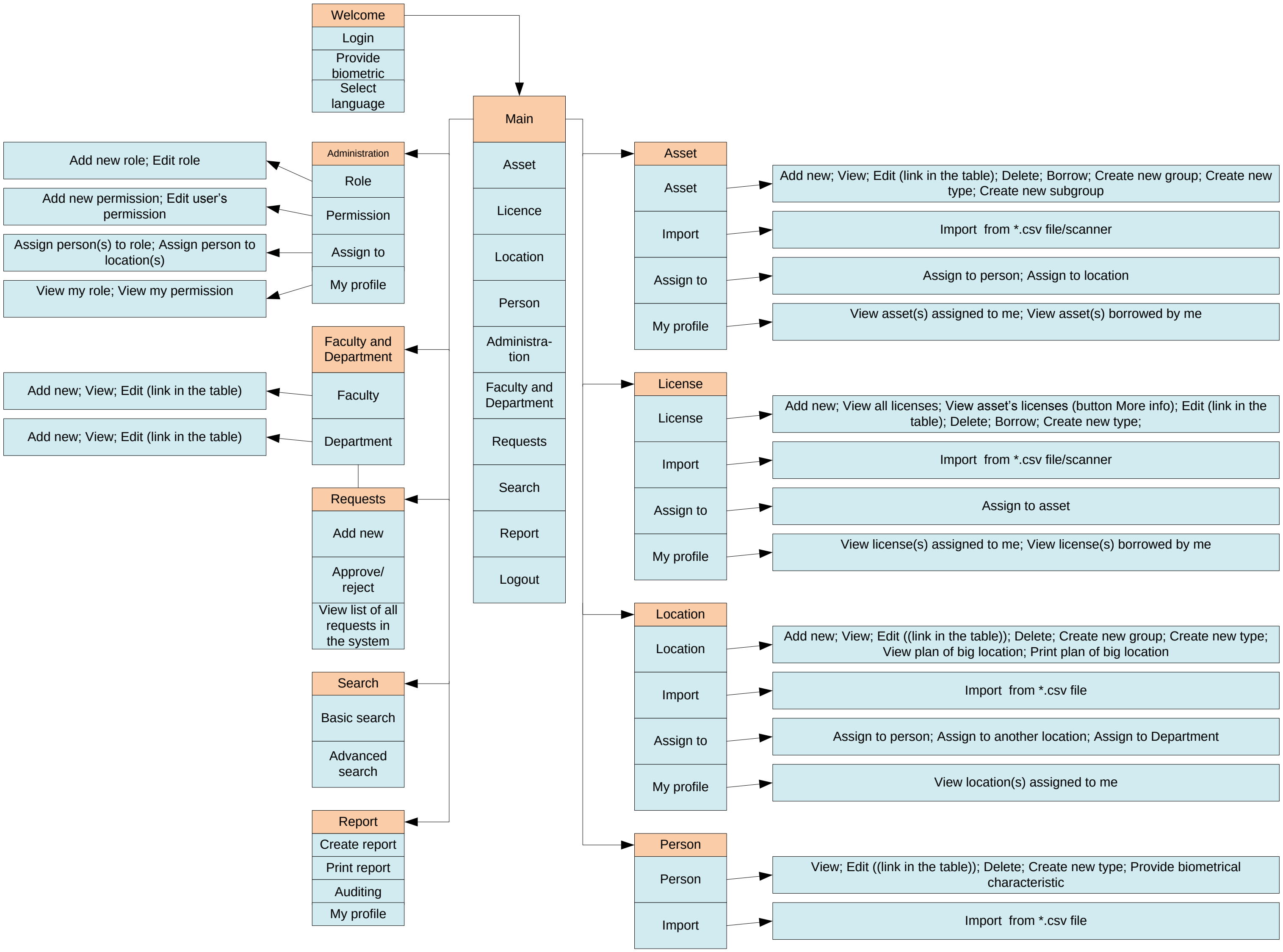

**Figure 33.** Scheme of the main web pages (pink rectangles) and main actions (blue rectangles) that can be performed on each page

## 6.1.2. User interface

Used GUI components are menus, submenus, buttons, text boxes, check boxes, down drop lists, links, and tables. The only means of access to the entire database, by all users, is through this UI.

### 6.1.2.1. Screen image

Some examples of UI are presented below:

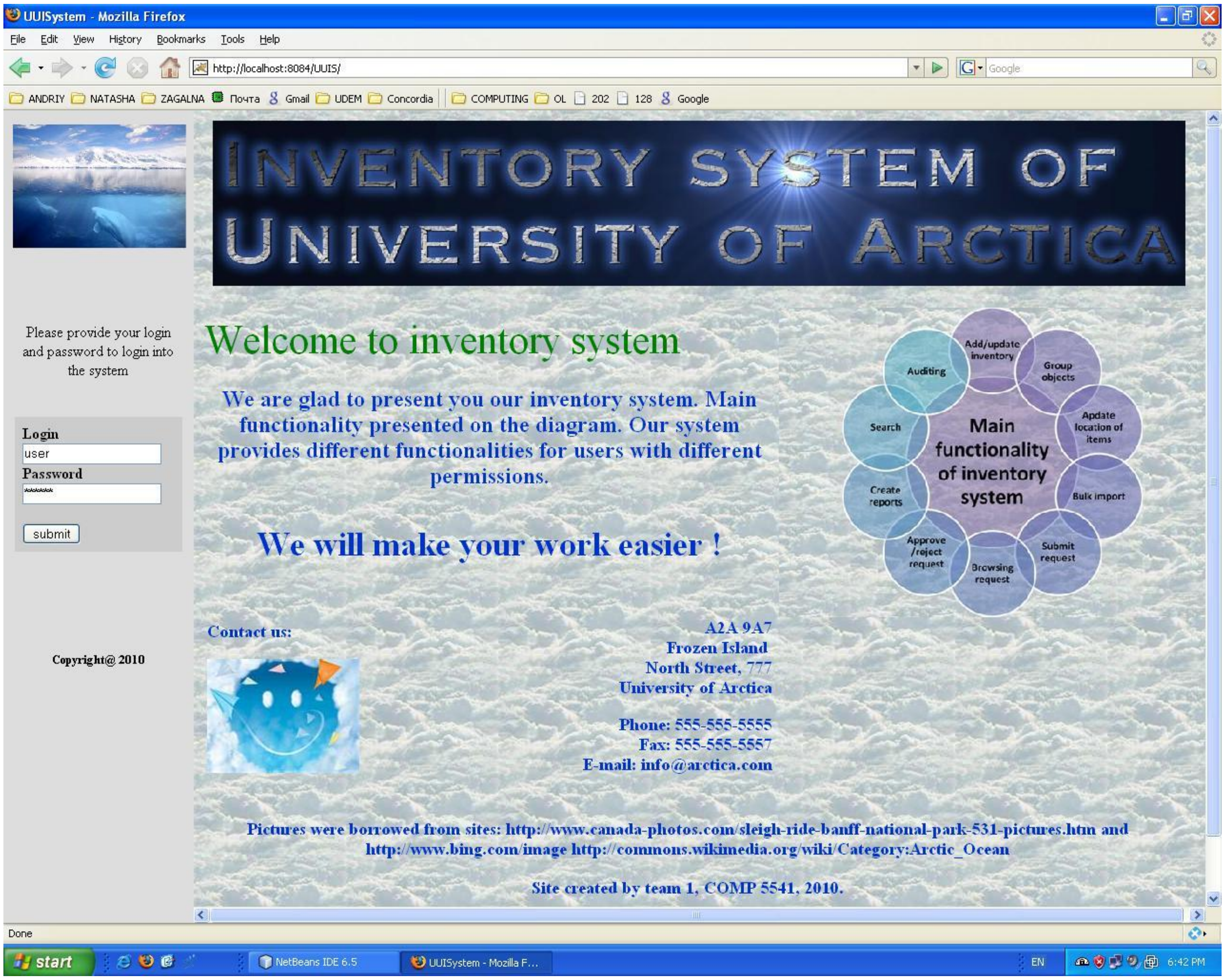

**Figure 34.** “Login” page

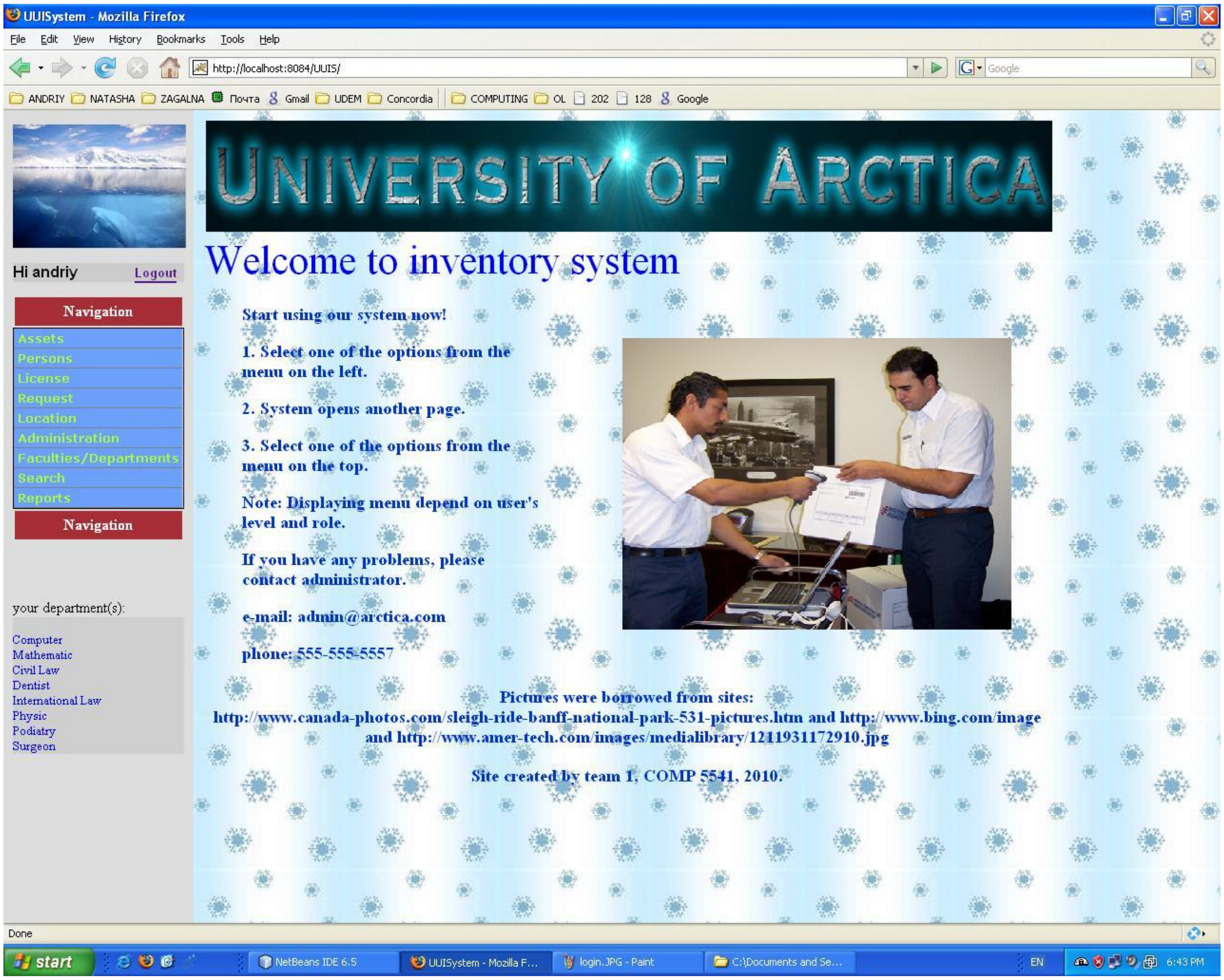

**Figure 35.** “Welcome” page after logging and main menu for administrator

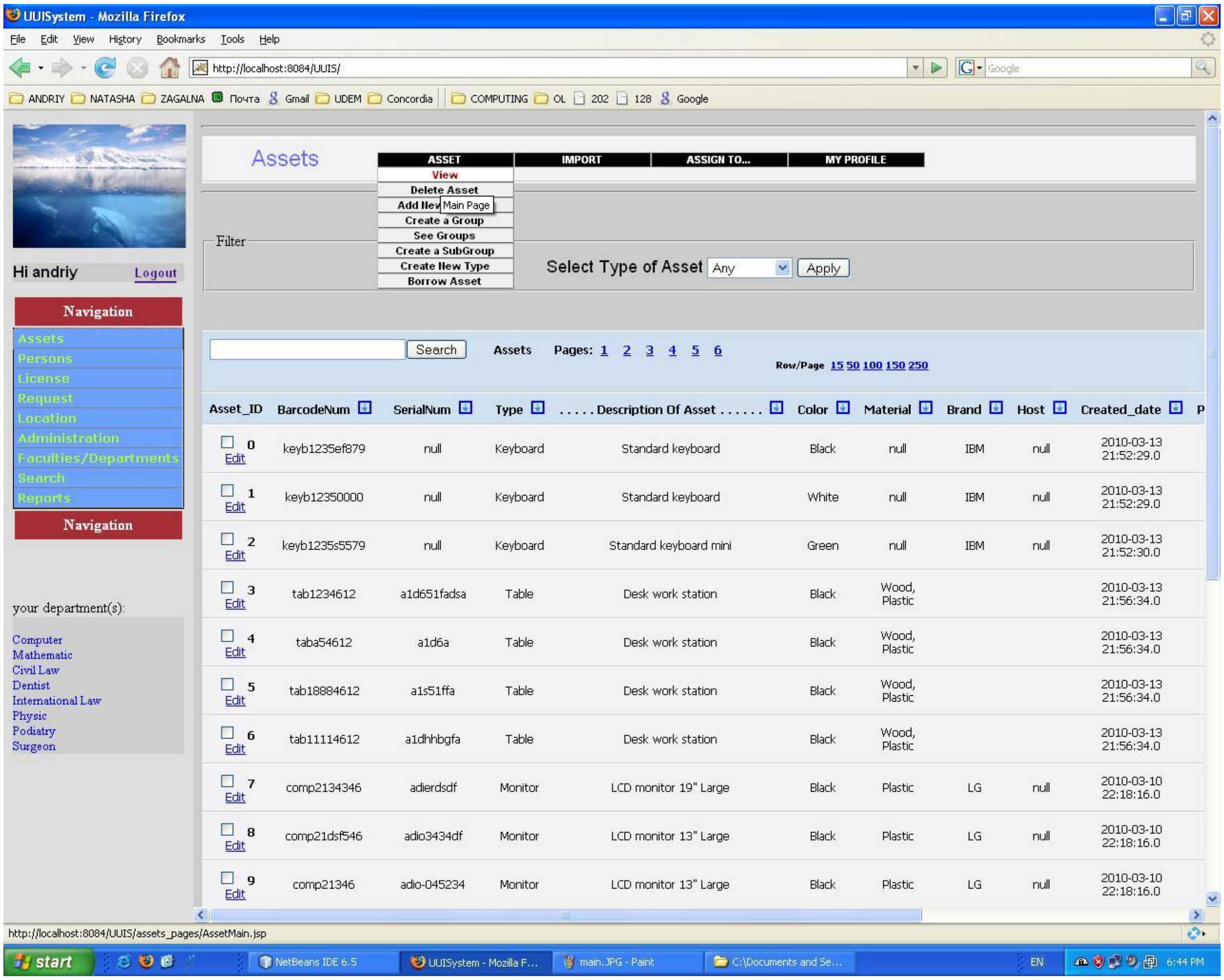

**Figure 36.** Page “Asset”

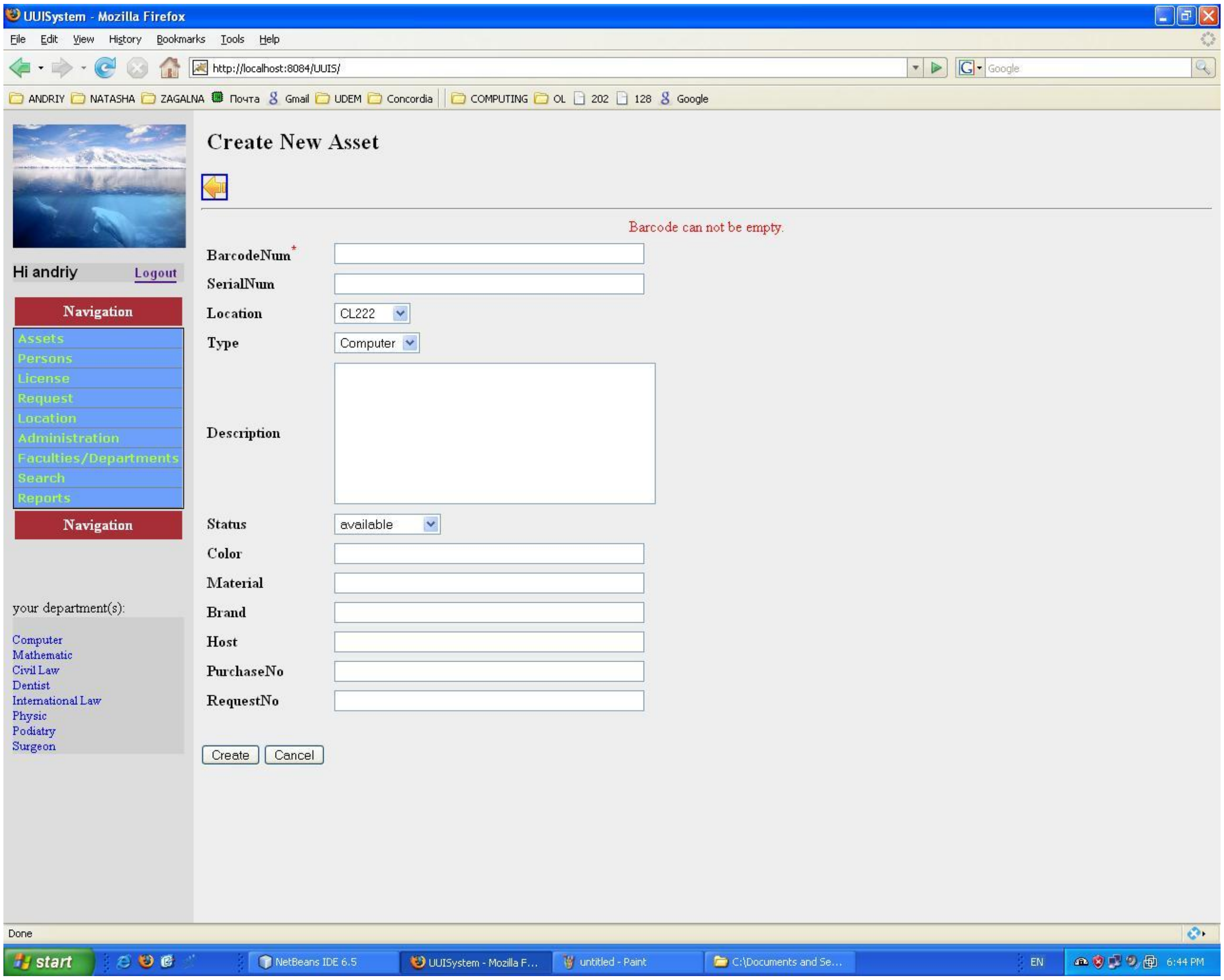

**Figure 37.** Add new Asset

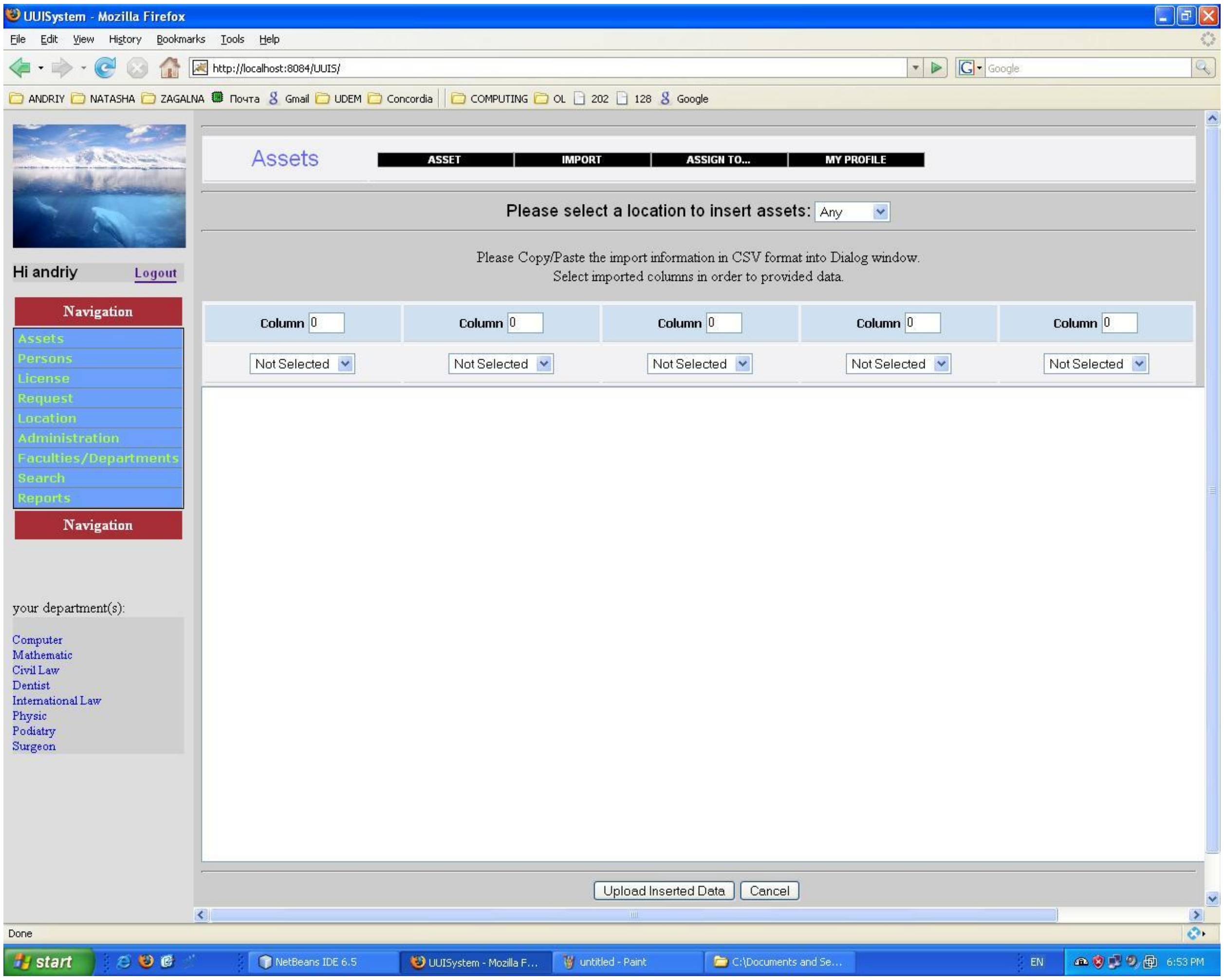

**Figure 38.** Import asset from *.csv file/scanner

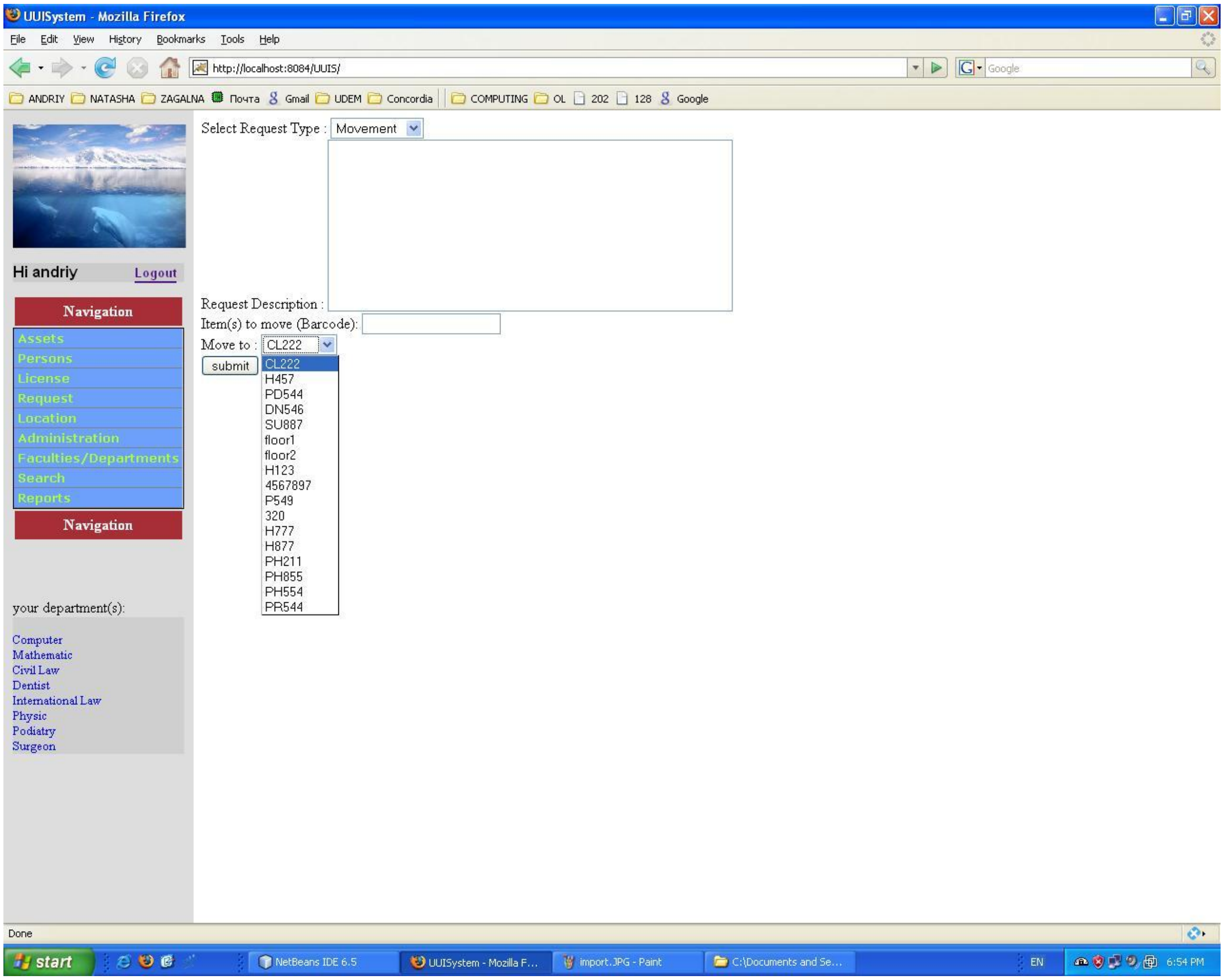

**Figure 39.** Add new request

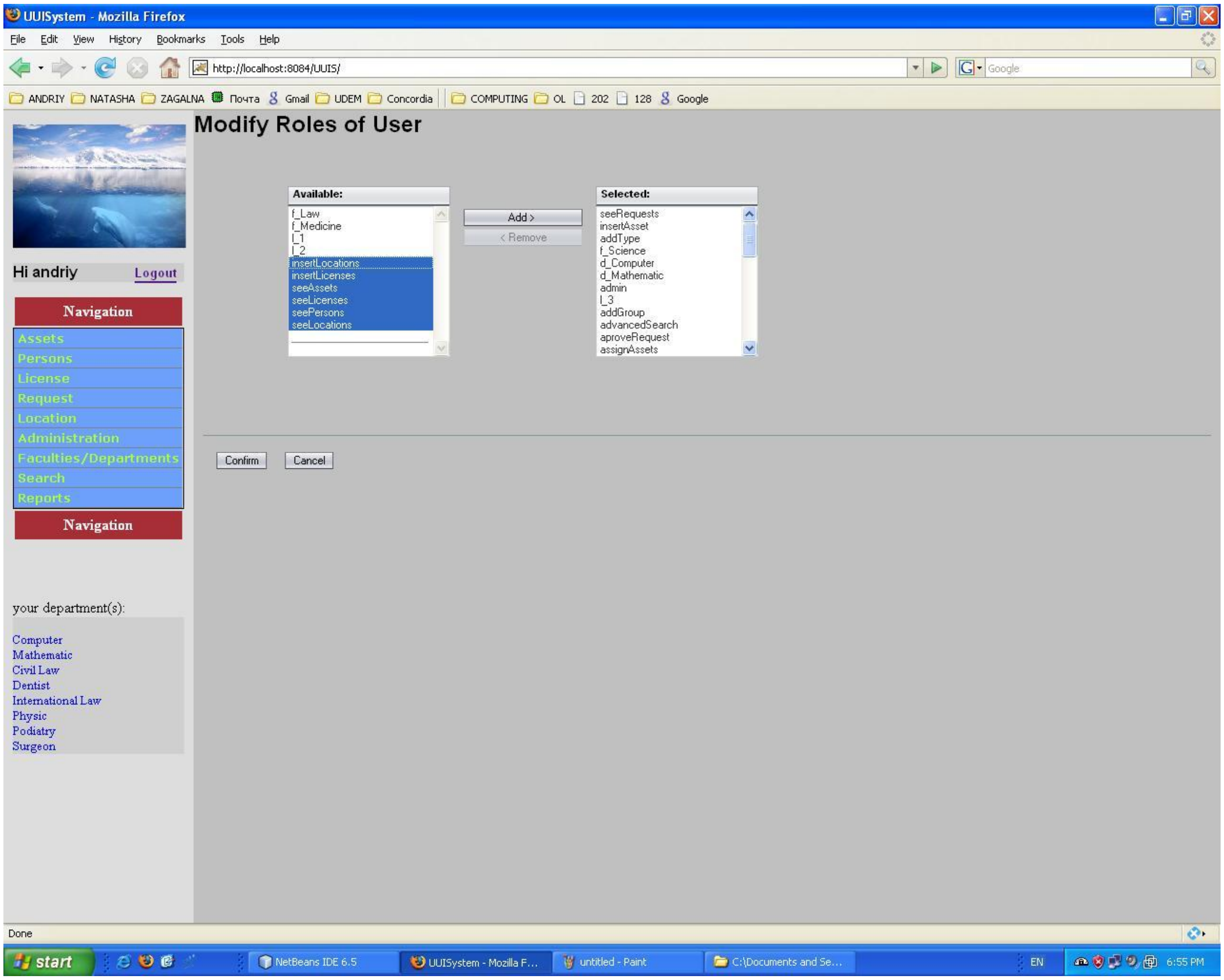

**Figure 40.** Modify roles of User

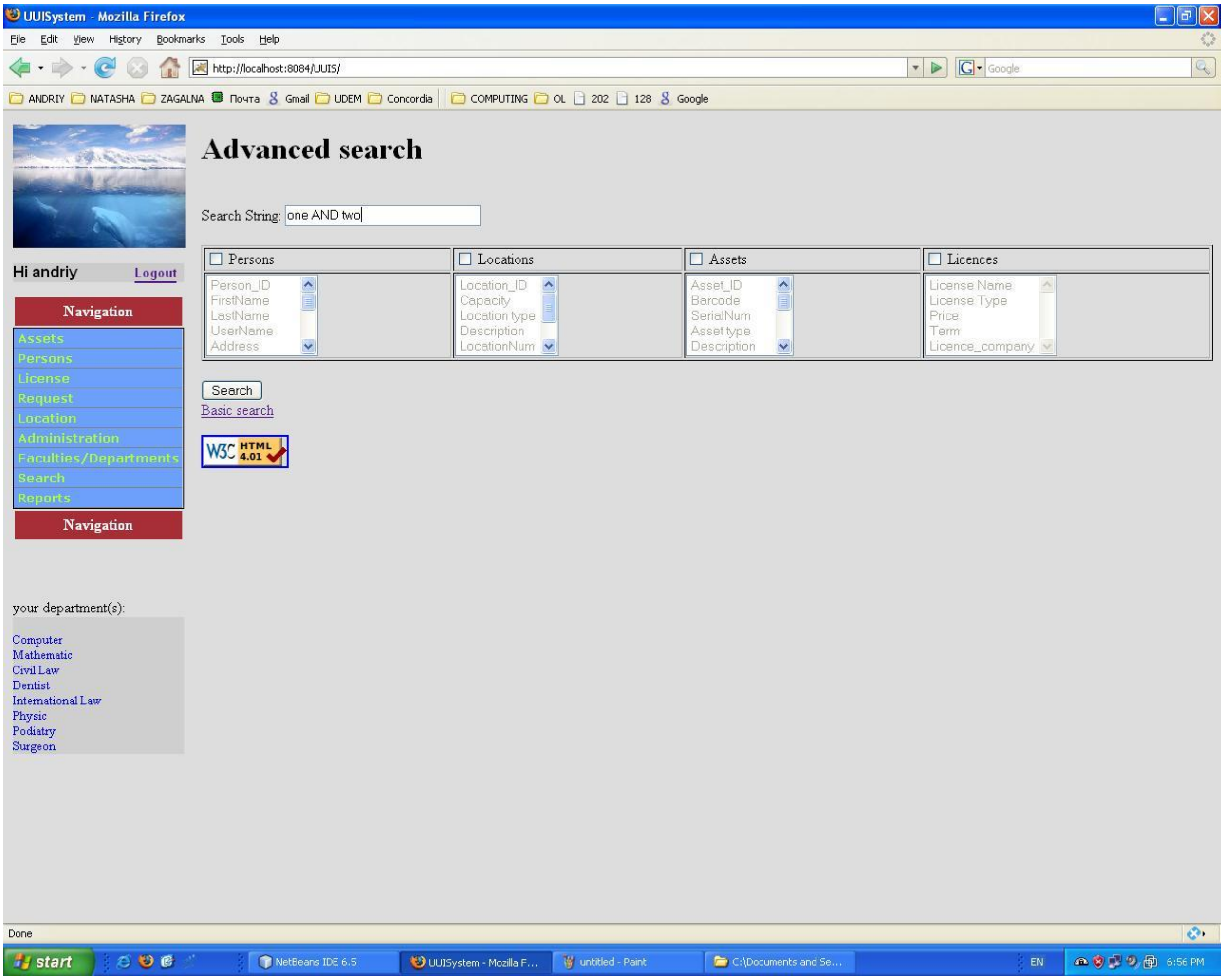

**Figure 41.** Advanced search

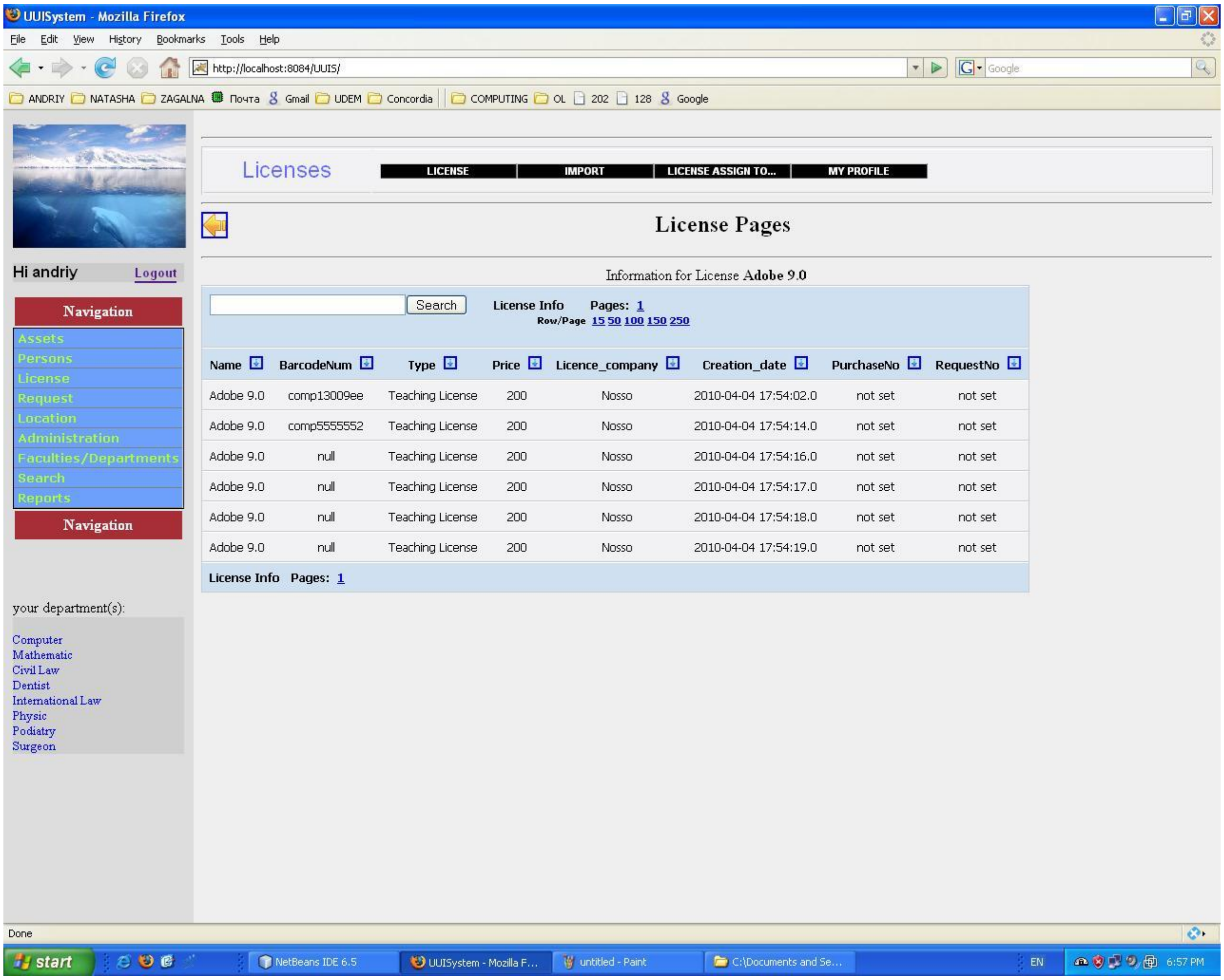

**Figure 42.** “License” page

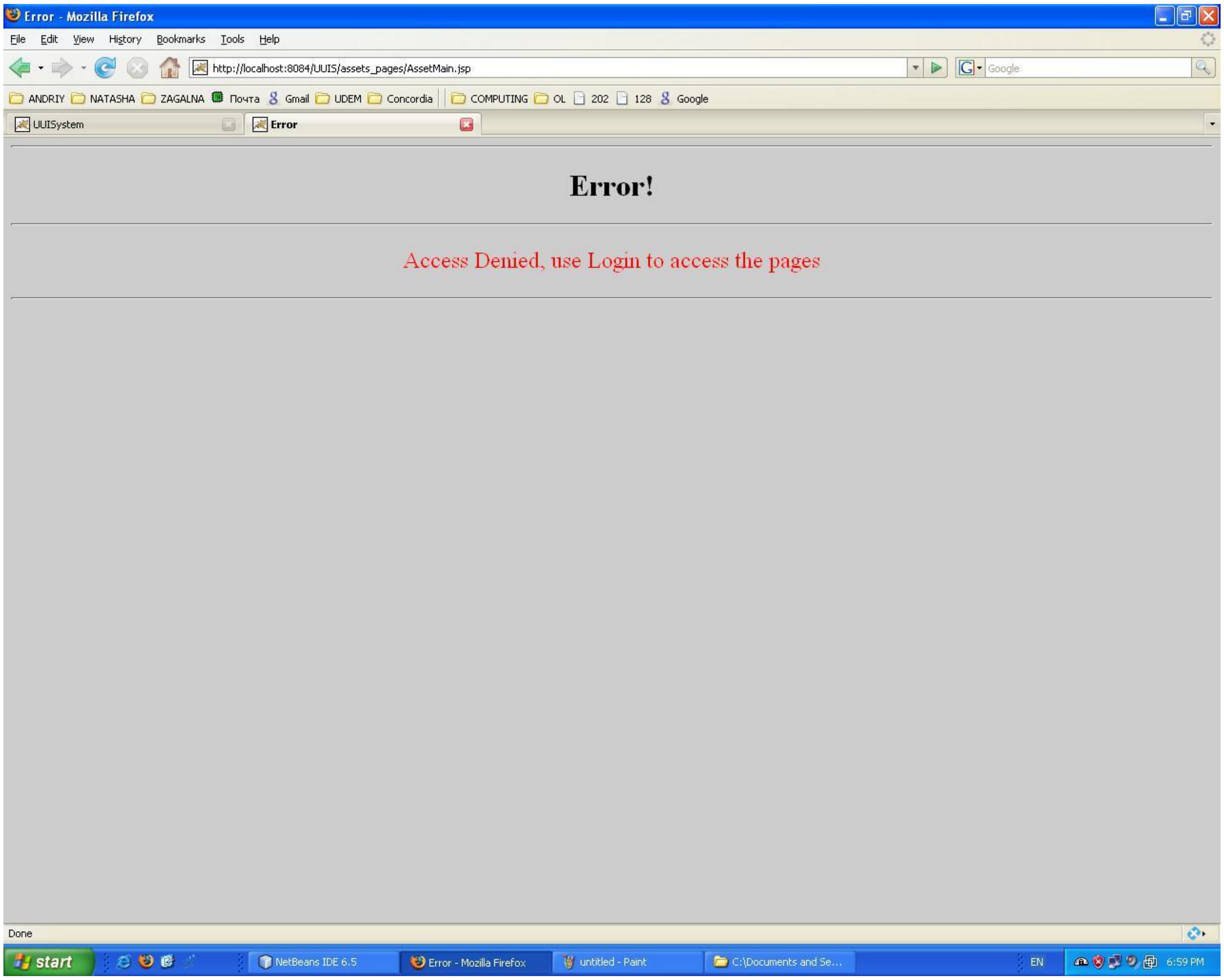

**Figure 43.** Provided security

## 6.2. Module interface design

Module design maintains MVC (Model – View - Controller) architecture. View is a UI. Through UI user inputs data which goes to Controller. Controller transfers data into Model. If data is incorrect Model shows error message. Otherwise it processes the request, prepares the result and sends it to the Controller. Finally, Controller transfers generated code into View. The user views the result.

General principle of MVC (Model – View - Controller) architecture can be described with sequential diagram (see in Figure 42) .

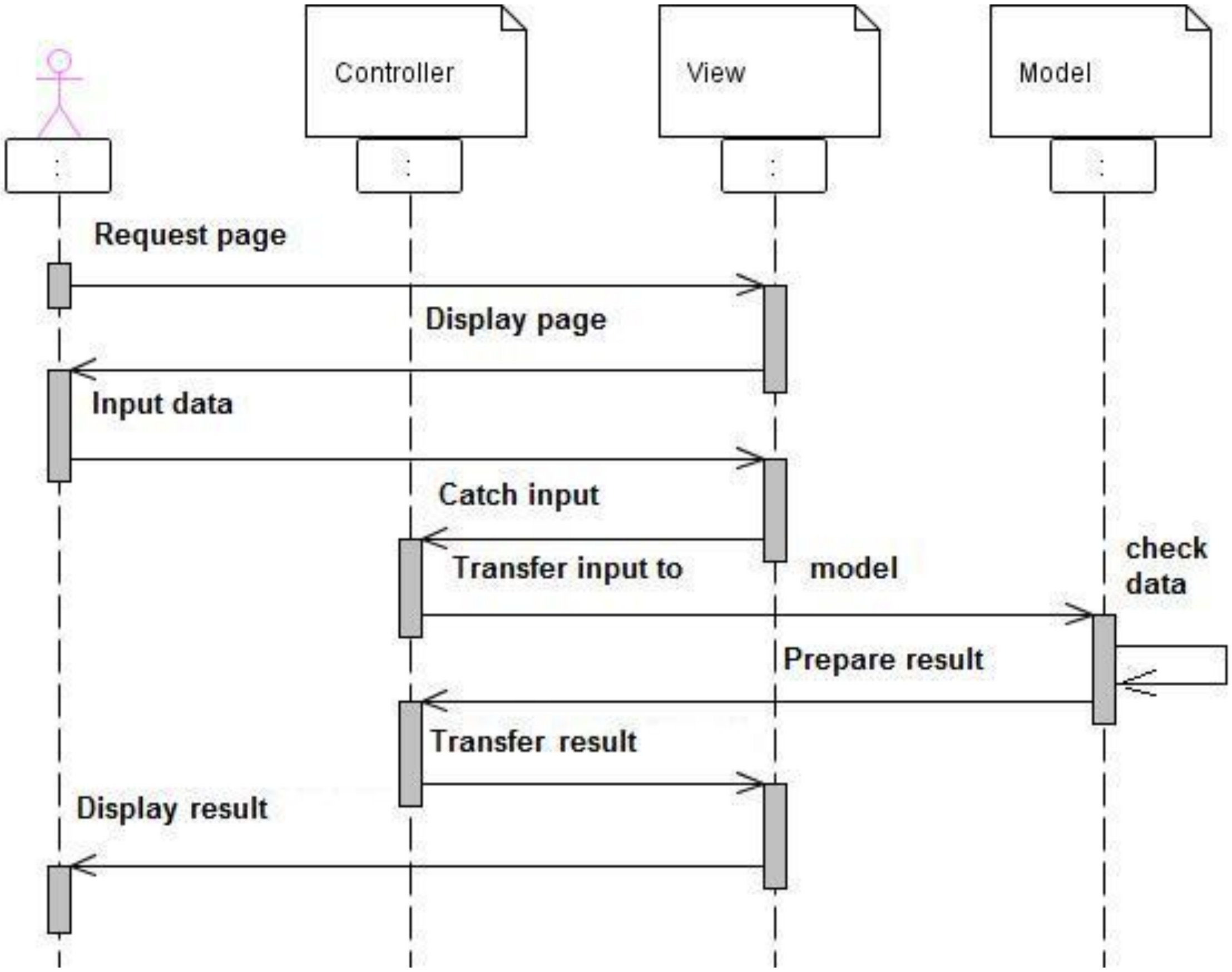

**Figure 44.** Sequence diagram of MVC (Model – View - Controller) architecture

# 7. Class diagrams

## 7.1. Basic folders for class diagram

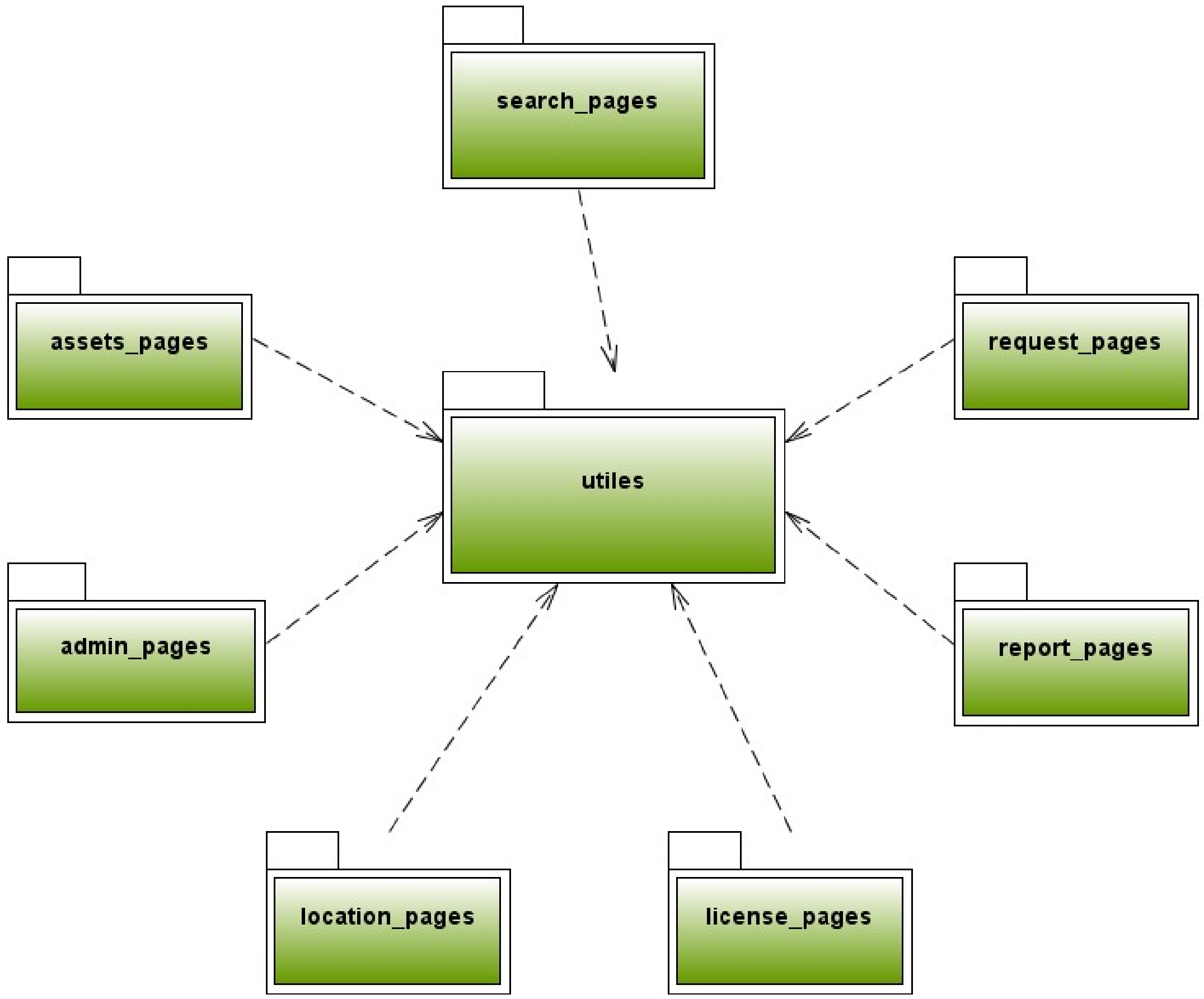

**Figure 45.** Basic folder for class diagram

## 7.2. Basic class diagram

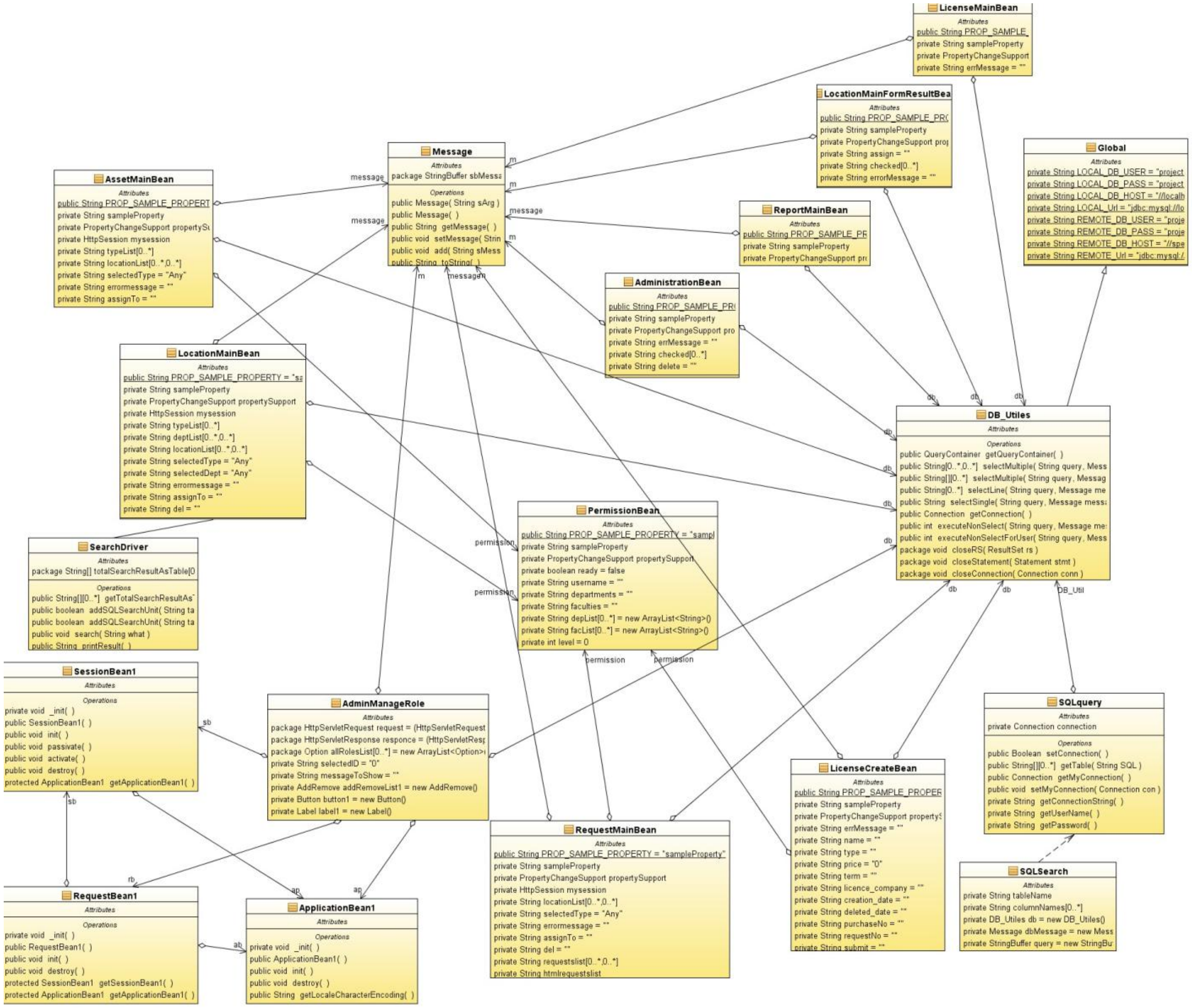

**Figure 46.** Basic class diagram

## 7.3. Full class diagram

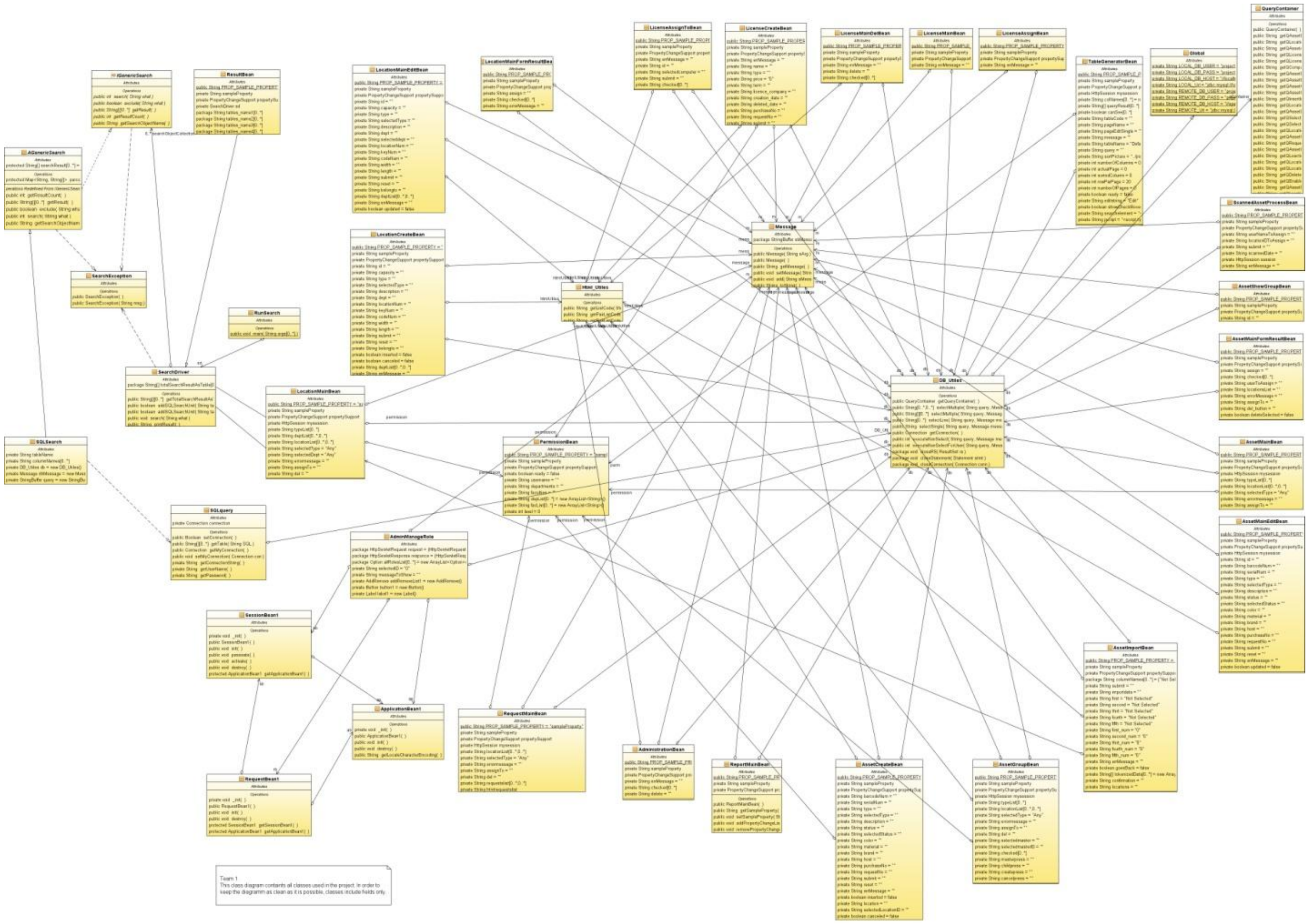

**Figure 47.** Full class diagram

## 7.4. Module Add new asset

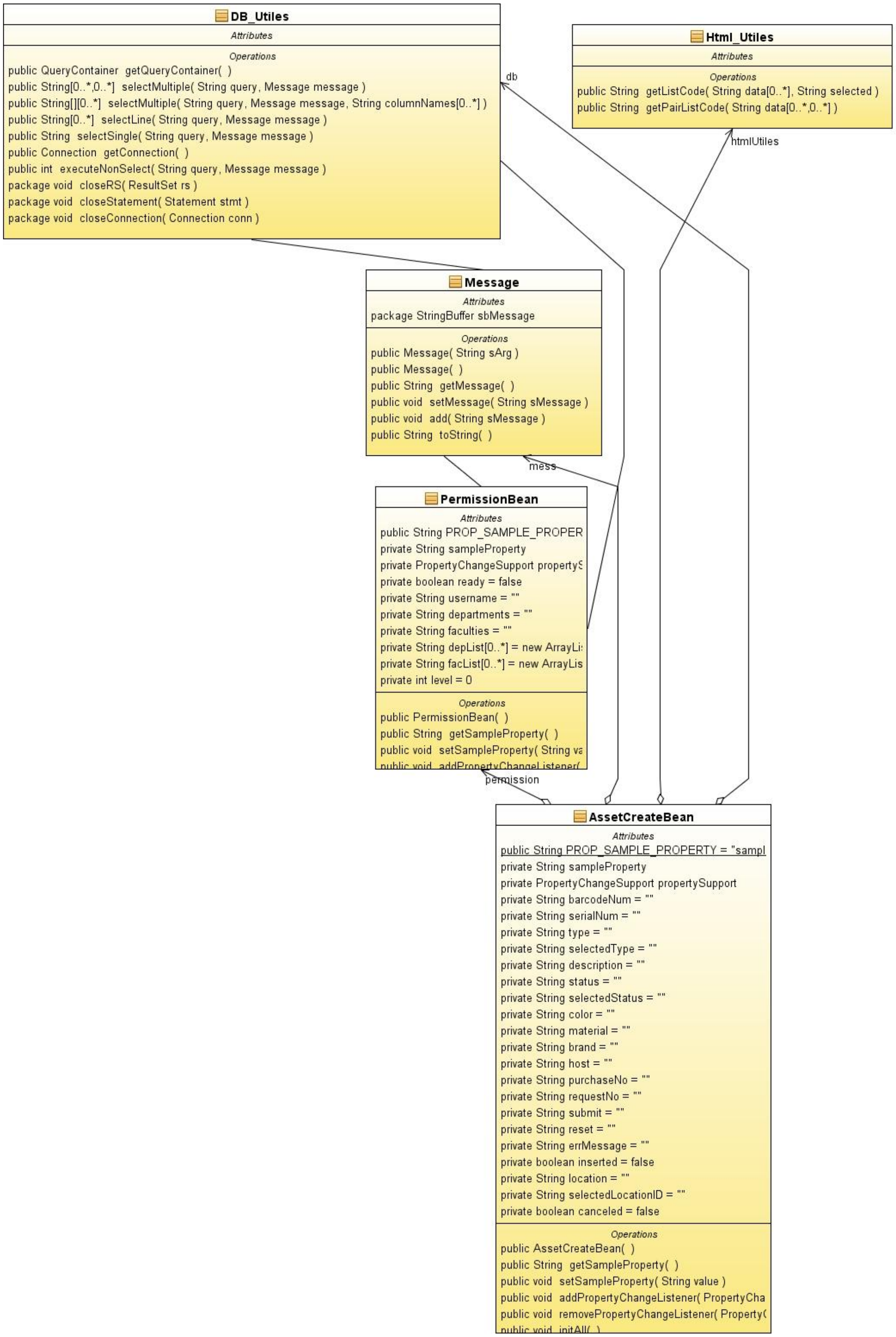

**Figure 48.** Class diagram for Add new asset

The Java class for adding a new asset is: "AssetCreateBean.java", it imports four other classes to use them for calling different methods. The four classes are : "DB_Utiles" , "Message", "PermissionBean" and "Html_Utiles".

"DB_Utiles" contains several methods to connect to the database, "AssetCreateBean" uses this class to select data from the Database, or Update the Database. The relationship between the two classes is a simple aggregation relationship. The parent (Container) class is "AssetCreateBean" and the child (Contained) class is "DB_Utiles"; the child class can still exist and function if the parent class is destroyed.

The main methods used by "AssetCreateBean" are : "selectMultiple()" → This method returns multiple rows; "selectLine()" → This method returns simple row; "selectSingle()" → This method gets a result from query that returns a single result; "executeNonSelect()"→ This method executes a query UPDATE, INSERT, DELETE

"Message" is used to accumulate a message as a mutable string. "AssetCreateBean" uses this class for error messages, to store and get messages in case an error occurred. The relationship between "AssetCreateBean" and "Message" is a simple aggregation relationship. The parent (Container) class is "AssetCreateBean" and the child (Contained) class is "Message"; the child class can still exist and function if the parent class is destroyed.

"PermissionBean" is used by "AssetCreateBean" to get the Permission of the user. The main method that "AssetCreateBean" calls from "PermissionBean" is : "loadPermissions()" → This method executes query and initializes all permissions. The relationship between "AssetCreateBean" and "PermissionBean" is a simple aggregation relationship. The parent (Container) class is "AssetCreateBean" and the child (Contained) class is "PermissionBean"; the child class can still exist and function if the parent class is destroyed.

"AssetCreateBean" uses "Html_Utiles" to get an array of options in a drop down list where one of the options can be chosen. The main method that "AssetCreateBean" calls from "Html_Utiles" is : "getListCode ()". The relationship between "AssetCreateBean" and "Html_Utiles" is a simple aggregation relationship. The parent (Container) class is "AssetCreateBean" and the child (Contained) class is "PermissionBean"; the child class can still exist and function if the parent class is destroyed.

There is an association relationship between "DB_Utiles" and "Message", and "PermissionBean". "DB_Utiles" uses "Message" to stores error messages if an error occurred while one of the methods of "DB_Utiles" is running, and this error can be retrieved from "Messages". "PermissionBean" uses "Message" in the same fashion. "PermissionBean" uses "DB_Utiles" to connect to the Data Base in order to get the Permission Level.

# 7.5. Module Add new request

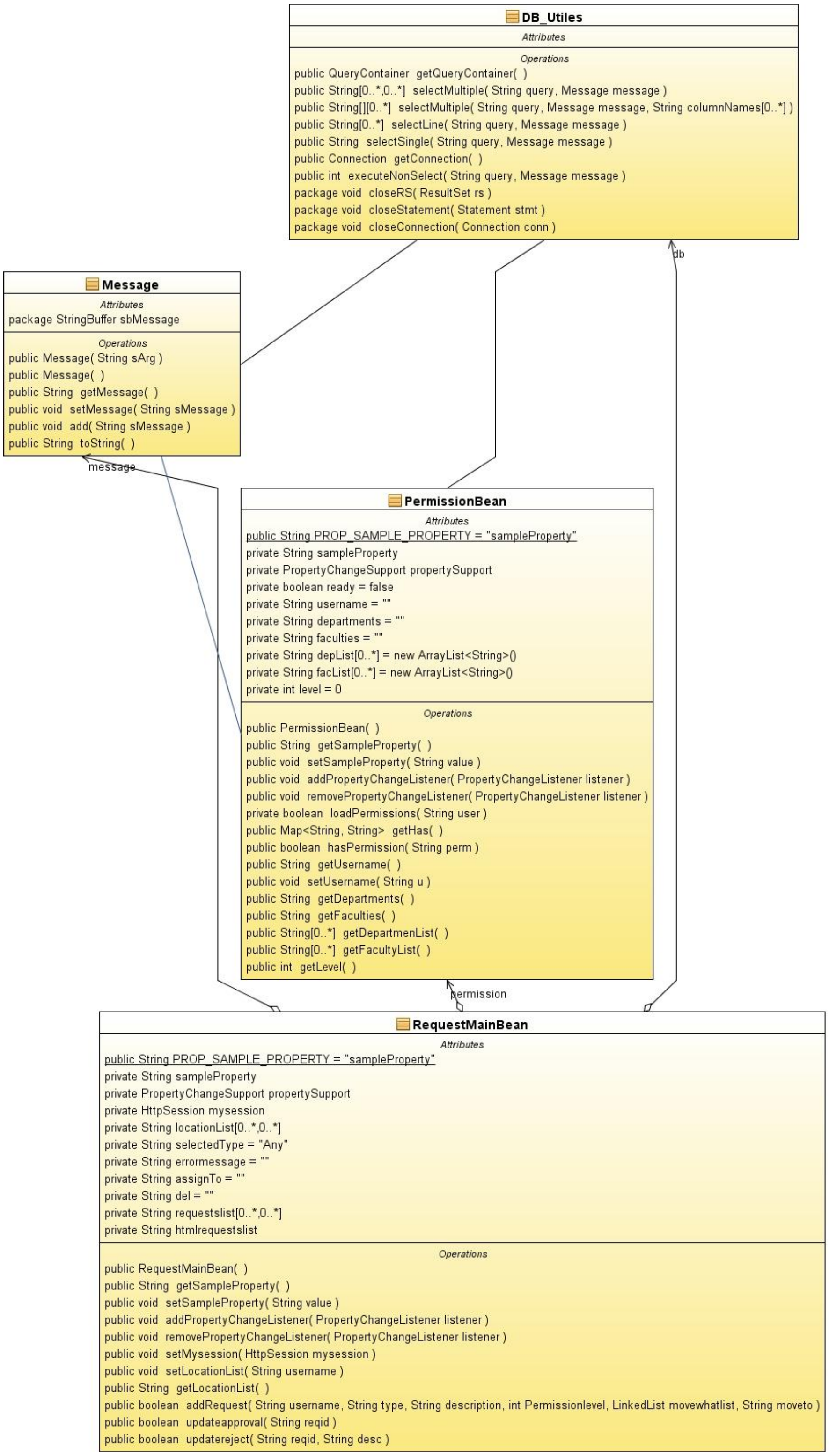

**Figure 49.** Class diagram for Add new request

The Java class for adding a new asset is : “RequestMainBean.java” , it imports three other classes to use them for calling different methods. The four classes are : “DB_Utiles” , “Message”, “PermissionBean”.

“DB_Utiles” contains several methods to connects to the database, “RequestMainBean” uses this class to select data from the Database, or Update the Database. The relationship between the two classes is a simple aggregation relationship. The parent (Container) class is “RequestMainBean” and the child (Contained) class is “DB_Utiles”; the child class can still exist and function if the parent class is destroyed.

The main methods used by “AssetCreateBean” are :

“selectMultiple()” → This method returns multiple rows

“selectLine()” → This method returns simple row

“selectSingle()” → This method gets a result from query that returns a single result

“executeNonSelect()”→ This method executes a query UPDATE, INSERT, DELETE

“Message” is used to accumulate a message as a mutable string. “RequestMainBean” uses this class for error messages, to store and get messages in case an error occurred. The relationship between “RequestMainBean” and “Message” is a simple aggregation relationship. The parent (Container) class is “RequestMainBean” and the child (Contained) class is “Message”; the child class can still exist and function if the parent class is destroyed.

“PermissionBean” is used by “RequestMainBean” to get the Permission of the user. The main method that “RequestMainBean” calls from “PermissionBean” is :

“loadPermissions()” → This method executes query and initializes all permissions. The relationship between “RequestMainBean” and “PermissionBean” is a simple aggregation relationship. The parent (Container) class is “RequestMainBean” and the child (Contained) class is “PermissionBean”; the child class can still exist and function if the parent class is destroyed.

There is an association relationship between “DB_Utiles” and “Message” , and “PermissionBean”. “DB_Utiles” uses “Message” to stores error messages if an error occurred while one of the methods of “DB_Utiles” is running, and this error can be retrieved from “Messages”. “PermissionBean” uses “Message” in the same fashion. “PermissionBean” uses “DB_Utiles” to connect to the Data Base in order to get the Permission Level.

## 7.6. Module Search

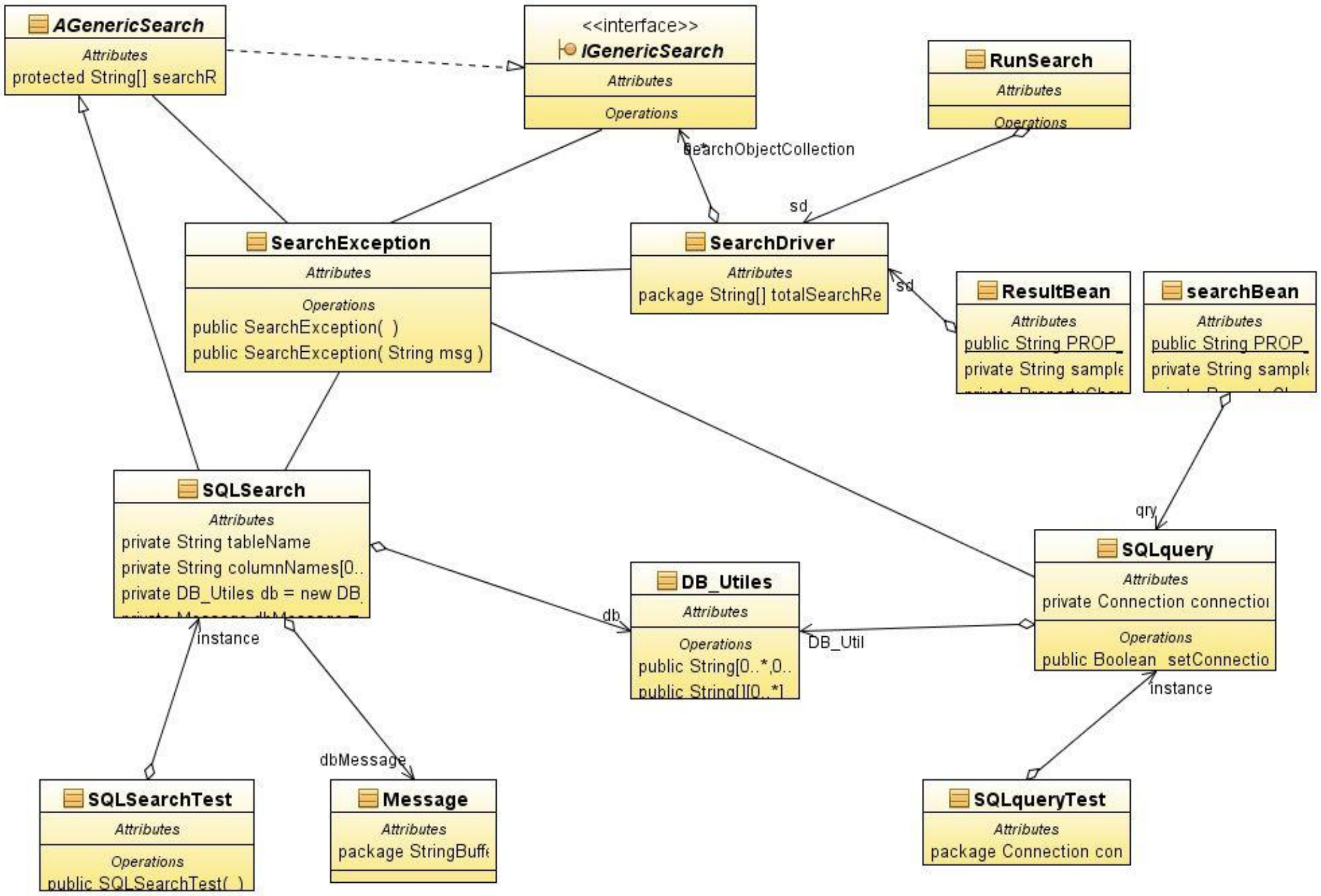

**Figure 50.** Class diagram for Search

Interface **IGenericSearch** – this interface contains names of universal methods which are applicable for any type of search.

Class **AGenericSearch** – this is abstract class contains abstract methods. Those methods performs particular task, but they don't know where data come from. Also this class inherits all methods from interface IGenericSearch.

Class **SQLSearch** – this is more concrete class, which inherits all methods from interface IGenericSearch and class AGenericSearch. This class performs search in specific area, in databases.

Class **DB_Utiles** – this class connects with DB and return result of query in different data structures.

Class **SQL_query** - general class to run any SQL query and receive a table as the result.

Class **Message** – this class transfers information (for example, about errors) from class to class. We choose message instead of String because it is more universal. String is not mutable, as soon as we create new String, we lose old information, while with message this problem is absent.

Class **SearchDriver** – this class shows in which way we perform search. It takes from user request what and where to search, creates necessary objects, after all another class performs search in those objects.

Class **SearchException** – this class and its subclasses are form of Throwable that indicates conditions that a reasonable application might want to catch.

Class **ResultBean** – contains getters for accessing Data Module (JAVA classes) from the View Module (HTML). Getters must have names with following strong rule: "get" + MethodName. Correspondent method will be automatically called.

Class **SearchBean** – contains getters for accessing Data Module (JAVA classes) from the View Module (HTML). Getters must have names with following strong rule: "get" + MethodName. Correspondent method will be automatically called.

Classes **RunSearch, SQLSearchTest, SQLQueryTest** used to run application or test some application's functions.

We implemented module SQLSearch, based on the data access method (SQL database), it is no class CollectionSearch. However modules design allows performing search in any collections in easy way. For this it is necessary to add concrete class (for example CollectionSearch) with specific methods (for example, getFromCollection()). The logic of search in this collection doesn't change, because this class will inherit methods from abstract file AGenericSearch and interface IGenericSearch.

So from this follows that in our design we use Abstract Factory Pattern. By definition Abstract Factory Pattern "provides an interface for creating families of related or dependent objects without specifying their concrete classes". In implementation it is no families (it is only one class SQLSearch), because as was mentioned earlier, in our project Basic search is a one of the cases of Advanced search. In addition all information in inventory system keeps in databases, so there is no necessity to perform search in another collections. However design is flexible and allows creating families (for example, SQLSearch, CollectionSearch) for more complicated task.

We didn't use Visitor pattern in our design, because of its main disadvantages:

arguments and the return type of visiting methods have to be known in advance;

visitor pattern breaks the Open-Closed Principle (Software entities should be open for extension but closed for modification);

writing additional code: the visitor class with one abstract method per class, and a accept method per class. If we add a new class, the visitor class needs a new method.

Description of modules is presented in the table below.

**Table 2.** Description of modules for search

| **interface IGenericSearch** | |
|---|---|
| **Attributes** | **Description** |
| None | none |
| **Operations** | **Description** |
| public int search(String what) | **param** what phrase that has to be found in any type of data<br><br>**return** number of found results<br><br>**throws** ca.concordia.encs.search.SearchException is generic exception that contain information about error |
| public boolean exclude(String what) | General method for any data type that excludes some rows from existing data based on argument.<br><br>**param** what is a word/phrase that has to be excluded from existing set of result from previously executed search(word) method<br><br>**return** true if some rows are excluded |
| public ArrayList<String[]> getResult() | Return the result of search.<br><br>**return** ArrayList<String[]> that contain the result of searching |
| public int getResultCount() | **return** number of found results |
| public String getSearchObjectName() | **return** Name of data where search was done |
| **Class AGenericSearch** | |
| **Attributes** | **Description** |
| protected ArrayList<String[]> searchResult | Array for keeping search results. |
| **Operations (redefined from** | **Descriptions** |
| **IGenericSearch)** | |
| public int getResultCount() | Return number of rows in result pool. |
| public ArrayList<String[]> getResult() | Return searchResult - pool for all result of search. |
| public boolean exclude(String what) | Exclude token NOT from existing result Set, do filter and update existing searchResult.<br><br>**param** what String that suppose to be excluded from result<br><br>**return** true if some rows are excluded |
| public abstract int search(String what) | Abstract method that has to be implemented by derived classes.<br><br>**param** what string to be found (can be complex, can contain AND, OR, NOT)<br><br>**return** number of found rows<br><br>**throws** SearchException |
| public abstract String getSearchObjectName() | Abstract method hast to be implemented by derived classes. Getter for name of actual searchable object.<br><br>**return** name of actual searchable object |
| **Class SQLSearch** | |
| **Attributes** | **Description** |
| private String tableName | The table name where to search |
| private String[] columnNames | List of column names where to search |
| private DB_Utiles db = new DB_Utiles() | DB_Utiles Instance to provide the SQL search (connection, run queries) |
| private Message dbMessage = new Message() | Create an error message, if any. |
| private StringBuffer query = new StringBuffer("") | A StringBuffer to build the SQL query. |

| Operations | Descriptions |
|---|---|
| public SQLSearch(String tableName, String[] columnNames) | Constructor for SQLSearch class<br><br>**param** tableName name of table where search has to be done<br><br>**param** columnNames name of columns where search will be done |
| public int search(String what) | Implementation for concrete type of abstract "search(...)" method. This method is adapted for SQL search.<br><br>**param** what any string to be found (can be complex and contain AND, OR, NOT)<br><br>**return** number of found rows<br><br>**throws** SearchException |
| public String getSearchObjectName() | Just return the name of table where actual object will be search.<br><br>**return** return the name of table where actual object will be search |
| public String toString() | **return** information about query of actual object and some additional information |

| **Class SQL_query** | |
|---|---|
| **Attributes** | **Description** |
| private Connection connection | SQL connection |
| **Operations** | **Descriptions** |
| public Boolean setConnection() | Opening an SQL connection. Return true, if the connection is established successfully. |
| public List<String[]> getTable(String SQL) | Getter for the result table |
| public Connection getMyConnection() | Getter for the connection |
| public void setMyConnection(Connection con) | Setter for the connection |
| private String getConnectionString() | Getter for the SQL connection string |
| private String getUserName() | Getter for the user name |
| private String getPassword() | Getter for the password |
| **Class SearchDriver** | |
| **Attributes** | **Description** |
| ArrayList<IGenericSearch> searchObjectCollection = new ArrayList<IGenericSearch>() | pool for any searchable (IGenericSearch) object that suppose to be searched |
| ArrayList<ResultHolder> totalSearchResult = new ArrayList<ResultHolder>() | pool for result of search from any searchable (IGenericSearch) object |
| ArrayList<String[]> totalSearchResultAsTable = new ArrayList<String[]>(1000) | Represents the result as ONE table |
| **Operations** | **Descriptions** |
| public ArrayList<String[]> getTotalSearchResultAsTable( ) | Get result of search as a table |
| public boolean addSQLSearchUnit(String tableName, String[] columnNames) | Method that add one SQL searchable element into pool of searchable object.<br>**param** tableName name of table or VIEW into DB<br>**param** columnNames colums into Table, search will be done only in provided colums<br>**return** true if new searchable object is added to pool |
| public void search(String | Execute search on all searchable objects into pool and |
| what) | result of every search is added to searchObjectCollection.<br><br>**param** what String what has to be searched. String can include AND OR NOT with different combination. Search will be done based on provided complex of word.<br><br>**Example of argument can be like:**<br><br>"green NOT black"   (search any row with green, exclude black)<br><br>"green AND black NOT red (search row with green and black, exclude red) |
| public String printResult() | Method print to console all results from existing searchable  objects to consol.<br><br>**return** same result in CVS format with table names. |
| **Class SearchException** | |
| **Attributes** | **Description** |
| None | none |
| **Operations** | **Descriptions** |
| public SearchException() | Creates a new instance of <code>SearchException</code> without detail message. |
| public SearchException(String msg) | Constructs an instance of <code>SearchException</code> with the specified detail message.<br><br>**param** msg the detail message. |
| **Class DB_Utiles** | |
| **Attributes** | **Descriptions** |
| none | none |
| **Operations** | **Descriptions** |
| public String[][] selectMultiple(String query, Message message) | This method returns multiple rows - result of query.<br><br>**param** query is a string of SQL query<br><br>**param** message argument which content is updated |
| | during query execution if some errors occurs<br><br>If no error during execution, then message = ""  //empty string.<br><br>**return** Strin[][] result of query<br><br>**Note:** always verify if returned String[][] is null and verity if message.toString() is equal ""<br><br>* <code><br><br>* if ((message.toString()).equals(""))<br><br>*   { no errors in query, work with returned result }<br><br>* </code> |
| public ArrayList<String[]> selectMultiple(String query, Message message, ArrayList<String> columnNames) | This method returns multiple row in ArrayList<String[]> - result of query.<br><br>**param** query is a string of SQL query<br><br>**param** message argument which content is updated during query execution if some errors occurs<br><br>If no error during execution , then message = ""  //empty string<br><br>**return** Strin[][] result of query<br><br>**Note:** allways verify if returned String[][] is null and verity if message.toString() is equal ""<br><br>* <code><br><br>* if ((message.toString()).equals(""))<br><br>*   { no errors in query, work with returned result }<br><br>* </code> |
| public String[] selectLine(String query, Message message) | This method returns simple row - result of query of type select name from table. This method take first element from any returned row.<br><br>**param** query is a string of SQL query<br><br>**param** message argument whose content is updated |
| | during query execution if some errors occur<br><br>If no error during execution , then message = "" //empty string<br><br>**return** Strin[] result of query<br><br>**Note:** allways verify if returned String[] is null and verity if message.toString() is equal ""<br><br>* <code><br><br>* if ((message.toString()).equals(""))<br><br>* { no errors in query, work with returned result }<br><br>* </code> |
| public String selectSingle(String query, Message message) | Method gets a result from query that return single result.<br><br>When multiple row or column, only first row, and first column is returned.<br><br>**param** query: a query to be executed<br><br>**param** message: if query done without errors, message="", otherwise it will contain the error message<br><br>return String - result of query or NULL if error was occurred<br><br>**Note:** always verify if returned String is null and verity if message.toString() is equal ""<br><br>* <code><br><br>* if ((message.toString()).equals(""))<br><br>* { no errors in query, work with returned result }<br><br>* </code> |
| public Connection getConnection() | Method that creates and returns Connection. Connection may be used for different purposes, but must be closed at the end<br />.Before use returned connection always verify if it not equal to NULL.<br><br>**return** Connection object or null if some errors happened |
| | with connection. |
| public int executeNonSelect(String query, Message (message) | Method that executes a query UPDATE, INSERT, DELETE<br>**param** query<br>**param** message<br>**return** int - number of affected rows<br>**Note**: allways verify if message.toString() is equal ""<br>* <code><br>* if ((message.toString()).equals(""))<br>* { no errors in query, work with returned result }<br>* </code> |
| void closeRS(ResultSet rs) | Method just to close ResultSet object. |
| void closeStatement(Statement stmt) | Method just to close ResultSet object. |
| void closeConnection(Connection conn) | Method just to close ResultSet object. |
| **Class Message** | |
| **Attributes** | **Description** |
| StringBuffer sbMessage | The message |
| **Operations** | **Description** |
| public Message(String sArg) | Constrictor with an initial string |
| public Message() | Default constructor |
| public String getMessage() | Getter for the message |
| public void setMessage(String sMessage) | Setter for the message |
| public void add(String sMessage) | Adding a line to the message |
| public String toString() | As string representation. |
| **Class searchBean.** | |
| **Attributes** | **Description** |
| String[] tables_name1;<br>String[] tables_name2;<br>String[] tables_name3;<br>String[] tables_name4; | columns for tables 1 - 4 |
| **Method** | **Description** |
| public String[] getTables_name1 | Returns columns for table 1 (same method is exists for tables 2,3 and 4) |
| public void setTables_name1(String[] tables_name1) | Set columns for the table 1 (same method is exists for tables 2,3 and 4) |
| String getColumnNamesPerson() | Return HTML-formatted string for columns for Person table |
| String getColumnNamesAssets() | Return HTML-formatted string for columns for Assets table |
| String getColumnNamesLocations() | Return HTML-formatted string for columns for Locations table |
| String getColumnNamesLicenses() | Return HTML-formatted string for columns for Licenses table |
| **Class ResultBean** | |
| **Attributes** | **Description** |
| String[] tables_name1;<br>String[] tables_name2; | columns for tables 1 - 4 |
| String[] tables_name3;<br><br>String[] tables_name4; | |
| **Method** | **Description** |
| public String[] getTables_name1 | Returns columns for table 1 (same method is exists for tables 2,3 and 4) |
| public void setTables_name1(String[] tables_name1) | Set columns for the table 1 (same method is exists for tables 2,3 and 4) |
| String getTest() | Prepare the result table for advanced search |
| String getBasicTest() | Prepare the result table for basic search |
| String createTable(ArrayList<String[]> data) | Return HTML-formatted result table |

## 7.7. Module Assign asset to location

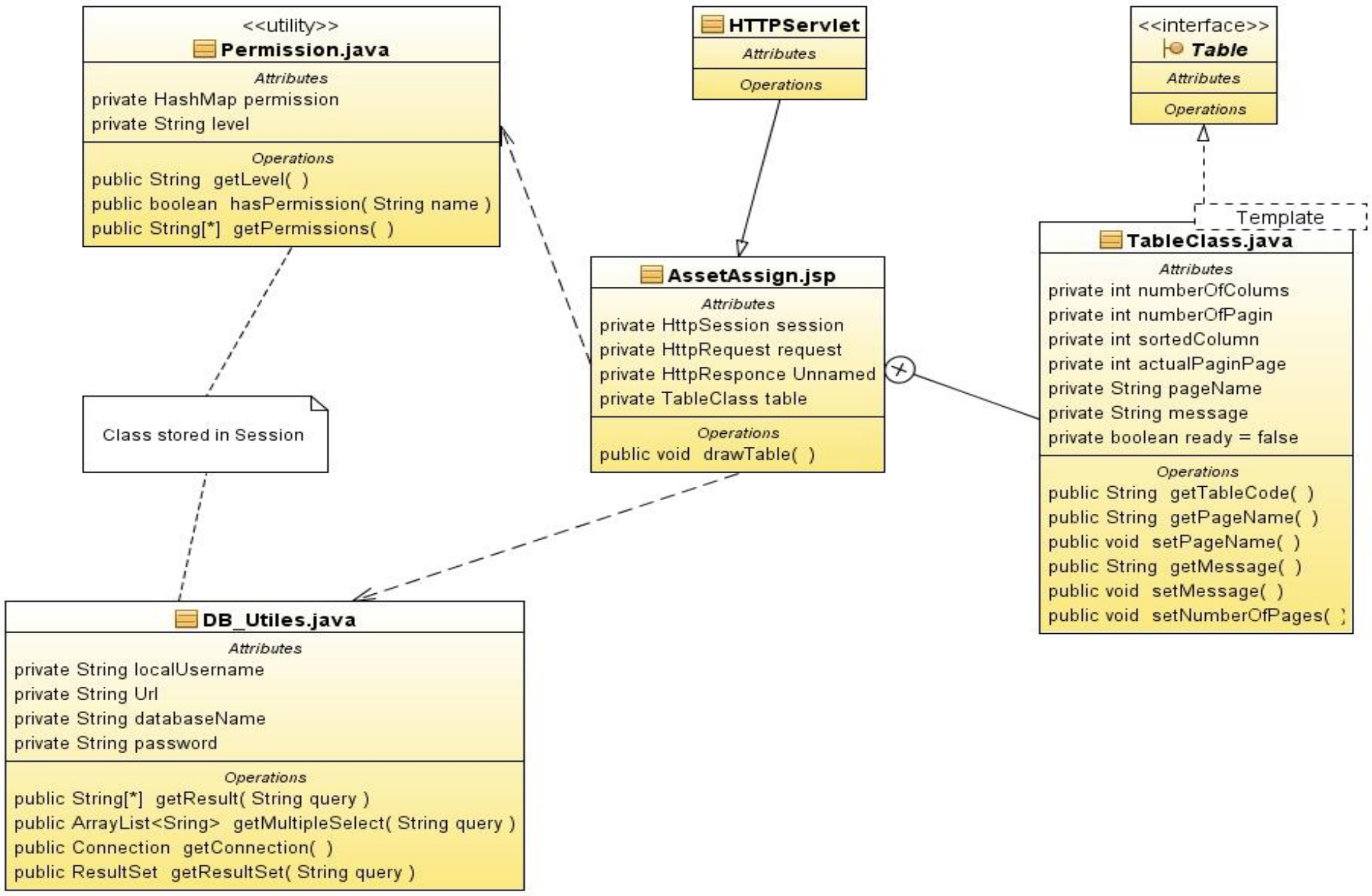

**Figure 51.** Class diagram for Assign asset to location

# 8. Appendices

## A. Setup and Configuration

Installation:

1. Install Linux Suse 11.2. It is available on: http://software.opensuse.org/112/en
2. Include MySQL database in installer of Suse11.2
3. After installation of OS following components have to be installed:
    1. Java JDK 6 (jdk1.6.0_11). Last version can be downloaded from: http://java.sun.com/javase/downloads/index.jsp
    2. Apache Tomcat 6 (ver. 6.0.24.) can be downloaded from http://tomcat.apache.org/download-60.cgi
    3. phpMyAdmin (ver.3.2.3) graphical tools to manage MySQL database can be downloaded from: http://downloads.sourceforge.net/project/phpmyadmin/phpMyAdmin

After installation of all components some configuration has to be done:

1. MySQL database configuration:
   - To start mySQL automatically, run in shell chkconfig --add mysql
   - Create a database with name “project” and Collation utf-8-general
   - CREATE DATABASE project
   - CHARACTER SET utf8
   - COLLATE utf8_bin;
   - Set password for root user: SET PASSWORD FOR 'root'@'localhost' = PASSWORD('yourPassword');
   - Create user in DB : CREATE USER project IDENTIFIED BY PASSWORD 'project12345'
   - Grand privileges to user: GRANT ALL ON project.* TO 'project'@'%';
   - Run script “project.sql” to populate database.

2. Tomcat 6.0.24 configuration:
   - Install Tomcat in /var/lib directory
   - Change password for root user, modify configuration file – server.xml, to provide a valid password.
   - Provide pa variables in shell: export CATALINA_HOME=/var/lib/apache-tomcat-6.0.24 and export CATALINA_BASE=/opt/tomcat-instance/project
   - To make Tomkat start automatically after reboot, export has to be done: chkconfig --add tomcat
   - Deploy the web application into tomcat, login as administrator and upload file UUIS.war

3. To configure Tomcat server SSL connection we need next steps:
   -install openSSL packages.
   -generate a key with keytool of java
   -modify configuration of server.xml file in %Catalina Home%conf directory
   <Connector port=”8443″
   ….
   keystoreFile=”../webapps/key.bin”
   keystorePass=”password”
   />
   -reboot Tomcat server.

## Deployment of the UUIS on the server:

On the server UUIS has following structure:

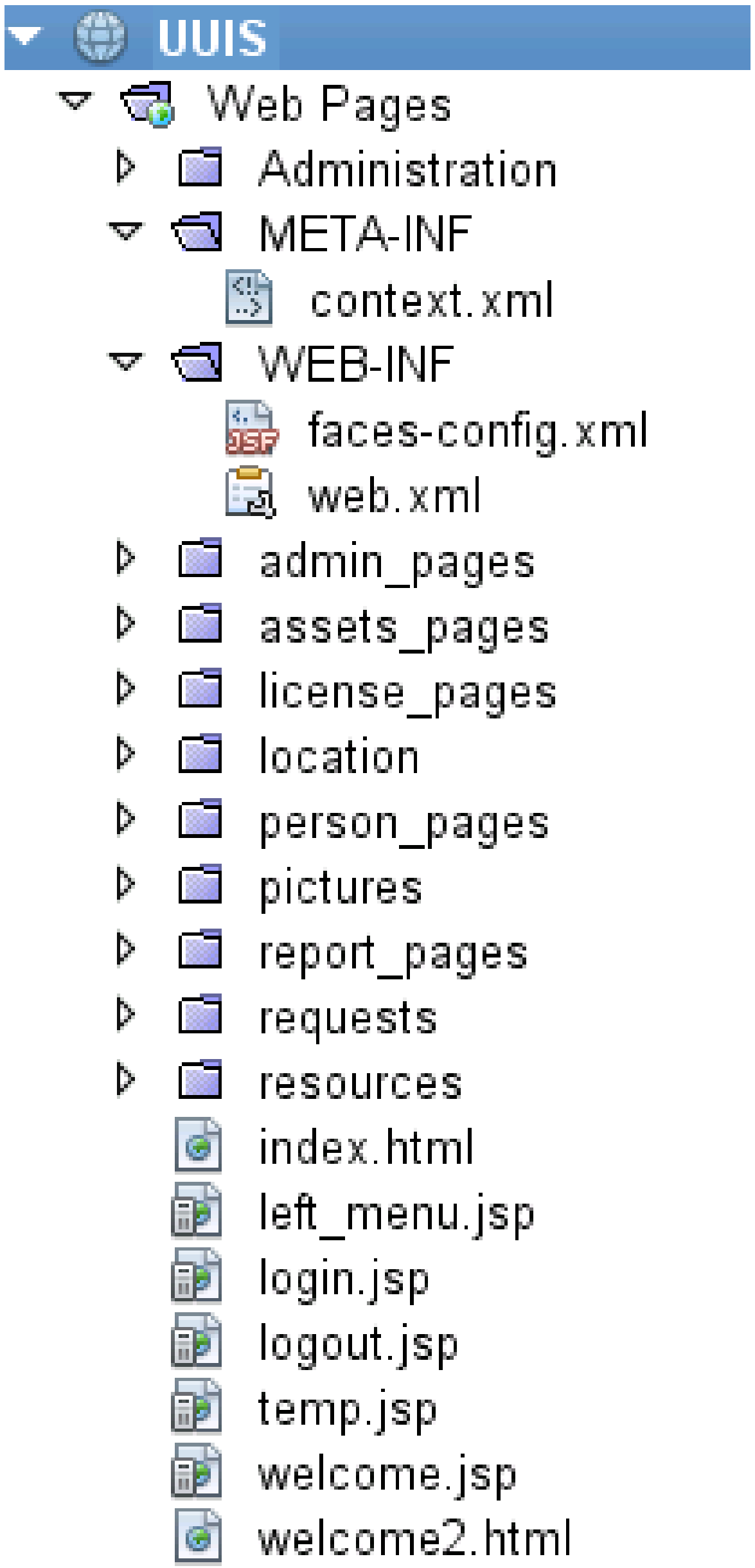

## B. Tool set and environment

**Table 3.** Tools and Environment

| | |
|---|---|
| Netbeans version 6.5 and 6.8 | Programming, testing, and making UML diagrams |
| Microsoft Visio | UML diagrams: block diagram, DT, Sequence diagrams, Activity diagrams |
| WampServer Version 2.0 | Admin MySQL through PHP MyAdmin |
| Microsoft Office Word | Documentation |
| Microsoft Office Power Point | Presentation |
| Microsoft Office Excel | Log sheets, cost estimation |
| MySQL | For database management |
| Mailing list :<br>http://samson.podzone.org/mail_list | Repository and communication |
| COCOMO II | COnstructive COst MOdel II (COCOMO II) for cost estimation |
| e-mail | Communication |
| CVS integrated into NetBeans | Checkout, update code |
| JUnit | Unit test |
| Dreamweaver | Welcome pages |
| FlamingText.com | Text for welcome page |
| Tomcat | As HTTP web server |

## C. Implementation list

The following table shows the list of requirements and indicates whether the requirement was fulfilled in UUIS V1.0.

**Table 4.** Implementation list

| Name | Implemented (+) / not implemented (-) | Note |
|---|---|---|
| 1. Asset | | |
| 1.1. Asset | | |
| 1.1.1. Add new | + | |
| 1.1.2. View | + | |
| 1.1.3. Edit | + | |
| 1.1.4. Delete | + | |
| 1.1.5. Borrow | + | with restrictions |
| 1.1.6. Create group | + | |
| 1.1.7. Create new type | - | |
| 1.1.8. Create new subgroup | + | |
| 1.2. Import | | |
| 1.2.1. Import from *.csv file/scanner | + | |
| 1.3. Assign to | | |
| 1.3.1. Assign to person | + | |
| 1.3.2. Assign to location | + | |
| 1.4. My profile | | |
| 1.4.1. View asset(s) assigned to me | + | |
| 1.4.2. View asset(s) borrowed by me | + | |
| 2. License | | |
| 2.1. License | | |
| 2.1.1. Add new | + | |
| 2.1.2. View all licenses | + | |
| 2.1.3. View asset's licenses | + | |
| 2.1.4. Edit | - | |
| 2.1.5. Delete | + | |
| 2.1.6. Borrow | + | with restrictions |
| 2.1.7. Create new type | - | |
| 2.2. Import | | |
| 2.2.1. Import from *.csv file/scanner | - | |
| 2.3. Assign to | | |
| 2.3.1. Assign to asset | + | |
| 2.4. My profile | | |
| 2.4.1. View license(s) assigned (borrowed) to me | + | |
| 3. Location | | |
| 3.1. Location | | |
| 3.1.1. Add new | + | |
| 3.1.2. View | + | |
| 3.1.3. Edit | + | |
| 3.1.4. Delete | - | |
| 3.1.5. Create group | - | |
| 3.1.6. Create new type | + | |
| 3.1.7. View plan of big location | + | with restrictions |
| 3.1.8. Print plan of big location | - | |
| 3.2. Import | | |
| 3.2.1. Import from *.csv file | - | |
| 3.3. Assign to | | |
| 3.3.1. Assign to person | - | |
| 3.3.2. Assign to another location | - | |
| 3.3.3. Assign to Department | + | |
| 3.4. My profile | | |
| 3.4.1. View location(s) assigned to me | + | |
| 4. Person | | |
| 4.1. Person | | |
| 4.1.1. View | + | |
| 4.1.2. Edit | - | |
| 4.1.3. Delete | - | |
| 4.1.4. Create new type | - | |
| 4.1.5. Provide biometrical characteristic | - | |
| 4.2. Import | | |
| 4.2.1. Import from *.csv file | - | |
| 5. Administration | | |
| 5.1. Role | | |
| 5.1.1. Add new role (package of permissions) | - | |
| 5.1.2. Edit role | + | |
| 5.2. Permission | | |
| 5.2.1. Add new permission | - | |
| 5.2.2. Edit user's permission | - | |
| 5.3. Assign to | | |
| 5.3.1. Assign role to person(s) | + | |
| 5.3.2. Assign permission to person(s) | - | |
| 5.4. My profile | | |
| 5.4.1. View my role | - | |
| 5.4.2. View my permission | - | |
| 6. Faculty and Department | | |
| 6.1. Faculty | | |
| 6.1.1. Add new | - | |
| 6.1.2. View | - | |
| 6.1.3. Edit | - | |
| 6.2. Department | | |
| 6.2.1. Add new | - | |
| 6.2.2. View | - | |
| 6.2.3. Edit | - | |
| 7. Requests | | |
| 7.1. Add new | + | |
| 7.2. Approve/Reject | + | |
| 7.3. View list of all requests in the system | + | |
| 8. Search | | |
| 8.1. Basic search | + | |
| 8.2. Advanced search | + | |
| 9. Report | | |
| 9.1. Create report "Teaching lab" | + | |
| 9.2. Create report "Research lab" | + | |
| 9.3. Create report "offices" | + | |
| 9.4. Print report | - | |
| 9.5. Auditing | + | with restrictions |
| 9.6. My profile<br>10. Select language<br>11. Login/Logout | +<br>-<br>+/+ | |

## D. Log sheet

Each team member contributed at least the hours indicated below, to finish the project.

**Table 5.** Logsheet

| | |
|---|---|
| Abirami Sankaran | 3:15 + 4:25 + 5:55 + 2:45 + 34:49 + 10:01 + 0:45 + 6:05 + 12:54 + 11:01 + 51:21 + 26:55 + 8:30 + 20:00 = 198:41 |
| Andriy Samsonyuk | 11:00 + 20:00 + 24:00 + 18:00 + 17:00 + 14:00 + 16:00 + 20:00 + 3:00 + 20:00 = 163:00 |
| Maab Attar | 3:45 + 3:45 + 6:45 + 6:00 + 10:30 + 12:45 + 17:00 + 2:30 + 25:00 + 14:00 = 102:00 |
| Mohammad Parham | 1:00 + 4:30 + 11:30 + 25:00 + 4:30 + 7:30 + 9:30 + 5:00 + 24:00 + 11:00 + 21:00 + 29:00 + 6:00 + 25:30 = 185:00 |
| Olena Zayikina | 8:00 + 16:00 + 15:00 + 11:30 + 15:30 + 17:00 + 31:00 + 25:00 + 40:00 + 16:00 + 14:00 = 210:00 |
| Omar Jandali Rifai | 4:30 + 5:30 + 9:00 + 8:00 + 38:30 + 25:00 + 12:00 + 6:00 = 108:30 |
| Pavel Lepin | 3:00 + 13:00 + 12:00 + 4:00 + 13:00 + 6:00 + 14:00 + 11:00 + 7:00 + 6:00 + 3:00 + 5:00 = 97:00 |
| Rana Hassan | 8:00 + 6:30 + 7:00 + 16:00 + 9:30 + 2:00 + 8:00 + 17:00 + 19:30 + 15:00 + 18:00 + 17:00 = 135:30 |

# E. Test Report

## E.1. General Test Steps & Tips

### Web Application Security Testing - General Tips:

Security aspects for any kind of applications are exiting enough to look for. But when it comes to security testing for web applications it is also something that needs to be taken seriously. The best way to be successful from vulnerabilities is to be prepared in advance and know what to look for. That is why we provide a basic checklist for our Web Application Security Plan:

#### Set the proper time and everyone's roles:

Problems are likely to be happened at the beginning and during the testing procedure. So do not be afraid of locked accounts, server reboots, performance failures and etc. Although it will be never an ideal time for testing, it is advised to select test dates and timeframes that will minimize the side effects.

Also it is of high importance to perform a security assessment once everyone who is being participated in the testing plane, is on the same page. Project sponsor, IT director, VP of audit, and others, they all must follow the person who is in charge of the testing.

#### Prepare and gather appropriate tools:

High-quality tools have a significant role in testing. In fact, the number of legitimate vulnerabilities discovered is directly proportional to the quality of the security tools. There are several open source tools like OWASP which we are going to use it. Good tools usually lead to more security flaws discovered, as well as less time and effort wasted to track them. Having reporting capabilities is an asset.

#### Look at it from every perspective:

Try to perform role of a hacker, at least a beginner. Using Google and its hacking tools to see what the world can see from your Web Application. You will never find vulnerabilities until you check them. Then, run a web vulnerability scanner as both, an unauthenticated and untrusted user from outside and also an authenticated and trusted user from inside. Web abuse has no boundaries. By looking at your application from various perspectives, you will increase the chance to undoubtedly find different types of vulnerabilities that can corrupt your system from outside or inside.

#### Test for underlying weaknesses:

Failing to scan the underlying operating system and installed applications is very common in testing planes. Some specific tools such as Nessus and QualysGuard designed in finding out missing patch and misconfiguration problems in the operating system that in turn leads to a web application leakage. If you want to complete your work, you need to also look at the back-end servers as well as databases. Related network hubs are from the same importance.

**Verifying your testing:**

Everything must be verified. Not only every required quality, but also verification by itself must be verified. So do not trust marketing machines when tend to show security testing tools are enough to prevent shortcomings. They are not! Try to figure out that the security weaknesses that the tool already discovered is legitimate.

Verification not only saves everyone time and effort in the run time, but also gives enough confidence to others to take the code and testing seriously.

**Manually check for weaknesses:**

Do not leave the job happily after a successful verification. Although prepared tools may discover many security weaknesses, there are likely several things left behind. This is where the human expertise and profession come to play. Try various aspects and poke the application a bit more to see if anything else can be done from another point of view.

**Test the source code:**

Source code security testing tools are flexible and mature enough to be used by everyone, not only developers. So trying security aspects of source code is a must to complete testing procedure. At least it is a priority for to-do lists after all the web application security tests.

## E.2. Security Testing

### SkipFish – Web Application Security Scanner

Skipfish is an automated web application security tool which is capable of testing the web application in terms of tens of various aspects. It works by carrying out recursive methods and dictionary-based probes. The result is then supplied by a number of security checks including a final report. The report is used as the security assessment fortunately not only for developers and security team but also ordinary users are able to read and understand the easy-written report.

1. The main server-side machine for our project is "spec111". Although it is wise to run skipfish scanner on a separate machine we will run it on our server. Otherwise our performance will be monitored in the university network and somebody from IT members comes after us!
2. In order to prepare appropriate tool, we decide to use skipfish scanner because of its:

- **High Performance:** Even under high-load pressure of hundreds of requests per second it does not lose working with an acceptable performance. This can be attributed to: Advanced HTTP/1.1, multiplexing single-thread, fully asynchronous network I/O model, using smart response caching and advanced server behaviour heuristics and finally being performance-oriented.

- **Ease of use:** Skipfish is highly adaptable and reliable according to handling of multi-framework sites where certain paths follow a completely different semantics, or are object to different filtering rules. Heuristic recognition of path and query based parameter handling is in addition to ease use of skipfish.

- **Well-designed security checks:** Since the tool is designed to reflect meaningful and easily-understandable results for everybody, it has been designed very well in terms of: Handling tricky scenarios, post-processing reports, using ratproxy-style to spot security problems like cross-site script inclusion, mixed content and etc.

3. In order to (at least try to) look at the testing from various perspectives, roughly the following specific tests are to be done in our project. **

- Explicit SQL-like syntax in GET and POST parameters.

- Format string vulnerabilities.

- Integer overflows vulnerabilities.

HTML pages may submit their parameters using either GET or POST method. If they use GET, all the names and values will appear in the URL string that user can see. And if the method is POST, information will be hidden in some sort of forms and sent to server-side machines as opaque data for attackers. However, still both methods are subject for spoofing. Integer overflow happens when an arithmetic numeric result is larger than the available storage space and format string occurs when an attacker uses specific characters (e.g. %s, %x in printf() - C programming language) to retrieve information from stack or possibly other locations in memory.

- Stored and reflected XSS vectors in document body (minimal JS XSS support present).

- Stored and reflected XSS vectors via HTTP redirects.

- Stored and reflected XSS vectors via HTTP header splitting.

- Attacker – supplied script and CSS inclusion vectors (stored and reflected).

Cross-site scripting (XSS) is a kind of attack when an attacker injects client-side script into web pages viewed by clients. Their impact varies from minor to significant risks depend on sensitivity of data handled by the hacked site as well as prevention methods used by the site's owner. These kinds of attacks are the most common kind of web site attacking.

While some secure authentication methods are available in the network and data link layers, cross-site scripting attacks are subject to the application layer and cannot be prevented by lower layer security tools.

- Redirection to attacker – supplied URLs (stored and reflected).
- Attacker-supplied embedded content (stored and reflected).
- HTTP credentials in URLs.
- Expired or not-yet- valid SSL certificates.
- SSL certificate host name mismatches.
- General SSH certificate information.
- Broken links.

Using digital certificates with SSL becomes important in order to add trust to online transactions. It is done by requiring web site operators to deal with a certificate authority (CA) in order to get an SSL certificate. As a result of any successful SSL connection, most users are not aware of whether the web site owner has been validated or not. It is the time that attackers are able to play role of the web site owner and steal users credential. That is why establishing SSL certificates intended to restore enough confidence among users that a web site owner has been legally established its business or organization with verifiable identity for further pursue in case of a fraud.

**Due to unacceptability of some Linux command lines, we have not been able to perform the actual tests on the Search Project. But we will provide the full result report from running skipfish scanner for this project along with the final report of the main UUIS project. However, a brief description of each specific test is provided.

Skipfish is able to test web applications by means of the following command line and options to catch such leakages:

$./skipfish -X –I

-X: Prevents matching URLs from being fetched. Also speeds up the scan by excluding /icons/, /doc/, /manuals/.

-S: Limits the scope of a scan (ignore links on pages where a substring appears in response body) by restricting it only to a specific protocol or port number.

-I: Limits the scope of a scan (only spider URLs matching a substring) by restricting it only to a specific protocol or port number.

-D: allows specifying additional hosts or domains to consider in-scope for the test.

-H: Inserts any additional and non-standard headers in order to customize favourite HTTP requests.

-F: Defines a custom mapping between a host and an IP in order to customize favourite HTTP requests for not-yet-launched or legacy services.

## E.3. GUI Testing

As a test plan for UUIS application, GUI testing is the process of application graphical user interface to ensure that it meets all the required features and specification.

In order to have a good GUI test plan, the test design must consider all the functionality of the system and fully exercises the GUI itself. That's why the scalability of the plan comes from a great importance as well as sequence of functionalities. As an example unlike a CLI (Command Line Interface) system, a GUI has many operations that need to be tested. Even a very small program could have a couple of possible GUI operations. Moreover, when dealing with some functionality (e.g. open a new file), user has to follow some complex sequence of GUI events. Obviously, increasing the number of possible functionalities increases the complexity of sequencing problem. This can become a serious issue In case tester designs test plans manually.

These issues have driven the GUI testing plans towards automation. In practice, manually GUI testing is not feasible. That is why many different techniques have been proved to perform automatically GUI test suites that are complete and that simulate were behaviour.

Typically, lots of techniques used to test GUI were based on techniques those used to test CLI programs. However, due on the scaling problem they are not applicable any more. As an example, finite state machine- where an application is worked as a finite state machine and software program is used to generate test cases to cover all states- can work perfectly for an application with CLI but many may become extremely complex for an application with GUI.

Since our application –UUIS- is not a complex multifunctional application, we derive some GUI test plans manually but scalable for further extensions automatically.

## E.4. Regression Testing

A professional computer program is composed of various modules. During the process of programming each module may developed and modified independently. In order to uncover software errors by partially testing individual modified path, regression testing is being used. The intention of this kind of testing is to ensure that no additional errors were introduced in the process of fixing other problems. Typical regression test methods include rerunning previously run tests and checking if previously fixed faults have no problem working with other parts. It is often too difficult for a programmer to figure out how a change in one part will echo in other parts and that is why regression test is useful.

According to previous experiences, where developers are using a revision control system it is likely that a fix gets lost through poor revision control practices or even a simple human error. In a software development process usually affix for a problem in one area causes a software bug in another area. Moreover, it has often been the cases that when a specification is redesigned the same mistakes that were made in the original design were made again in the redesign.

Consequently, it is advised that when a bug is getting fixed, a test that exposed the bug is recorded and regularly reused after a set of changes to the program.

Regression testing should be an integral part of the whole programming software development. It is efficient once repeatable and automated testing packages at every stage in the software development cycle like are being replaced by design documents.

## E.5. Functional Testing

**artf_001**

Title: String name, spelling errors.

Date: 08/04/2010

Author: Abirami Sankaran

Status: Bug

Description: In Welcome2 page there is an spelling error, Please verify the second number in the list.

1). easier

instead of

easer

2). Bye.

instead of

Bay. (when user make logout)

Solution: fixed in commit from 11/04/2010

**artf_002**

Title: Child asset is not assigned to a selected list even if it is correctly selected in the table.

Date: 09/04/2010

Author: Olena Zayikina

Status: Bug

Description: When I create a group of asset, I need to select Master asset first, and then Child assets, but when i select child asset, there is a message "At least one asset has to be selected". The message doesn't corresponds actual situation, because asset was previously selected.

Step by step:

1. Go to Asset

2. Press on Asset/Add new Asset button

3. Select a master element in the table, pres "Add Master" button

4. Select a child element in the table, press "Add Child" button; in my case I have a message: "At least one asset has to be selected". The message is wrong.

Solution: fixed in commit from 22:00 12/04/2010

**artf_003**

Title: After duplicate barcode is entered, drop down list: Title, Status, and Location is empty.

Date: 06/04/2010

Author: Olena Zayikina

Status: Bug

Description: When I create a new Asset, I insert a barcode that already exist in the system. System correctly shows the message that this is a duplicate barcode number. But after the message all dropdown lists on the page are empty, it is impossible to continue operations.

Step by step:

1. Go to Asset/Asset/Add new Asset

2. Type some date, in place of barcode - insert an barcode that already exist in the system.

3. Press Create button.

4. Error message is shown ... Duplicate barcode in the system.

5. Observe all dropdown lists, there is no value inside!

Solution: fixed in commit from 11/04/2010

**artf_004**

Title: Too many options in menu "My Profile".

Date: 11/04/2010

Author: Olena Zayikina

Status: Bug

Description: In top menu of Asset, Licenses, Locations, there are too many sub options in "my profile" division.

Following the Specification, it supposes to be:

Assets/My profile --> Assets Assigned to me

Assets/My profile --> My Borrowed assets

Licenses/My Profile --> My Licenses

Locations/My Profile --> My locations.

Solution: fixed in commit from 11/04/2010

**artf_005**

Title: Asset assign to location doesn't work

Date: 11/04/2010

Author: Olena Zayikina

Status: Bug

Description: When I assign asset to location, it shows the message that asset was correctly assigned, bat in table of asset the information is not updated.

Step by step:

1. Go to Asset menu,

2. Press Assign To.../Assign To Location button

3. Select an asset, press "Assign" button.

4. Message about operation is shown, but in table of asset the information is not changed.

Solution: It work fine, the problem is that after assigning - assigned asset change the place and seems to be not assigned.

Note: in Asset table the number in the first column is not Asset_ID, it is only Sequential number that change with any page update or sorting.

**artf_006**

Title: UUIS can not run with SSL protocol

Date: 01/04/2010

Author: Andriy Samsonyuk

Status: Bug

Description: Following specification of “Client” UUIS application has to run on SSL protocol. Actually UUIS can run only on standard HTTP1.1 protocol.

Solution: fixed by server configuration 10/04/2010

**artf_007**

Title: Asset table become unavailable after operation is cancelled.

Date: 12/04/2010

Author: Olena Zayikina

Status: Bug

Description:

Asset -->Create a group -->Add Master --> Add Children --> user selects some children but decide to cancel operation --> Cancel --> after this all table Asset not available anymore, even if you go again Asset -->Create a group… or Asset -->View … It starts work again only after "Logout", so problem with code for "Cancel" button (the last one, after add master and add children).

Solution: For me it works fine. I think the problem is – when you browse to other pages then first page.

Example: Before you click "Cancel" you are in page 5 with all assets. When you cancel operations you automatically redirected to page of Computers, to select "Master", but your table still in page 5, but there are not enough computers to be shown in page 5. You have to go to page 1 of table.

**artf_008**

Title: In AssetImport page: Text desapears in Text box, after error is shown.

Date: 12/04/2010

Author: Olena Zayikina

Status: Bug

Description:

Asset --> Import --> From file/scanner--> User paste content of file, puts # of columns, selects name of columns, but forget select location --> Error message "Select location" --> but user should repeat paste content of file in text box. Problem: content of text box disappear, after error message.

Solution: Error was fixed in commit 20:44 12/04/2010

Two more error are fixed with this commit.

**artf_009**

Title: Asset table become unavailable after operation is cancelled.

Date: 12/04/2010

Author: Olena Zayikina

Status: Bug

Description:

Asset -->Create a group -->Add Master --> Add Children --> user selects some children but decide to cancel operation --> Cancel --> after this all table Asset not available anymore, even if you go again Asset -->Create a group… or Asset -->View … It starts work again only after “Logout”, so problem with code for “Cancel” button (the last one, after add master and add children).

Solution: For me it works fine. I think the problem is – when you browse to other pages then first page.

Example: Before you click “Cancel” you are in page 5 with all assets. When you cancel operations you automatically redirected to page of Computers, to select “Master”, but your table still in page 5, but there is not enough computer to be shown in page 5.You has to go to page 1 of table.

**artf_010**

Title: UserName disapears, when I brows in Asset table.

Date: 12/04/2010

Author: Olena Zayikina

Status: Bug

Description:

Assets-->Assign to --> person --> write user name--> when I move between the pages in DB Asset user name disappear, so it is necessary to type user name if I go to another page.

Solution: Can not be fixed. Additional note is added on the page.

**artf_011**

Title: Licence-->Edit license doesn’t work.

Date: 12/04/2010

Author: Olena Zayikina

Status: Bug

Description:

Licence-->Edit license doesn't work.

Solution: Page not implemented yet. Will be implemented after additional requirement.

**artf_012**

Title: We allow to login with the same username and PW to several persons, it should be unique.

Date: 12/04/2010

Author: Olena Zayikina

Status: Bug

Description:

We allow to login with the same username and PW to several persons, it should be unique.

## E.6. Standard compliance test

Using online Mark up validation Service, validation by direct input was performed for Basic search.

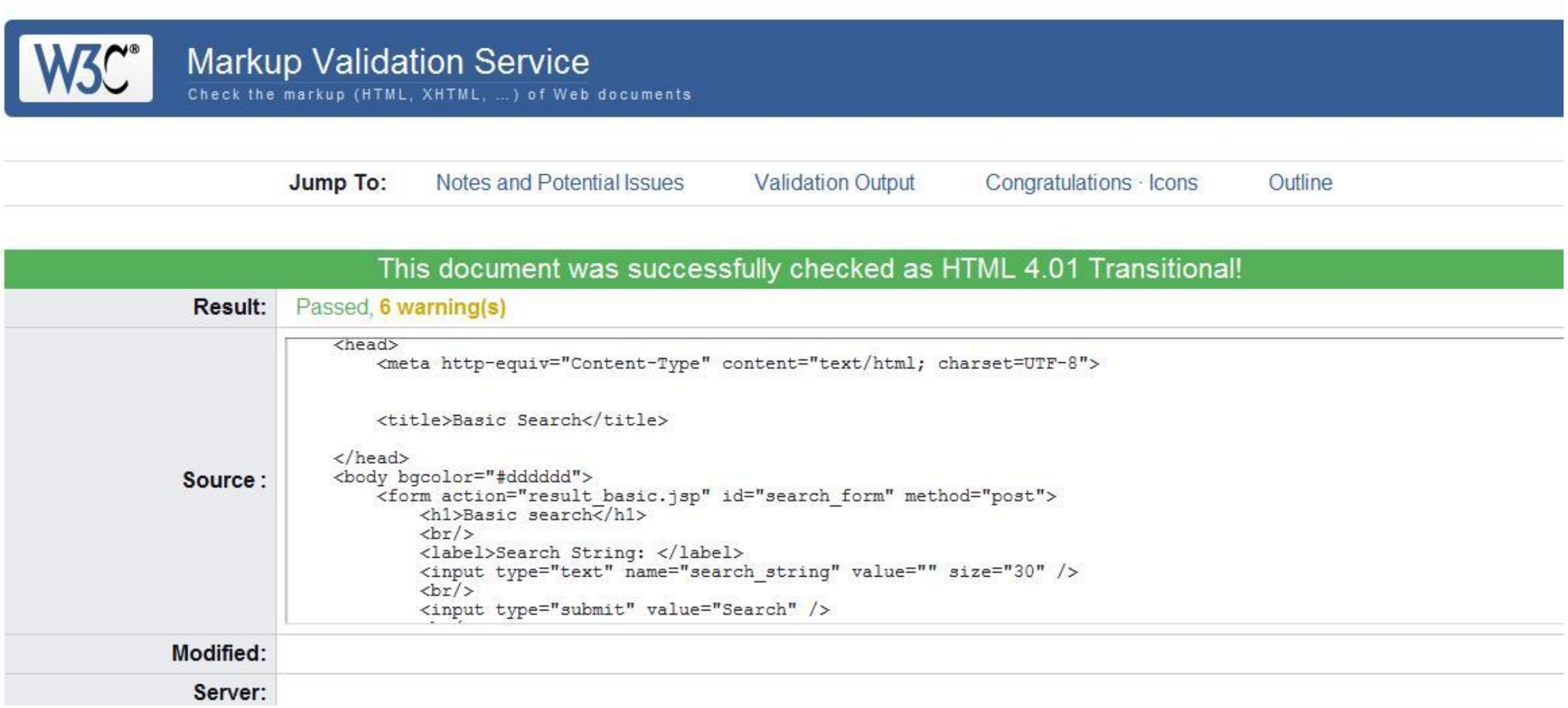

**Figure 52**. Results of standard compliance test for Basic search

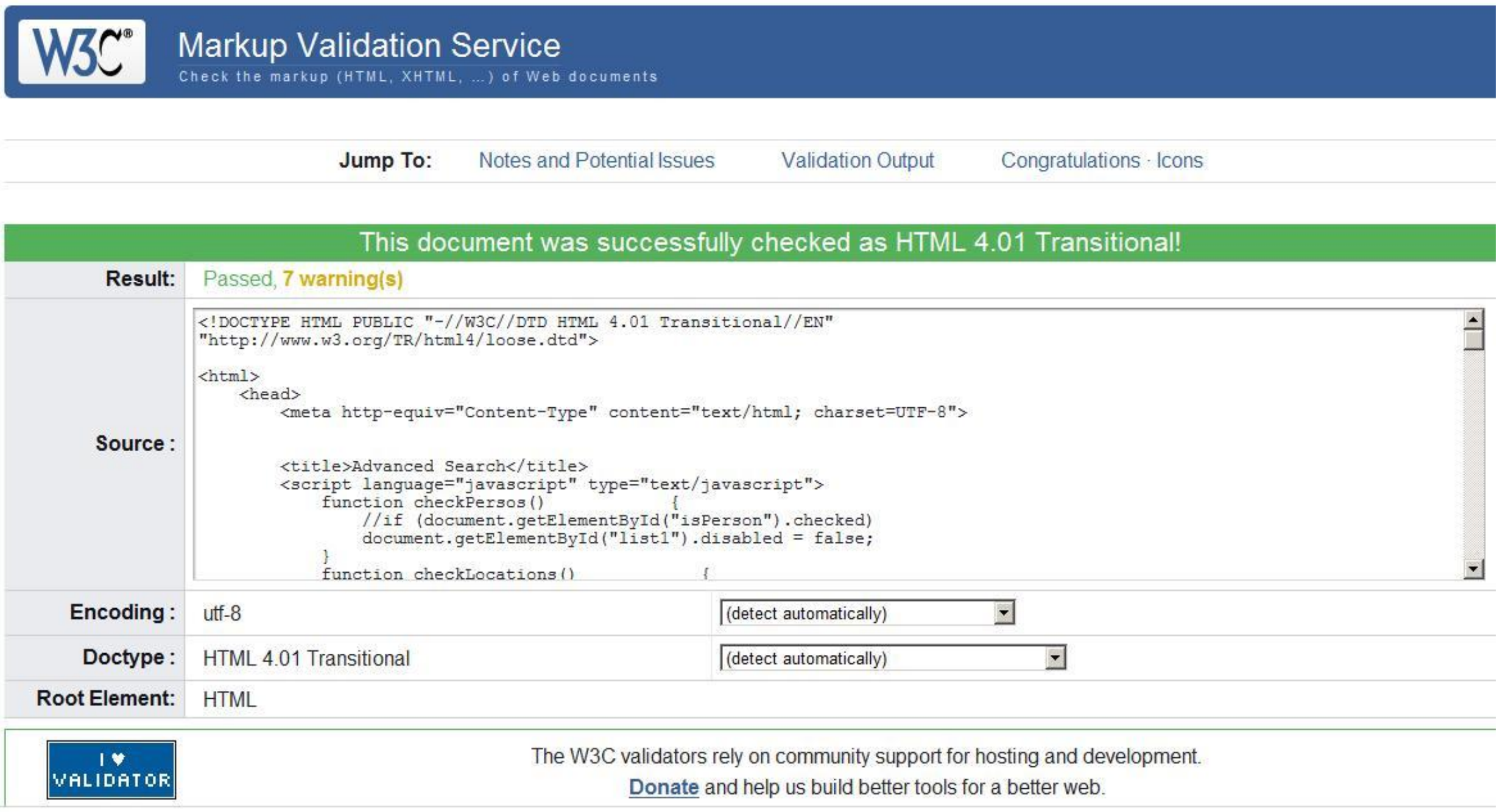

**Figure 53.** Results of standard compliance test for Advanced search.

## E.7. JUnit Testing

**Class SQLSearch.**

Unit tests for class SQLSearch is presented below:

```
package ca.concordia.encs.search;

import com.sun.net.httpserver.Authenticator.Success;
import org.junit.After;
import org.junit.AfterClass;
import org.junit.Before;
import org.junit.BeforeClass;
import org.junit.Test;
import static org.junit.Assert.*;
/**
* @author Lepin
 */
public class SQLSearchTest
{

    public SQLSearchTest()
{
  }

    /**
     * Test of search method, of class SQLSearch.
```

```
   * Tests how many lines in the result of search (can be zero or positive, can't be negative)
     */
    @Test
    public void testSearch() throws Exception
{
        System.out.println("search");
        String what = "a";

        String[] colums = {"FirstName"};
        SQLSearch instance = new SQLSearch("person", colums);
        int result;
        try{
            result = instance.search(what);
        }
        catch(Exception e){
            System.out.println(e.getMessage());
            result = -1;
        }
        assertTrue(result>-1);


    }

    /**
     * Test of getSearchObjectName method, of class SQLSearch.
     * Tests in which DB and which field to perform search (for example, if sets DB "person"
     *and field – first name, test will be passed).
     */
    @Test
    public void testGetSearchObjectName() {
        System.out.println("getSearchObjectName");
        String[] colums = {"FirstName"};
        SQLSearch instance = new SQLSearch("person", colums);
        String expResult = "person";
        String result = instance.getSearchObjectName();
        assertEquals(expResult, result);
    }

    /**
     * Test of toString method, of class SQLSearch.
     * If string starts "The query is: SELECT * FROM person WHERE" than test will be
passed.
     */
    @Test
    public void testToString() throws SearchException {
        System.out.println("toString");
        String what = "a";

        String[] colums = {"FirstName"};
```

```
        SQLSearch instance = new SQLSearch("person", colums);

        instance.search(what);
        String expResult = "The query is: SELECT * FROM person WHERE";
        String result = instance.toString();

        assertTrue(result.contains(expResult));

    }

}
```

Results of unit tests for class SQLSearch presented below.

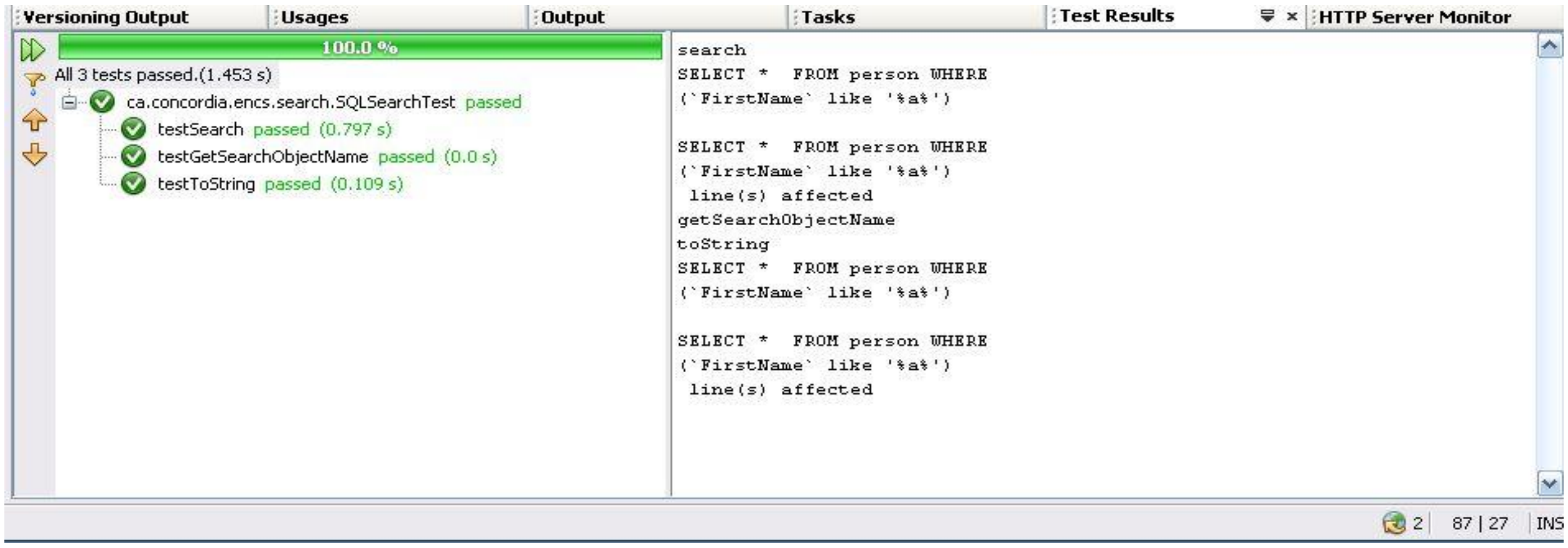

**Figure 54.** Results of unit tests for class SQLSearch

**Class SQLqueryTest.**

```
package ca.concordia.encs.search;

import java.sql.Connection;
import org.junit.Before;
import org.junit.Test;
import static org.junit.Assert.*;

/**
 *
 * @author PavelLepin
 */
public class SQLqueryTest
{

    public SQLqueryTest()
    {
    }
```

```
    SQLquery instance;
    Connection connection;
    @Before
    public void beforeSetConnection()
    {
        instance = new SQLquery();
    }
    /**
     * Test of setConnection method, of class SQLquery.
     */
    @Test
    public void testSetConnection()
     {
        System.out.println("setConnection");
        Boolean expResult = true;
        Boolean result = instance.setConnection();
        assertEquals(expResult, result);

    }

    /**
     * Test of getTable method, of class SQLquery.
     */
    @Test
    public void testGetTable()
      {
        System.out.println("getTable");
        String SQL = "SELECT TABLE_NAME from INFORMATION_SCHEMA.TABLES
where TABLE_NAME = 'SearchinPersons'";
        String result = "";
        for (String[] row : instance.getTable(SQL))
      {
            result += row[0];
        }
        String expResult = "SearchinPersons";
        System.out.println(result);
        System.out.println(expResult);
        assertEquals(expResult, result);
        // TODO review the generated test code and remove the default call to fail.
    }

    /**
     * Test of getMyConnection and setMyConnection methods, of class SQLquery.
     */
    @Test
    public void testGetMyConnection()
      {
        System.out.println("getMyConnection");
        connection = null;
```

```
        instance.setMyConnection(connection);
        Connection result = instance.getMyConnection();
        assertEquals(connection, result);

    }
}
```

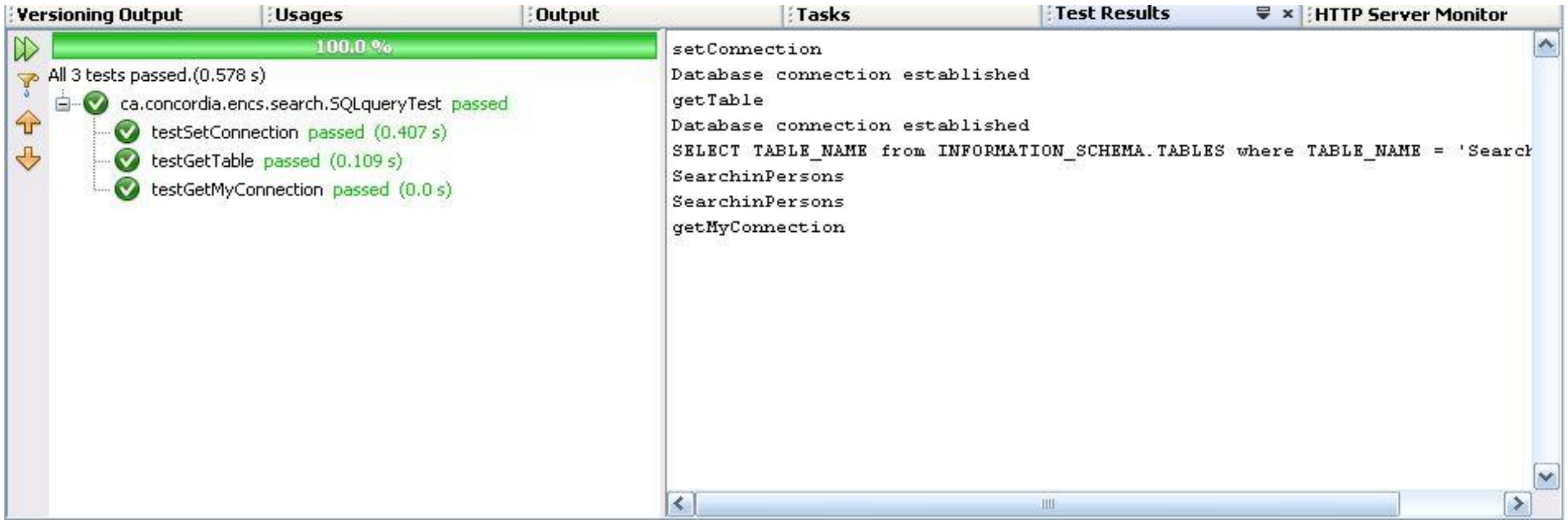

**Figure 55.** Results of unit tests for class SQLqueryTes

### Add new asset

Two methods "testSetBarcodeNum" and "testGetBarcodeNum" in the Create Asset Module were tested by giving a specific value to the input variable "barcodeNum" , writing the expected result and then comparing it to the actual result by calling "assertEquals (expResult, result)". No error occurred for both methods, and thus the JUnit test results were : "Passed".

The Test codes :

```
    public void testSetBarcodeNum() {

        System.out.println("setBarcodeNum");

        String barcodeNum = "compp5216843";

        instance.setBarcodeNum(barcodeNum);

        assertEquals (barcodeNum,"compp5216843") ;


        // TODO review the generated test code and remove the default call to fail.

    }

    @Test
```

```
public void testGetBarcodeNum() {
    System.out.println("getBarcodeNum");
    String expResult = "compp5216843";
    instance.setBarcodeNum("compp5216843");
    String result = instance.getBarcodeNum();
    assertEquals(expResult, result);
    // TODO review the generated test code and remove the default call to fail.
}
```

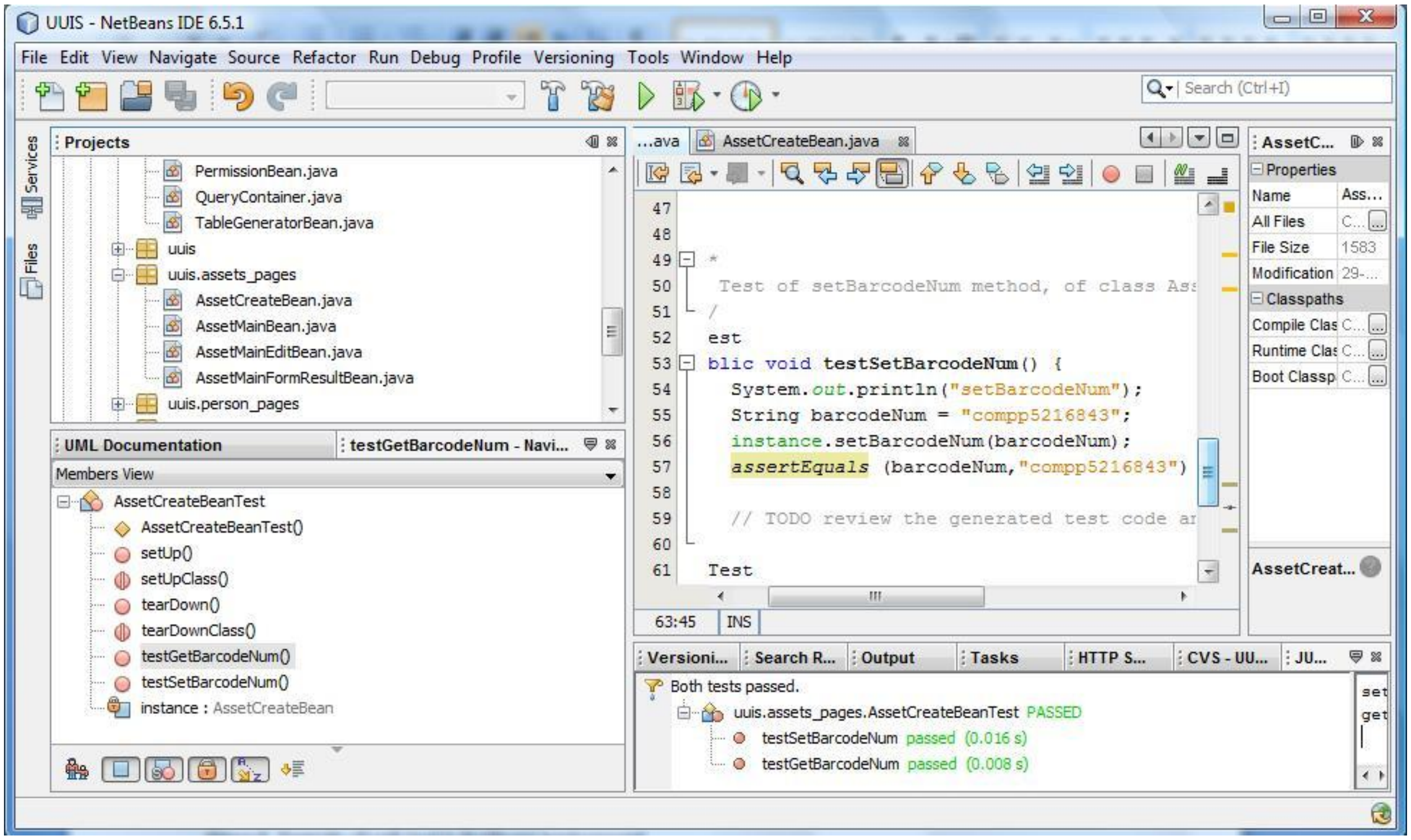

**Figure 56.** Unit test for method “testSetBarcodeNum()” in module “Add new asset”.

### Add new request

A method “testSetLocationList” in the Request Module was tested by giving a specific value to the input variable “username” , writing the expected result and then comparing it to the actual result by calling “assertEquals (expResult, result)”. No error occurred, and thus the JUnit test results were : “Passed”.

The test Code is:

```
@Test
```

```
public void testSetLocationList() {
    System.out.println("setLocationList");
    String username = "test";
    RequestMainBean instance = new RequestMainBean();
    String[][] expResult = new String[1][2];
    expResult[0][0] = "12";
    expResult[0][1]  ="65466";
    String[][] result = instance.setLocationList(username);
    assertEquals(expResult, result);
    // TODO review the generated test code and remove the default call to fail
}
```

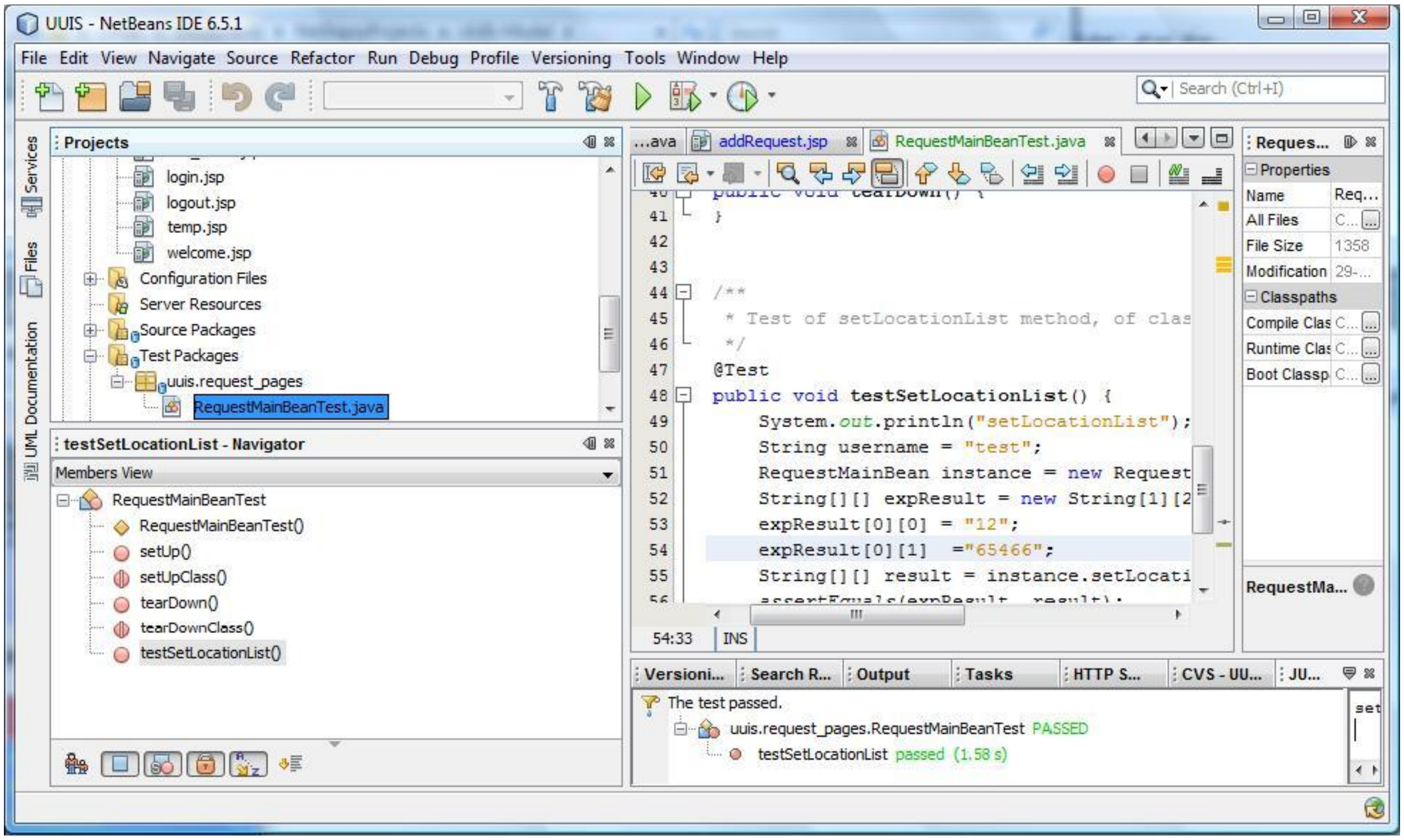

**Figure 57.** Unit test for method “testSetLocationsList()” in module “Add new request”.

# F. Test Cases

## F.1. Operational Testing

**Table 6.** Test case 1.1

| | |
|---|---|
| Test Case Number | 1.1 |
| Test Case Name | *Login* |
| Test Case Description | This test case verifies if a user can properly log into the system and see the menu. |
| Preconditions | N/A |
| Test Case Input | name: string<br>password: string |
| Test Case Expected Output | Display Home page & List of menu |
| Test Case Steps: | |
| Login | |
| | 1: Input:<br>name: string<br>password: string<br>Output:<br>Permission: string |
| Show Menu | |
| | 2: Input:<br>username: string<br>permission: string<br>Output:<br>List of menu |

**Table 7.** Test case 1.2

| | |
|---|---|
| Test Case Number | 1.2 |
| Test Case Name | *Report* |
| Test Case Description | This test case |
| Preconditions | User logged in. Report page. |
| Test Case Input | Username<br>Permissions<br>Name of report to be built |
| Test Case Expected Output | Report as a HTML table |
| Test Case Steps: | |
| Report | |
| | 1: Input:<br>Username<br>Permissions<br>Name of report to be built<br>Output:<br>Report as a HTML table |

**Table 8.** Test case 1.3

| | |
|---|---|
| Test Case Number | 1.3 |
| Test Case Name | *Advanced Search Model* |
| Test Case Description | This test case verifies if a user searches for a string the system shows all the items in Database including that string. |
| Preconditions | User logged in. Advanced search page. |
| Test Case Input | String to be found |
| | List of tables & List of columns |
| Test Case Expected Output | HTML table |
| Test Case Steps: | |
| Query | |
| | 1: Input:<br>String to be found<br>List of tables with<br>List of columns<br>Output:<br>SQL query: string |
| Connection | |
| | 2: Input:<br>address of SQL server: string<br>Output:<br>TRUE if connected |
| Retrieve Data | |
| | 3: Input:<br>Data from SQL<br>Output:<br>Array of string |
| Display HTML Table | |
| | 4: Input:<br>Array of string<br>Output:<br>HTML table |

## F.2. Data Integrity Testing

**Table 9.** Test case 2.1

| | |
|---|---|
| Test Case Number | 2.1 |
| Test Case Name | *Saved Data Integrity* |
| Test Case Description | This test case verifies if the data saved when the window is closed by double clicking on the close box? |
| Preconditions | Logged in user. |
| Test Case Input | Closing a window. |
| Test Case Expected Output | Saving data in that window |
| Test Case Steps: | |
| Saved Data Integrity | |
| | 1: Input:<br>Closing a window.<br><br>Output:<br>Saving data in that window. |

**Table 10**. Test case 2.2

| | |
|---|---|
| Test Case Number | 2.2 |
| Test Case Name | *Maximum character length* |
| Test Case Description | This test case checks the maximum field lengths to ensure that there are no truncated characters. |
| Preconditions | Logged in user needed to enter a string. |
| Test Case Input | Enter a large-length string. |
| Test Case Expected Output | Temporarily saved large-length string for further actions. |
| Test Case Steps: | |
| Maximum Character Length | |
| | 1: Input:<br>Enter a large-length string in the search text box.<br>Output:<br>System could keep it in a variable to search for it without losing any substring. |

**Table 11.** Test case 2.3

| | |
|---|---|
| Test Case Number | 2.3 |
| Test Case Name | *Default value preserving* |
| Test Case Description | This test case verifies where the database requires a value (other than null) then this should be defaulted into fields. The user must either enter an alternative valid value or leave the default value intact. |
| Preconditions | User logged in. |
| Test Case Input | Leave a field with default value. |
| Test Case Expected Output | Display the proper default value. |
| Test Case Steps: | |
| Default Value Preserving | |
| | 1: Input:<br>User does not enter any value for a field which is supposed to load with a default value.<br>Output:<br>Field is displayed with its default value. |

**Table 12.** Test case 2.4.1

| | |
|---|---|
| Test Case Number | 2.4.1 |
| Test Case Name | *Numeric value range* |
| Test Case Description | This test case checks maximum and minimum field values for numeric fields. |
| Preconditions | User logged in. |
| Test Case Input | Enter a numeric value not greater than the maximum limit. |
| Test Case Expected Output | System can handle the value without any problem. |
| Test Case Steps: | |
| Maximum Numeric value | |
| | 1: Input:<br>Enter a large numeric value not greater than maximum limit<br>Output:<br>System can handle the value (Variable type is defined properly). |

**Table 13.** Test case 2.4.2

| | |
|---|---|
| Test Case Number | 2.4.2 |
| Test Case Name | *Numeric value range* |
| Test Case Description | This test case checks maximum and minimum field values for numeric fields. |
| Preconditions | User logged in. |
| Test Case Input | Enter a numeric value not less than the minimum limit. |
| Test Case Expected Output | System can handle the value without any problem. |
| Test Case Steps: | |
| Minimum Numeric value | |
| | 1: Input:<br>Enter a small numeric value not less than the minimum value.<br>Output:<br>System can handle the value (variable type is defined properly). |

**Table 14.** Test case 2.5.1

| | |
|---|---|
| Test Case Number | 2.5.1 |
| Test Case Name | *Avoid truncation of string and rounding of numeric value.* |
| Test Case Description | This test case verifies If a particular set of data is saved to the database check that each value gets saved fully to the database. i.e. Beware of truncation (of strings) and rounding of numeric values. |
| Preconditions | User logged in. |
| Test Case Input | Enter an asset attributes. |
| Test Case Expected Output | Displays the modifications in the Database. |
| Test Case Steps: | |
| Avoid truncation of string values | |
| | 1: Input: Enter an asset attributes.<br>Output:<br>Displays the modifications in the database without losing any character. |

**Table 15.** Test case 2.5.2

| | |
|---|---|
| Test Case Number | 2.5.2 |
| Test Case Name | *Avoid truncation of string and rounding of numeric value.* |
| Test Case Description | This test case verifies If a particular set of data is saved to the database check that each value gets saved fully to the database. i.e. Beware of truncation (of strings) and rounding of numeric values. |
| Preconditions | User logged in. |
| Test Case Input | Enter a room size |
| Test Case Expected Output | Displays the modification in the Databse. |
| Test Case Steps: | |
| Avoid rounding of numeric values | |
| | 1: Input:<br>Enter a set of numeric values for a room size.<br>Output:<br>Displays all the actual values in the Database without losing anything. |

## F.3. Graphical User Interface Testing

**Table 16.** Test case 3.1

| | |
|---|---|
| Test Case Number | 3.1 |
| Test Case Name | *Application Compliance Standard* |
| Test Case Description | Closing the application should result in an "Are you Sure" message box |
| Preconditions | User logged in. |
| Test Case Status | Failed for main page. |
| Test Case Attempt: | A user logged in as a test user. By clicking on the close box, the application terminates without any notice! |

**Table 17.** Test case 3.2

| | |
|---|---|
| Test Case Number | 3.2 |
| Test Case Name | *Help Menu* |
| Test Case Description | All screens should have a Help button, F1 should work doing the same. |
| Preconditions | User logged in. |
| Test Case Status | Failed. |
| Test Case Attempt: | Pressing F1 in all the pages has no result! Moreover, there is no Help button. |

**Table 18.** Test case 3.3

| | |
|---|---|
| Test Case Number | 3.3 |
| Test Case Name | *List Box Color* |
| Test Case Description | List boxes are always white background with black text whether they are disabled or not. All others are grey. |
| Preconditions | User logged in. |
| Test Case Status | Passed. |
| Test Case Attempt: | In the search menu, there are list boxes with the correct color. Moreover, in the main page database attributes of each entity are written in black color with white background. |

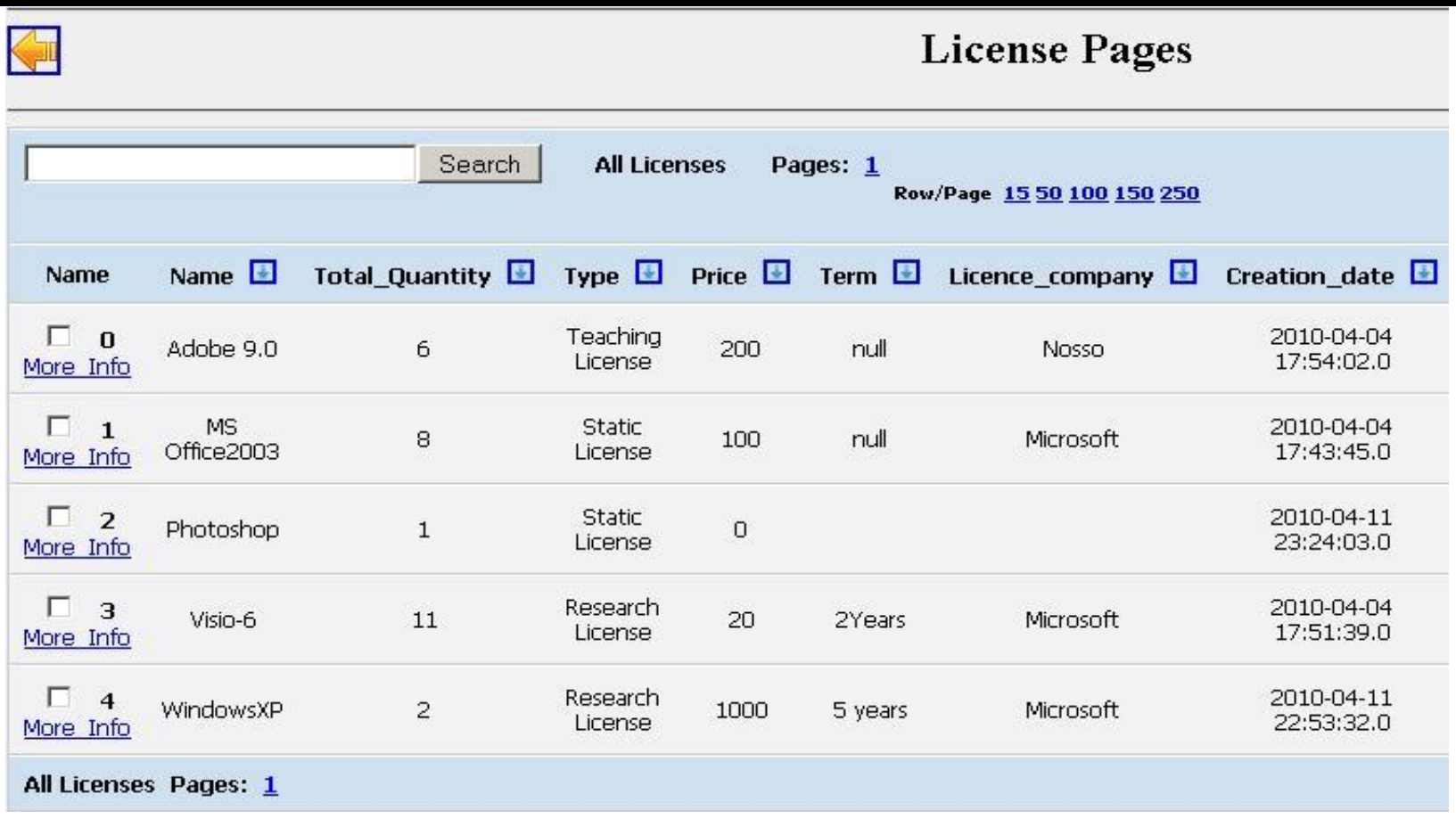

**Table 19.** Test case 3.4

| | |
|---|---|
| Test Case Number | 3.4 |
| Test Case Name | *Curser over an enterable text box* |
| Test Case Description | Move the Mouse Cursor over all Enterable Text Boxes.<br>Cursor should change from arrow to Insert Bar.<br>If it doesn't then the text in the box should be grey or non-updateable. Refer to previous page. |
| Preconditions | User logged in. |
| Test Case Status | Passed. |
| Test Case Attempt: | All over the search text box, curser always changes from arrow to insert bar.<br>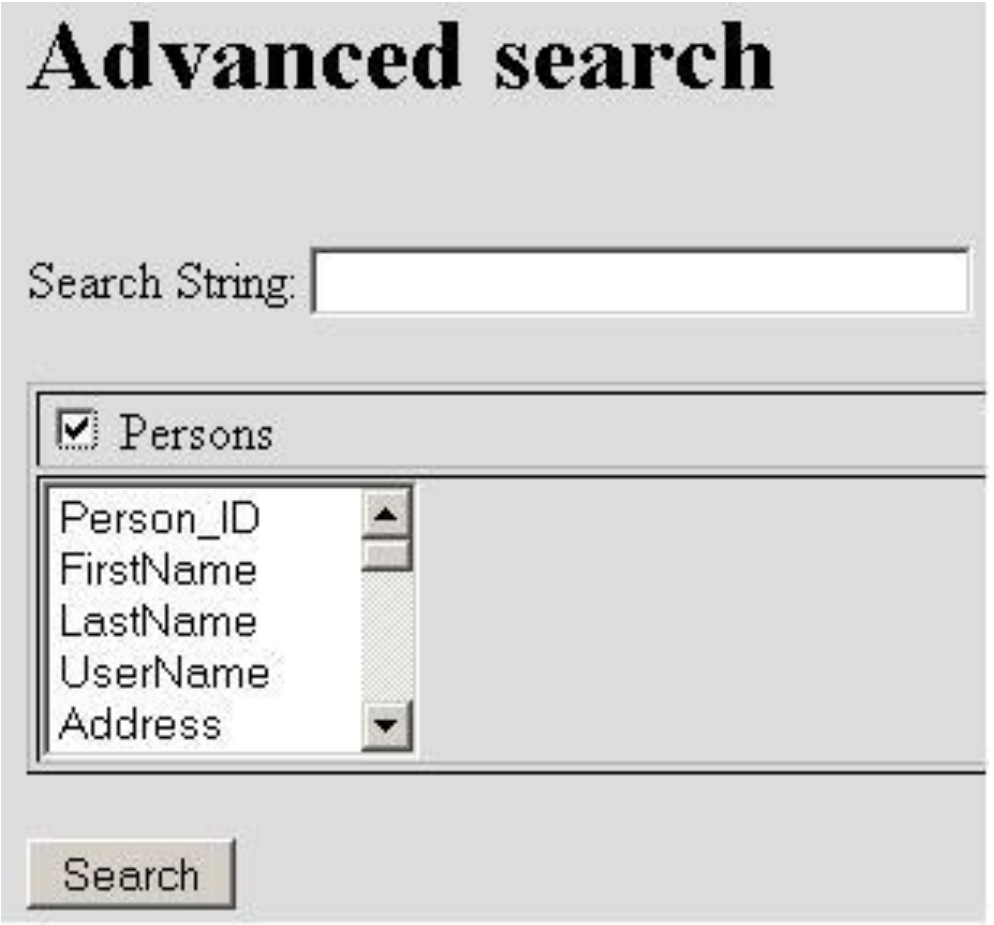 |
| | 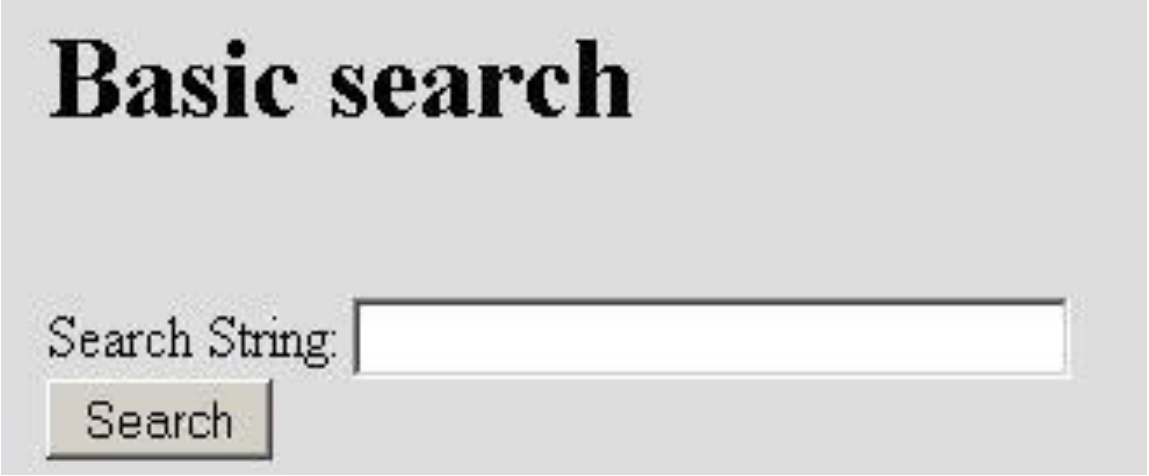  |

**Table 20.** Test case 3.5

| | |
|---|---|
| Test Case Number | 3.5 |
| Test Case Name | *Text Box Characters* |
| Test Case Description | Try to overflow the text by typing to many characters - should be stopped Check the field width with capitals W.<br>Enter invalid characters - Letters in amount fields, try strange characters like + , - * etc. in All fields.<br>SHIFT and Arrow should Select Characters. Selection should also be possible with mouse. Double Click should select all text in box. |
| Preconditions | User logged in. |
| Test Case Status | Passed, Failed, Passed. |
| Test Case Attempt: | In some of the fields as attributes of entities in the database, strange characters are accepted. |

**Table 21.** Test case 3.6

| | |
|---|---|
| Test Case Number | 3.6 |
| Test Case Name | *Color validation* |
| Test Case Description | Is the general screen background the correct colour?<br><br>Are the field prompts the correct colour?<br><br>Are the field backgrounds the correct color?<br><br>In read-only mode, are the field prompts the correct color?<br><br>In read-only mode, are the field backgrounds the correct color? |
| Preconditions | User logged in. |
| Test Case Status | Passed. |
| Test Case Attempt: | There is no strange color used in the screen, background and text. Read-only and deactive colors are from less sharpness. |

**Table 22.** Test case 3.7

| Test Case Number | 3.7 |
|---|---|
| Test Case Name | *Text validation* |
| Test Case Description | Are all the field prompts spelt correctly?<br><br>Are all character or alpha-numeric fields left justified? This is the default unless otherwise specified.<br><br>Are all numeric fields right justified? This is the default unless otherwise specified.<br><br>Is all the microhelp text spelt correctly on this screen?<br><br>Is all the error message text spelt correctly on this screen?<br><br>Is all user input captured in UPPER case or lower case |
| | consistently? |
| Preconditions | User logged in. |
| Test Case Status | Passed. |
| Test Case Attempt: | All the prompts are spelt correctly.<br>Alpha-numeric characters are left justified.<br>Numeric fields are right justified.<br>User inputs are captured in lower case. |

**Table 23.** Test case 3.8

| Test Case Number | 3.8 |
|---|---|
| Test Case Name | *Alignment* |
| Test Case Description | Are all the field prompts aligned perfectly on the screen?<br>Are all the field edit boxes aligned perfectly on the screen?<br>Are all group boxes aligned correctly on the screen? |
| Preconditions | User logged in. |
| Test Case Status | Passed. |
| Test Case Attempt: | Very likely all the users enjoy the simple and user-friendly interface of the application along with the right placement of options, buttons and boxes. |

# F.4. Web Application Security Testing

## F.4.1. Web Application Security Test Case # 1:

### Command Line

./skipfish -o result https://spec111.encs.concordia.ca:8443/UUIS/

This is the most basic command line in the skipfish window. It scan the application with the most general default settings.

By this command, skipfish scanner saves the results of scan in the ‘result’ folder which comes after ‘-o’ in the command line.

-o dir - write output to specified directory (required)

Options are followed by the target URL of the application in the command line.

### Statistics

Once executing the command line of skipfish scanner starts scanning and saving the results in the specified folder. Meanwhile, real-time statistics of scanning are shown in the Linux command line including the following information:

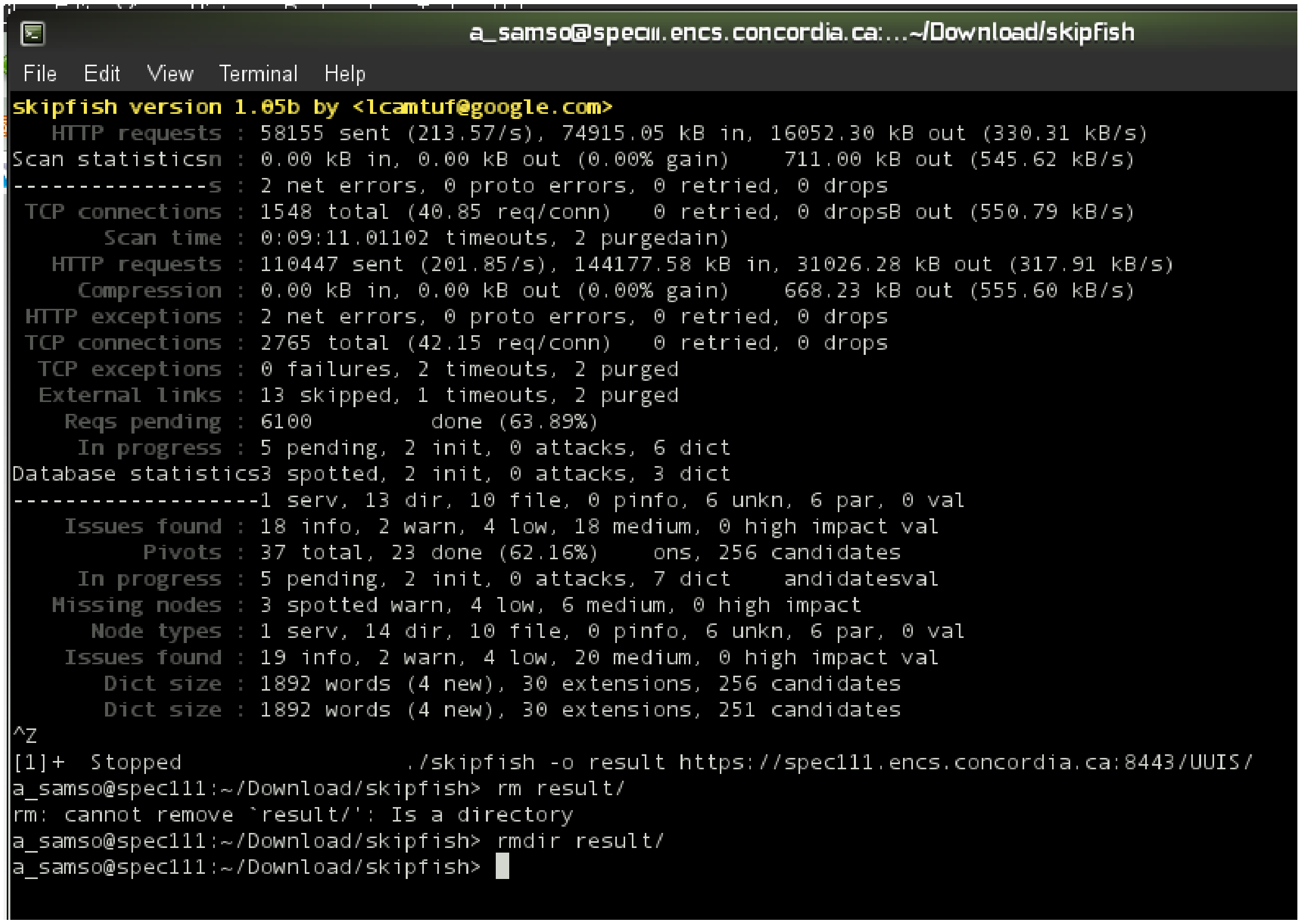

**Figure 58.** Statistics for Security Test Case#1

As the first try, having all the default settings first scan case lasts for more than nine minutes (shown in the above figure) without achieving any result. It seems based on the default settings; maximum number of HTTP requests to send is 100,000,000 requests totally.

For the first try as shown in the figure 4.1, we decided to stop scanning process once almost 60,000 requests had been sent. (58,155). That is why we do not have the HTML result for this very first try but from the statistics shown we reached the following information those are pretty promising:

Only 2 net errors are found totally.

Out of 2,765 TCP requests, there is no failure so far along with two timeouts and two purged.

Complexity of the application in terms of handling HTTP requests seems linear because almost 64% of all the requests have reached results and only 6,100 are pending after more than 9 minutes. Of course this comes from the skipfish efficiency as well (as a Google application).

## F.4.2. Web Application Security Test Case # 2:

**Command Line**

./skipfish -g 10 -m 3 –t 5 -o result https://spec111.encs.concordia.ca:8443 /UUIS/

Having changed some of the settings, we have tried the above command line as our second try. This time we changes parameters responding to performance settings as follow:

-g max_conn -It specifies maximum number of simultaneous TCP connections (default 50).

-m host_conn -It specifies maximum number of simultaneous connections per target IP(def. 10).

-t req_tmout -It specifies total request response timeouts(default 20s).

As a result, the above command line tries maximum of 10 simultaneous TCP connections for the whole application along with maximum of three per IP target.

Although the above settings facilitates scanning procedure and does not wait more than 5 seconds for pending requests (as opposed to 20 in the previous try) still the total number of requests are high enough to force us interrupt scanning and see the statistics without results.

## Statistics

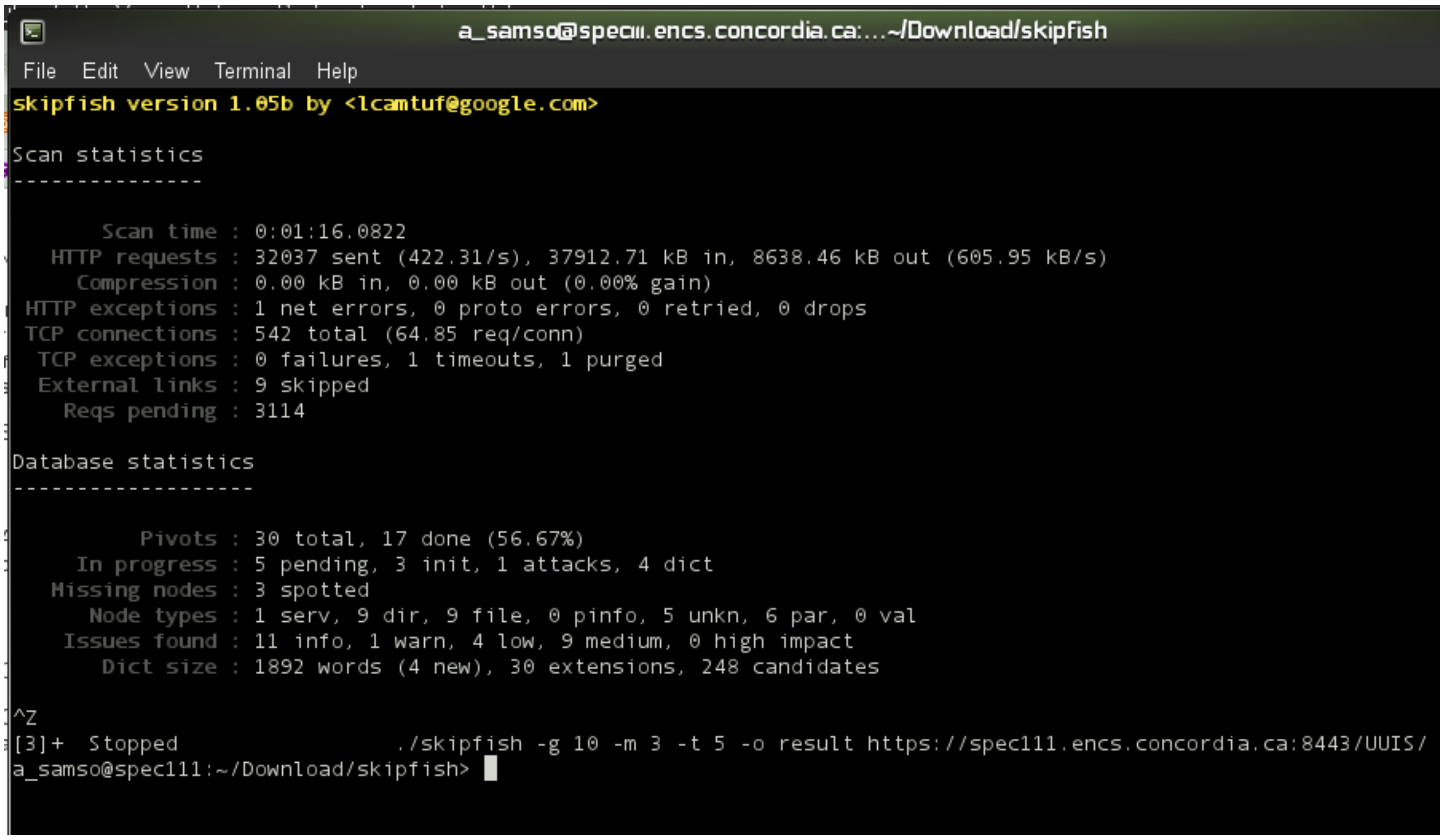

**Figure 59.** Statistics for Security Test Case# 2

Total number of 32,837 requests during only 1:16 sec time shows clearly that skipfish scanner could reach 100,000,000 requests very sooner than the previous try since it reached almost 58,155 requests in 9:11 sec. But still it needed to wait a long time to finish given the 5 seconds timeout that defined for pending requests in this try.

### F.4.3. Web Application Security Test Case # 3:

#### Command Line

./skipfish -A user:test -d 16 -c 1024 -r 200 -o result1 https://spec111.encs.concordia.ca:8443/UUIS/

In the test case three, in order to achieve HTML results we limited maximum number of requests to send to only 200 requests! Crawl options that have been defined are as follow in the above command line:

-d max_depth - It specifies maximum crawl tree depth (default:16).

-c max_child - It specifies maximum children to index per node (default:1024).

-r r_limit -It specifies max total number of requests to send (100,000,000).

Skipfish has the capability to specify authentication and access information for the scanner. As an option we specified a user for the scanner to scan along with that user. As a predefined user we have 'test' who has a role of administration and granted permissions of university administration (a level three of highest administration level).

-A user: pass - use specified HTTP authentication credentials.

**Statistics**

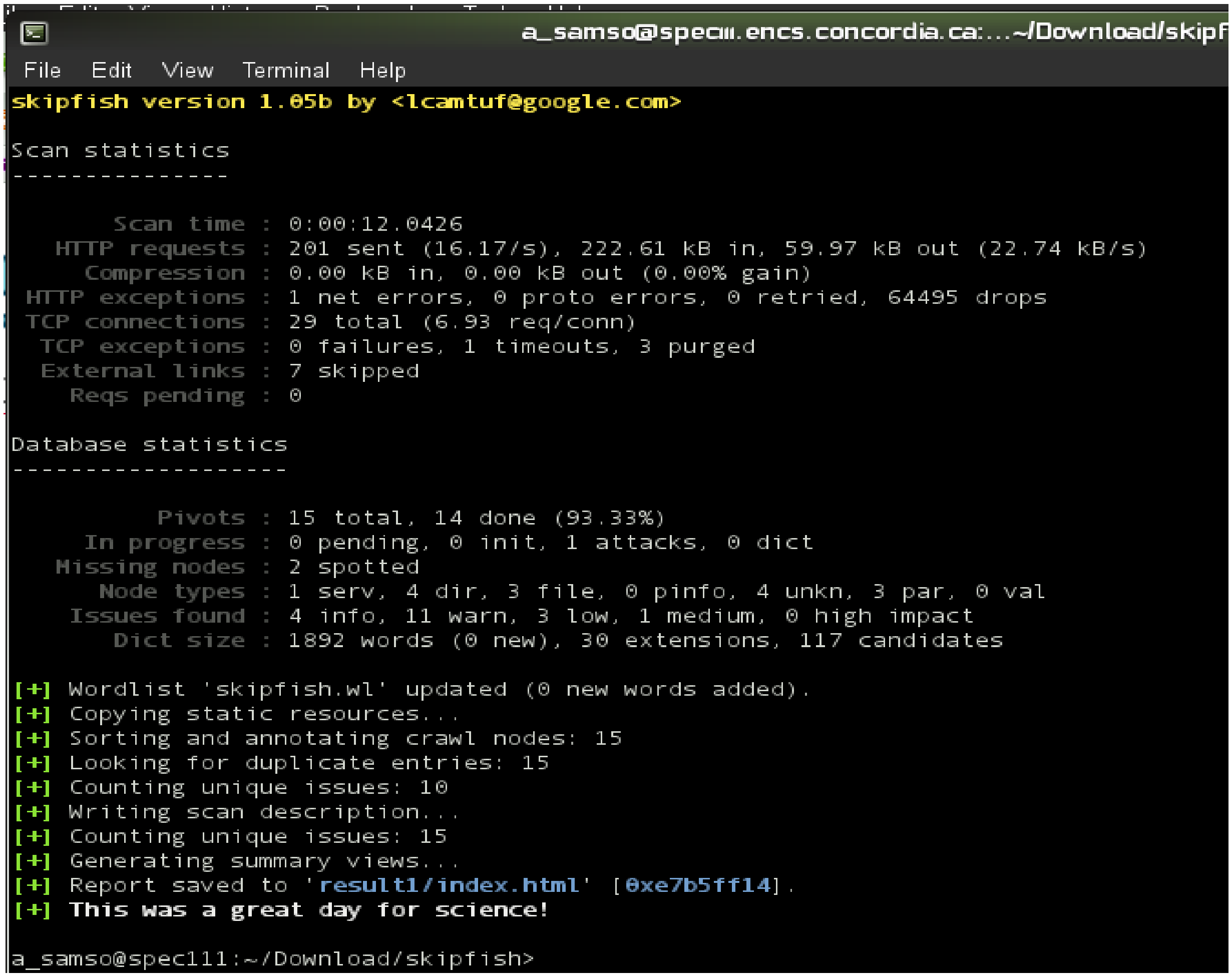

**Figure 60.** Statistics for Security Test Case# 3

This was a great day for science!

As specified in the command line 200 requests (shown in the statistics page 201 requests) have been sent in 12 seconds from which 222.6 KB of data has been entered to the system as inputs and 59.97KB of data has been received as outputs.

One network error has been found and totally 29 TCP connection requests have been sent. Given total number of 200 HTTP requests, 6.93 requests per TCP connection have been sent out from which no failure occurred.

Statistics related to database are pretty meaning full showing a correct and perfect database design for the application.

Total number of 14 pivots out of 15 is done in 12 seconds. The one which has not been complete yet is considered as an attack.

Each pivot might be considered as informative, warning, low-risk, medium-risk and/or high-risk issue. In this test case, 4 tested pivots are reported informative, 11 are reported warnings. 3 low-risks, 1 medium-risk and no high-risk issues have been reported in terms of their impact on database as an attack. Database dictionary has a total number of 1892 words.

**Result**

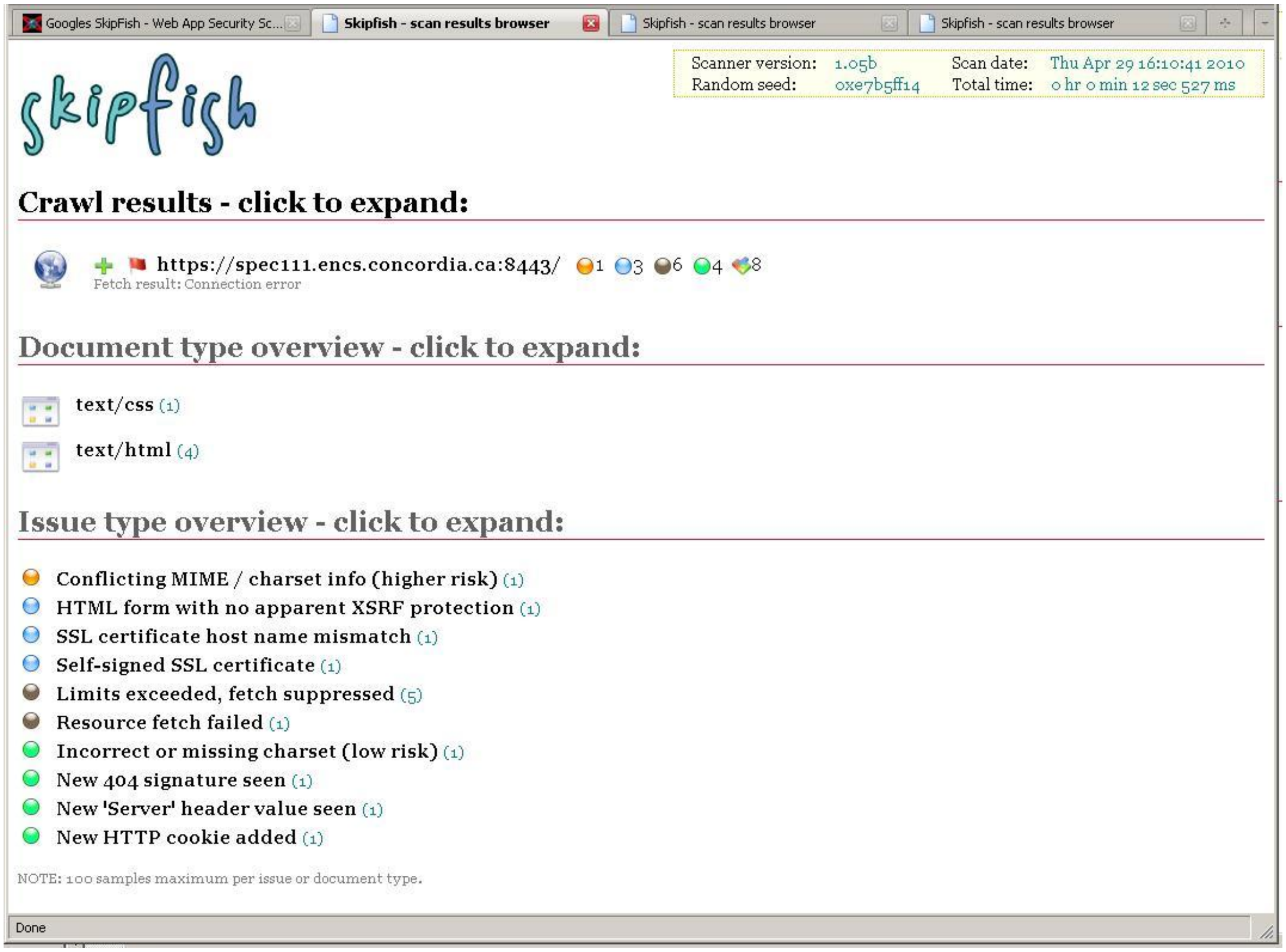

**Figure 61.** Results for Security Test Case# 3

General view of the Crawl result includes scanner version, scanning date and time, random seed and total scanning time at the top right part of the page.

The connection error specified in the statistics page is shown here as a 'Fetch result: Connection error ' in the first part of the HTML page. It shows also number of each type of warnings, errors and risks of the last part.

In the last part of the HTML report it is shown that one incorrect or missing character set from low risk of attack impact is found as well as one SSL certificate host name mismatch. The former was expected since the application has not been registered officially. Figure below shows the expansion of the scan report in this area:

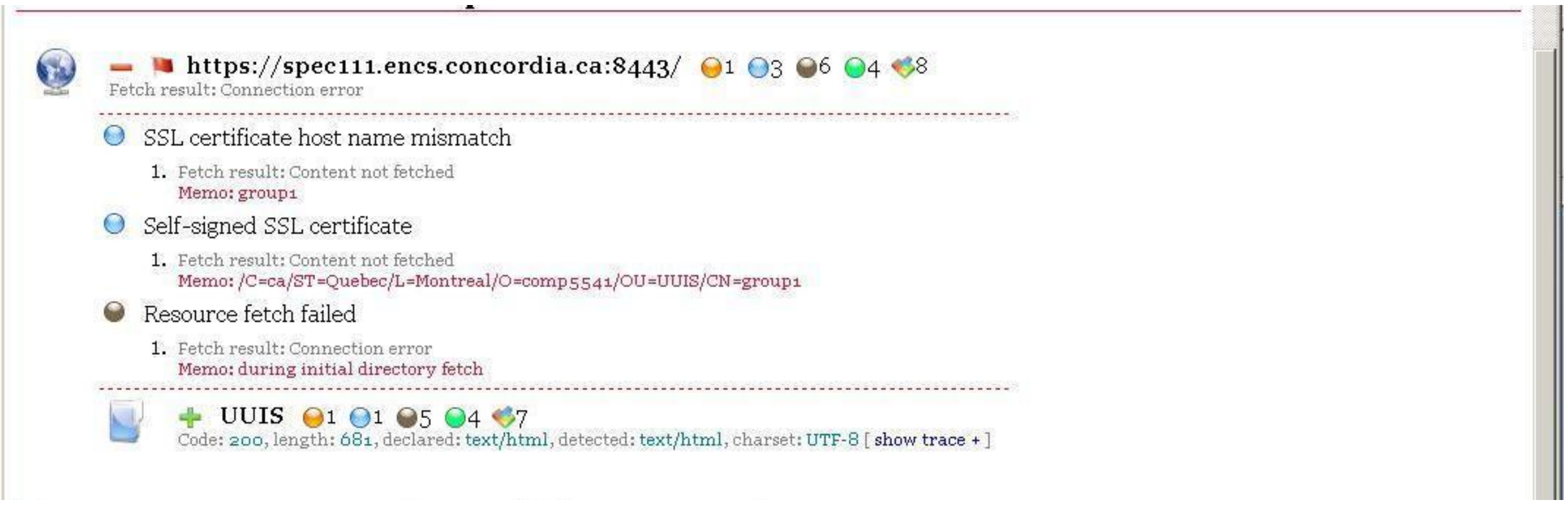

**Figure 62.** Expansion of the scan report

A total number of five limits exceeding and/or fetch suppressed are found that shows the main vulnerable part of the application (shown in figure below).

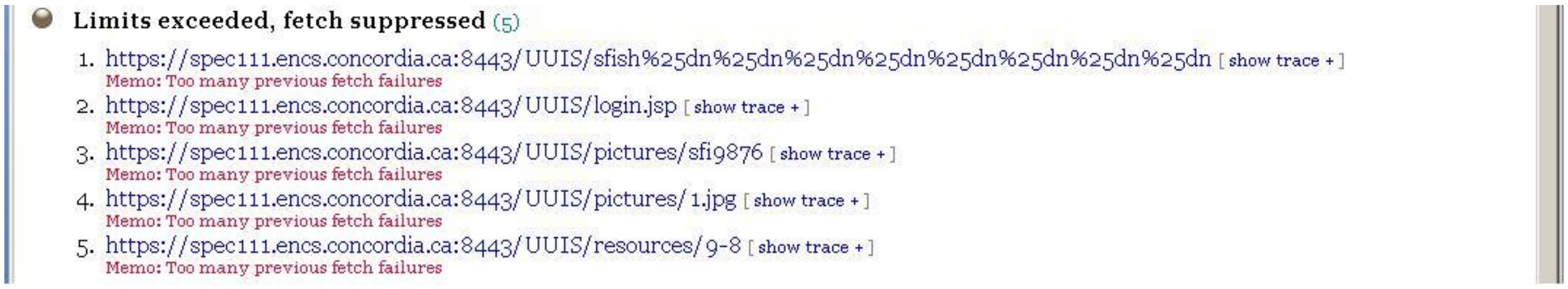

**Figure 63.** Main vulnerable part of the application

## F.4.4. Web Application Security Test Case # 4:

### Command Line

./skipfish -J -r 500 -o result3 https://spec111.encs.concordia.ca:8443/ UUIS/

In the 4th try, we decided to try more requests along with default settings which seem the most common real life situations and have results in folder result 3.

That is why a total number of 500 requests have been set to send.

As a reporting option we added "–J" to the command line:

-J - It specifies to be less noisy about MIME / character set mismatches in the report.

**Statistics**

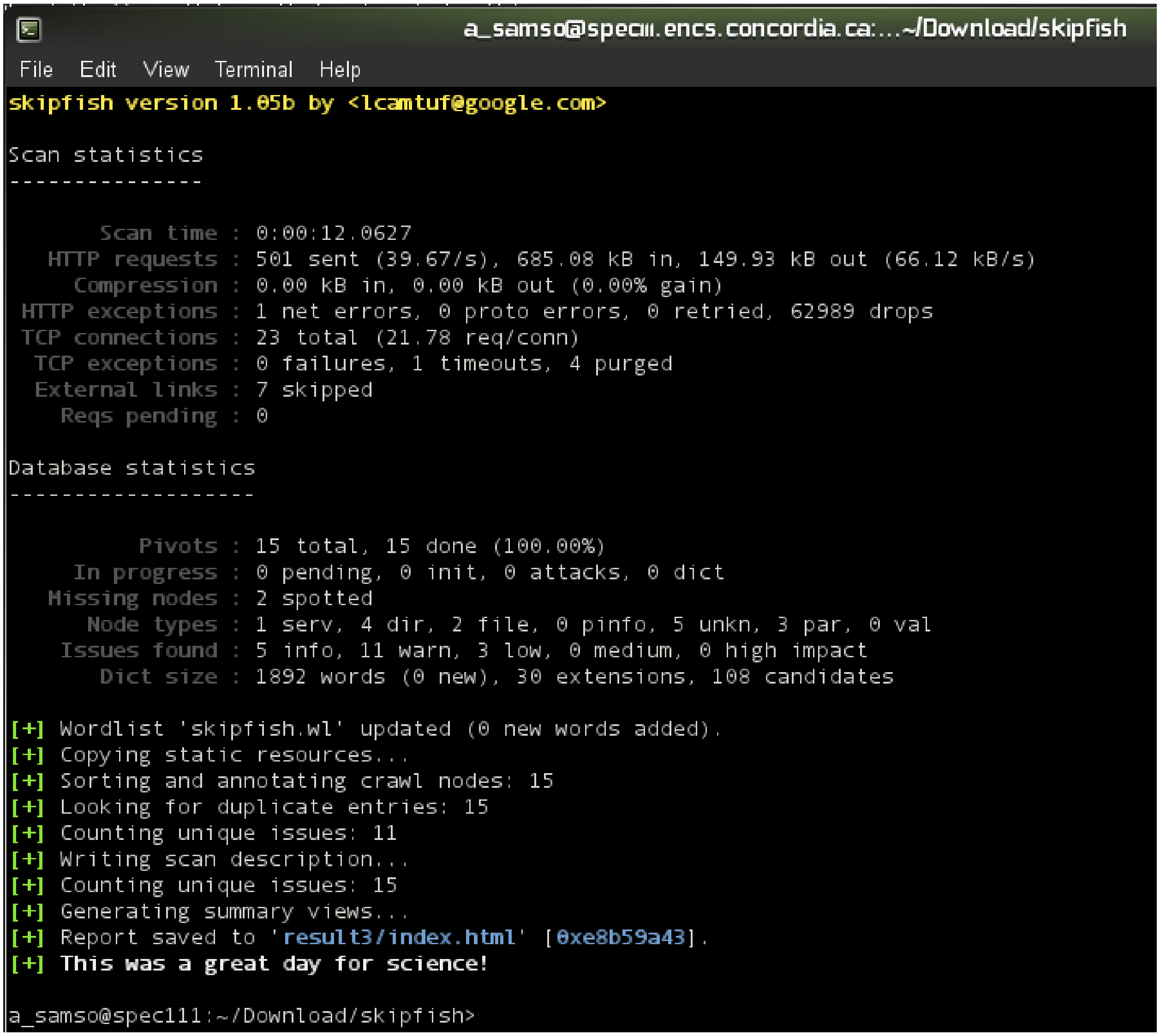

```
a_samso@specııı.encs.concordia.ca:...~/Download/skipfish
File  Edit  View  Terminal  Help
skipfish version 1.05b by <lcamtuf@google.com>

Scan statistics
---------------

      Scan time : 0:00:12.0627
  HTTP requests : 501 sent (39.67/s), 685.08 kB in, 149.93 kB out (66.12 kB/s)
    Compression : 0.00 kB in, 0.00 kB out (0.00% gain)
HTTP exceptions : 1 net errors, 0 proto errors, 0 retried, 62989 drops
TCP connections : 23 total (21.78 req/conn)
 TCP exceptions : 0 failures, 1 timeouts, 4 purged
 External links : 7 skipped
   Reqs pending : 0

Database statistics
-------------------

         Pivots : 15 total, 15 done (100.00%)
    In progress : 0 pending, 0 init, 0 attacks, 0 dict
  Missing nodes : 2 spotted
     Node types : 1 serv, 4 dir, 2 file, 0 pinfo, 5 unkn, 3 par, 0 val
   Issues found : 5 info, 11 warn, 3 low, 0 medium, 0 high impact
      Dict size : 1892 words (0 new), 30 extensions, 108 candidates

[+] Wordlist 'skipfish.wl' updated (0 new words added).
[+] Copying static resources...
[+] Sorting and annotating crawl nodes: 15
[+] Looking for duplicate entries: 15
[+] Counting unique issues: 11
[+] Writing scan description...
[+] Counting unique issues: 15
[+] Generating summary views...
[+] Report saved to 'result3/index.html' [0xe8b59a43].
[+] This was a great day for science!

a_samso@spec111:~/Download/skipfish>
```

**Figure 64.** Statistics for Security Test Case# 4

According to the above statistics in comparison with previous tries, here we have a better speed of achieving results (500 requests in 12 seconds) and no failure is reported in TCP connections. Still we have one HTTP connection error.

Dictionary size stays similar to previous cases but all the pivots are fully performed. Since no medium-risk impact is reported, again database correct design is being confirmed by the test.

### Result

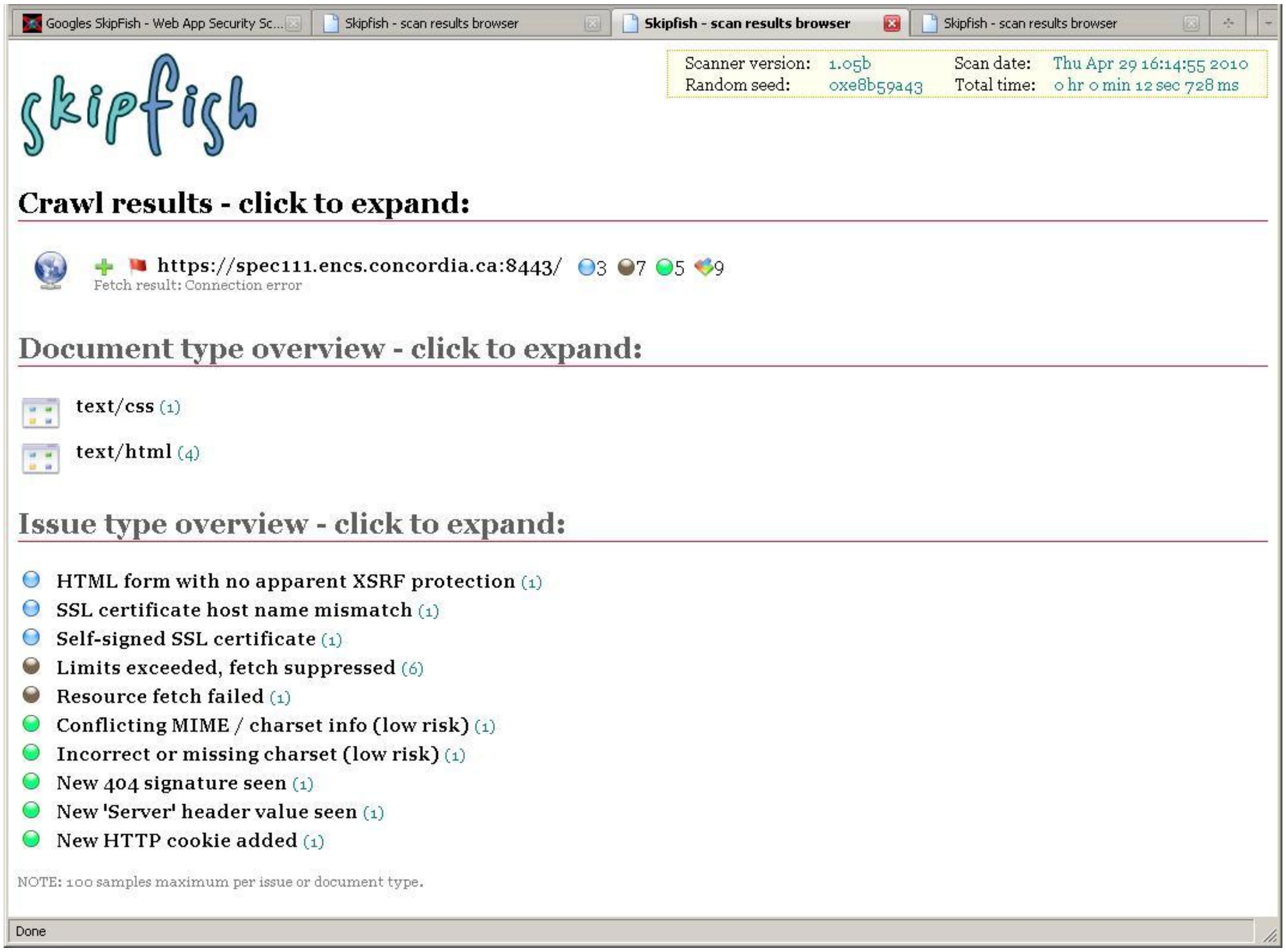

**Figure 65.** Result for Security Test Case# 4

## F.4.5. Web Application Security Test Case # 5:

### Command Line

./skipfish -J -L -V -r 1000 -o result4 https://spec111.encs.concordia.ca :8443/UUIS/

In the final try, we increased the number of requests to 1000 requests totally. But as dictionary management options we focused on the following options:

-L - do not auto-learn new keywords for the site.

-V - do not update wordlist based on scan results.

-J - It specifies to be less noisy about MIME / character set mismatches in the report.

**Statistics**

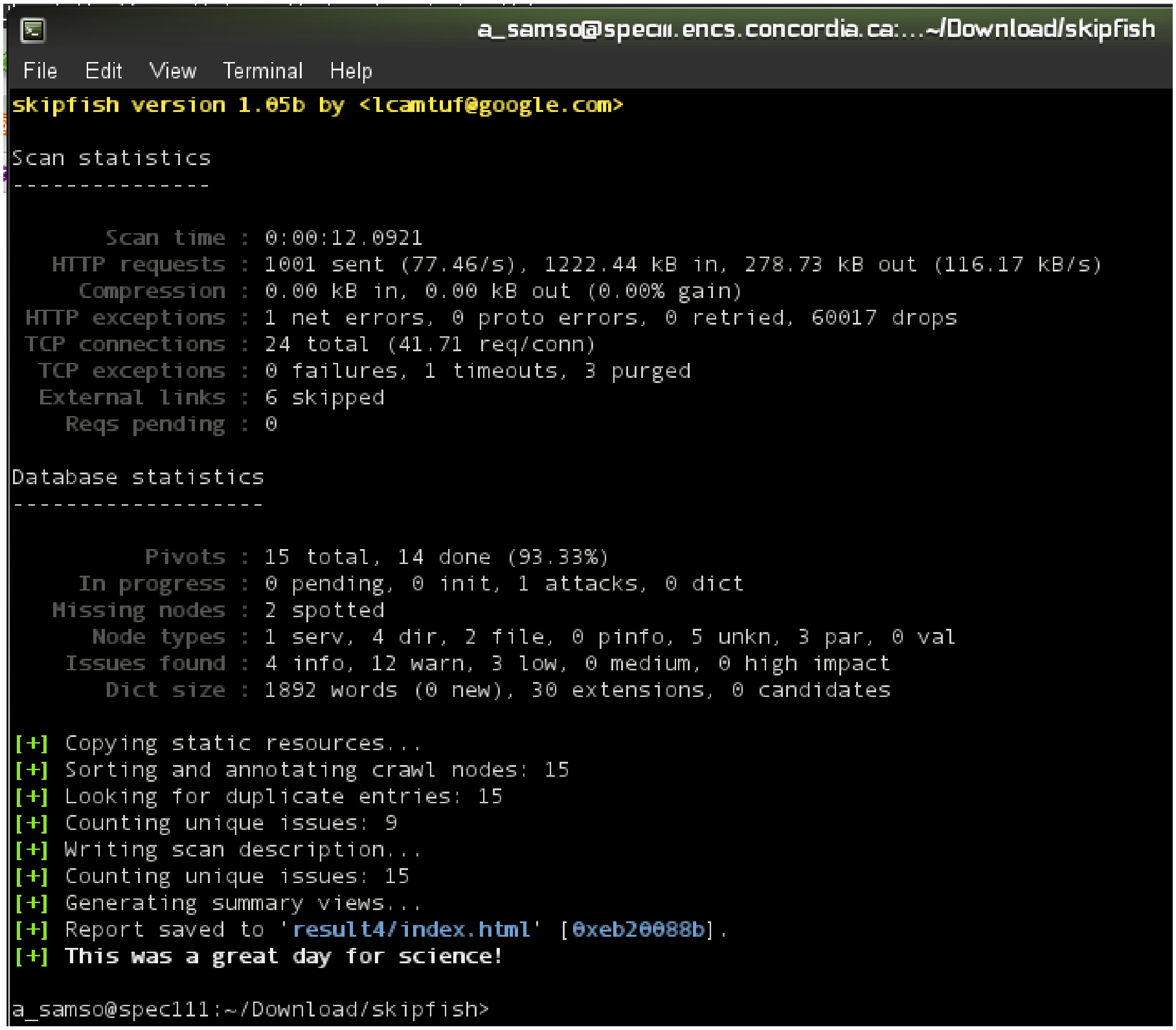

**Figure 66.** Statistics for Security Test Case# 5

According to the options for this command line, database is the main target for this test case.

Total number of 1000 requests has been sent to the application and one network error is reported.

In order to test the database 14 out of 15 pivots have been fully done from which one is reported as an attack.

In the database design, generally 4 informative issues, 12 warnings and 3 low-risk issues have been reported while no medium-risk neither high-risk issue is reported.

Since we added to dictionary management options of "-L" and "-V", decreasing the number of candidates for extensions of the dictionary from 117 in the 3rd test case and 108 in the 4th test case up to zero in this test case was expected. Total size of 1892 words for the dictionary stayed constant in this test case.

**Result**

Still we have one "Fetch result: Connection error". We have to do something for it!

But we do not have the "text/css" failure anymore.

Failures of the last test case are shown in the figure below.

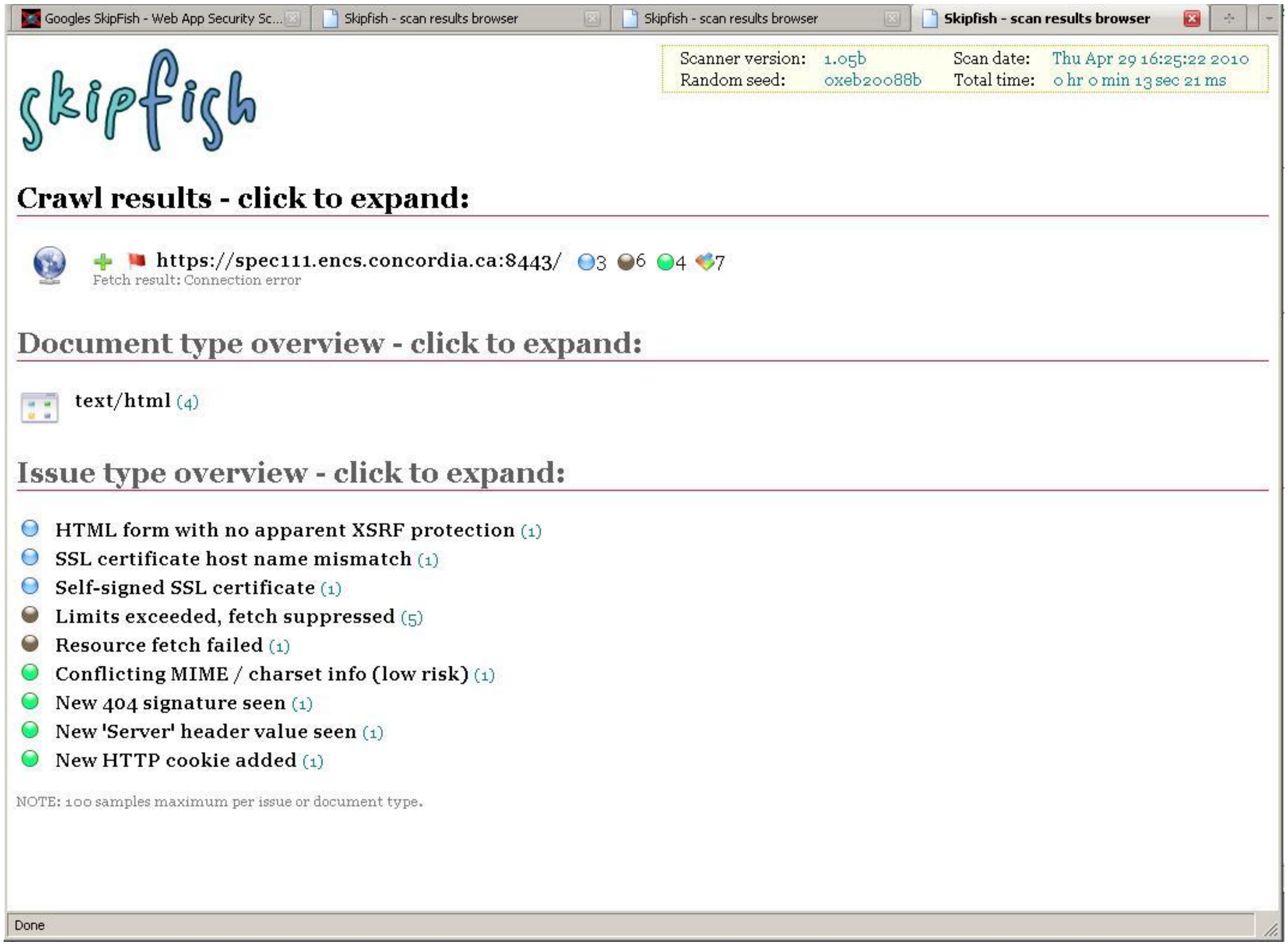

**Figure 67.** Result for Security Test Case# 5

# Glossary

**Table 24.** Glossary

| | |
|---|---|
| COCOMO II | COnstructive COst MOdel II (COCOMO II) is a model that allows one to estimate the cost, effort, and schedule when planning a new software development activity. |
| Dreamweaver | Dreamweaver is a professional visual editor for creating and managing Web sites and pages |
| FlamingText.com | FlamingText.com is a logo design tool where you can create custom logos and buttons |
| GUI | Graphical user interface |
| JUnit | Allows creating unit test for code, written on Java |
| MVC | Model – View – Controller |

# BIBLIOGRAPHY

Abstract Factory Pattern, UML diagram, Java Example. In *A&P Web Consulting.* Retrieved Mar. 25, 2010, from http://www.apwebco.com/gofpatterns/creational/AbstractFactory.html

Advantages and Disadvantages of Collection Framework. In *RoseIndia*. Retrieved Mar. 25, 2010, from http://www.roseindia.net/java/jdk6/collection-advantages-disadvantages.shtml

Design Pattern Variations: A Better Visitor. In *InformIt*. Retrieved Mar. 26, 2010, from http://www.informit.com/articles/article.aspx?p=1278987

Googles SkipFish – Web App Security Scanner. In *a4apphack*. Retrieved on March 28th 2010 from http://a4apphack.com/index.php/featured/googles-skipfish-web-app-security-scanner

IEEE 1016-1.pdf. In *essex*. Retrieved Mar. 25, 2010, from http://courses.essex.ac.uk/CE/CE201/restricted/projectDocuments/ieee1016-1.pdf

J2SE. In *Sun developer network*. Retrieved Mar. 26, 2010, from http://java.sun.com/j2se/1.5.0/docs/api/java/lang/Exception.html

MCV for beginners. In chtivo. Retrieved Apr. 5, 2010, from http://chtivo.webhost.ru/articles/mvc.php

Samples. In *Concordia University*. Retrieved Feb. 3 and 13, 2010, from http://users.encs.concordia.ca/~c55414/samples/

SDD. In *cmcrossroads*. Retrieved Mar. 25, 2010, from http://www.cmcrossroads.com/bradapp/docs/sdd.html

SDD. In *pdfqueen.* Retrieved Mar. 25, 2010, from

http://www.pdfqueen.com/html/aHR0cDovL3dpbG1hLnZ1Yi5hYy5iZS9+c2UxXzA4MDkvZG9jdW1lbnRzL2ZpcnN0Y3ljbGVfY29weS9TREQucGRm

Skipfish. In *Google*. Retrieved on March 25, 2010, from http://code.google.com/p/skipfish/

Skipfish: An Active Web Application Security Reconnaissance Tool! In *PenTestIT*. Retrieved on March 27, 2010 from http://pentestit.com/2010/03/20/skipfish-active-web-application-security-reconnaissance-tool/

SkipfishDoc. In *Google*. Retrieved on March 25, 2010 from http://code.google.com/p/skipfish/wiki/SkipfishDoc

Skipfish Updates In *Google*. Retrieved on April 12th, 2010 from http://code.google.com/p/skipfish/updates/list

Std_public. In *IEEE Standards Association.* Retrieved Mar. 25, 2010, from http://standards.ieee.org/reading/ieee/std_public/description/se/1016-1998_desc.html

User Interface Design Tips, Techniques, and Principles. In *ambysoft*. Retrieved Apr. 5, 2010, from http://www.ambysoft.com/essays/userInterfaceDesign.html

Visitor Pattern Versus Multimethods. In *Sourceforge.* Retrieved Mar. 26, 2010, from http://nice.sourceforge.net/visitor.html